\documentclass[12pt,a4paper]{iopart}
\usepackage[dvips]{graphicx}
\usepackage{float}
\usepackage{xcolor}

\usepackage{harvard}

\newcommand{\mean}[1]{\langle #1\rangle}
\newcommand{\eq}[1]{Eq.\ (\ref{#1})}
\newcommand{\Wcmsq}{\mbox{W cm}^{-2}}
\newcommand{\gcmsqu}{\mbox{g cm}^{-2}}
\newcommand{\gcmcub}{\mbox{g cm}^{-3}}

\newcommand{\taumelt}{\tau_\mathrm{melt}}
\newcommand{\taumin}{\tau_\mathrm{min}}
\newcommand{\Tmelt}{T_\mathrm{melt}}

\begin{document}
\title[Fast Electron Transport for FI]{Theory of Fast Electron Transport for Fast Ignition}
\author{A.P.L. Robinson$^1$, D.J. Strozzi$^2$, J.R. Davies$^3$, L. Gremillet$^4$, J.J. Honrubia$^5$,
  T. Johzaki$^6$, R.J. Kingham$^7$, M. Sherlock$^7$ and A.A. Solodov$^8$}
\address{$^1$ Central Laser Facility, STFC Rutherford-Appleton Laboratory, Harwell Science and Innovation Campus, OX11 0QX, UK}
\address{$^2$ Lawrence Livermore National Laboratory, University of California, P.O. Box 808, Livermore, California 94550, USA}
\address{$^3$Fusion Science Center for Extreme States of Matter, Laboratory for Laser Energetics and Mechanical Engineering, University of Rochester, Rochester, New York 14623, USA}
\address{$^4$ CEA, DAM, DIF, F-91297 Arpajon, France}
\address{$^5$ School of Aerospace Engineering, Technical University of Madrid, Plaza Cardenal Cisneros 3, 28040-Madrid, Spain}
\address{$^6$ Institute of Laser Engineering, Osaka University, 2-6 Yamadaoka, Suita, Osaka 565-0871, Japan}
\address{$^7$ Plasma Physics Group, Blackett Laboratory, Imperial College London, Prince Consort Rd, London, SW7 2BZ, UK}
\address{$^8$ Laboratory for Laser Energetics, University of Rochester, Rochester, New York 14623, USA}

\begin{abstract}
Fast Ignition Inertial Confinement Fusion is a variant of inertial fusion in
which DT fuel is first compressed to high density and then ignited by a
relativistic electron beam generated by a fast ($<$ 20~ps) ultra-intense laser
pulse, which is usually brought in to the dense plasma via the inclusion of a
re-entrant cone.  The transport of this beam from the cone apex into the dense
fuel is a critical part of this scheme, as it can strongly influence the overall
energetics.  Here we review progress in the theory and numerical simulation of
fast electron transport in the context of Fast Ignition.  Important aspects of
the basic plasma physics, descriptions of the numerical methods used, a review
of ignition-scale simulations, and a survey of schemes for controlling the
propagation of fast electrons are included.  Considerable progress has taken
place in this area, but the development of a robust, high-gain FI `point design'
is still an ongoing challenge.
\end{abstract}

\section{Introduction}

Since its proposal by Tabak and co-workers \cite{tabak-fi-pop-1994} in 1994 the concept of Fast Ignition (FI) Inertial Confinement Fusion (ICF) has attracted considerable attention \cite{tabak-fi-pop-2005}.  This advanced ICF concept is appealing because of its ability to achieve high energy gains ($G >$ 100) while reducing both the total laser energy and the hydrodynamic demands on the fuel assembly.  One of the new challenges in this concept is the need to efficiently couple the ignitor pulse energy via the relativistic ({\it fast}) electrons to a hot spot in the compressed fuel.  Within this, there are two parts - the absorption of laser light into fast electrons, and then the propagation and stopping of the fast electrons.  This review is concerned with the latter of these, i.e. {\it fast electron transport} (FET).

The fast electron transport aspect of FI is challenging for at least three reasons.  Firstly there is an issue that would exist even if fast electron propagation were purely ballistic.  The size of the hot spot is comparable to the size of the fast electron source (i.e. the laser spot), but the two are separated by a distance which is several times their size.  Therefore any appreciable angular spread in the fast electrons must either be mitigated or controlled, as a reduction in the coupling efficiency will otherwise occur.  Secondly there is the possibility that various instabilities might disrupt the beam propagation which in turn would impair the coupling efficiency.  Thirdly, any solution to the first and second problem must be compatible with achievable fuel assemblies and the fast electron parameters required to achieve stopping in the hot spot.  Yet another problem is the source characteristics as a function of the laser parameters: currently it would appear that the fast electron energy spectrum is too hard to allow for all the fast electrons to be deposited in an ideal hot spot.

Ultimately it is hoped that an overarching solution to these problems can be found which is still attractive and feasible, i.e.\ a `point design' for Fast Ignition.  Currently it is not possible to build an ignition-scale facility purely for the purposes of investigating the feasbility of FI or solving the problems associated with fast electron transport by purely iterative empirical methods.  Therefore detailed numerical simulation and a thorough understanding of the underlying theory are essential parts of realizing FI.  Hence the importance of the subject matter of this review.

In this review of the area, we will cover the following aspects of the theory of fast electron transport in FI:
\begin{enumerate}
\item{{\it Basic Physics}: The fundamental physical processes including scattering and stopping of fast electrons, the role of resistively generated fields, and key beam-plasma instabilities and phenomena.}
\item{{\it Simulation Methods}: The different simulation methods that have been applied to this problem and their relative strengths and weaknesses.}
\item{{\it Review of Ignition-Scale Calculations}: A review of simulation studies of the full-scale problem, and how this has informed the overall view of the current challenges that FI is facing.}
\item{{\it Concepts for Controlling Transport}: A survey of the various ideas that have been proposed to overcome the limits on the coupling efficiency that are imposed by realistic fast electron divergence angles, namely: fast electron self-collimation by resistively-generated magnetic fields (due to beam profile or resistivity gradients), electrostatic confinement by a vacuum gap (double-cone target), and imposed axial magnetic fields.}
\item{{\it Prospects for a Point Design}: What future FET studies will have to address in order to move closer to a FI point design.}
\end{enumerate}
In addition, we provide a very brief pr\'{e}cis of the requirements for fast electron heating to reach ignition in section \ref{require}.

Our review will draw attention to the considerable effort that has gone into both understanding the fundamental aspects of the problem and developing numerical tools that are suitable for studying fast electron transport.  The calculations that have been performed under conditions close to full-scale FI so far clearly show that the scheme must be adapted in some way given realistic fast electron beam parameters.  Our review also indicates there are potentially feasible ways to `control' fast electron transport and thus achieve a viable point design.

\section{Ignition via Rapid Heating of Compressed Fuel}
\label{require}
So that the fast electron transport problem can be put in context, we briefly summarize the objectives that must be achieved in order to obtain ignition and gain from the rapid heating of a particular region of highly compressed DT fuel.  Fast ignition is an isochoric mode of ignition, where a region of fuel of relatively constant density is heated to much higher temperatures and pressures.  It contrasts with isobaric ignition modes, such as central hot-spot ignition. The requirements for FI are determined by hydrodynamics and burn physics. Estimates of the optimal parameters that minimize the ignitor pulse energy have been obtained by Atzeni using both analytic calculation and 2D hydrodynamic simulations \cite{atzeni-fi-pop-1999}.  The resulting optimal fast electron energy ($E_{ign}$), fast electron intensity ($I_{ign}$), pulse duration ($t_{ign}$), and hot spot radius ($r_{hs}$)are:
\begin{equation}
\label{atzeni_energy}
E_{ign} = 140 \rho_{100}^{-1.85}\mbox{ kJ},
\end{equation}

\begin{equation}
\label{atzeni_intensity}
I_{ign} = 2.4\times10^{19} \rho_{100}^{0.95}\mbox{ W cm}^{-2},
\end{equation}

\begin{equation}
\label{atzeni_time}
t_{ign} = 54\rho_{100}^{-0.85}\mbox{ ps},
\end{equation}
 
\begin{equation}
\label{atzeni_spot}
r_{hs} = 60\rho_{100}^{-0.97}\,\mu\mbox{m}.
\end{equation}
where $\rho_{100}=\rho/$100 $\gcmcub$. A typical FI scenario will involve the assembly of a quasi-spherical DT mass reaching peak densities in the range 300 $< \rho <$ 1000 g~cm$^{-3}$.  Assuming that a re-entrant cone-guided FI scheme is being followed, the distance between the tip of the cone and the geometric centre of the DT mass (the `stand-off' distance) is typically 100 $\mu$m.  In FI schemes that employ `hole-boring' to create a path for the ignitor pulse, there will still be a substantial stand-off distance of at least 100 $\mu$m.  The DT density around the cone tip depends on the detailed hydrodynamics of the fuel assembly scheme, but is usually on the order of at least a few $\gcmcub$.  The fuel temperature at stagnation is usually around 200--300~eV. The ignition requirements were generalized in Ref.\ \cite{atzeni-fastig-pop-2007} to include effects like $r_{hs}$ exceeding the optimal value and fast electrons not fully stopping in the DT fuel. More recently, ignition requirements based on realistic, PIC-based fast electron sources have been found \cite{bellei1}.

The objective of fast electron transport theory is to ensure that a hot spot is produced within the constraints of equations (\ref{atzeni_energy}--\ref{atzeni_spot}) given conditions that do not differ too greatly from those outlined in the preceding paragraph.  As Direct Drive ICF with central hot spot ignition may be possible with total laser energies of 1--2~MJ, and advanced Indirect Drive ICF may be feasible with similar total laser energy, the FI concepts ideally aims to operate using not much more than 100~kJ of ignitor pulse energy.
\section{Basic Physics of Fast Electron Transport in FI}
\label{basic}

%
%

We now look at the basic physical phenomena that affect the propagation of the fast electron beam from the source to the compressed core. 
We will assume that the characteristics of the fast electron beam (FEB) at the source are given, and concentrate on the theoretical models describing propagation of fast electrons.  In this section, phenomena are considered in isolation, concentrating on the fundamental equations and models of each phenomenon.
Naturally, the interaction of these phenomena does occur, is rather complex, and requires use of simulation codes for quantitative prediction.  However these fundamental elements are the `building blocks' of fast electron transport theory, and are essential for understanding the simulation codes.

\subsection{Fast Electron Parameters}
\label{basic:beam}
The physics of the absorption of high power, high intensity laser light into the plasma and generation of the fast electron beam is a whole topic in its own right and is covered in detail elsewhere in this special issue.  Such details will not be covered, and we limit ourselves to a few general remarks.  Broadly speaking, the fast electrons are injected from the laser-plasma interaction region (e.g.~inner cone surface) towards the compressed fuel core with a broad distribution of energies and a significant degree of anisotropy.  A simple model that is sometimes used in transport calculations is,
\begin{equation}
  \label{basic:injection_DF}
  f_{inj}(E,\theta) \propto \exp(-E/\bar{E})\exp\left(-\frac{\theta^2}{2\theta_{inj}^2}\right).
\end{equation} 
$E=m_ec^2(\gamma-1)$ is the fast electron kinetic energy, and $\theta$ the angle between $\vec v$ and the nominal direction of beam propagation.

The mean fast electron energy is often taken to be close to the ponderomotive potential energy, $U_{pond}=[1+0.73I_{18}(\lambda_L)_{\mu\mathrm{m}}^2]^{1/2}-1$ ($I_{18}=I/10^{18}$ W cm$^{-2}$), in such simple models of the fast electron distribution.  Fast electron transport calculations that aspire to have good predictive capability need to include either a self-consistent, laser-generated fast electron source or take a detailed fast electron distribution from a PIC LPI calculation.  The ponderomotive scaling does provide a rough indication of the expected fast electron mean energy or temperature with intensity and wavelength, but the detailed scalings are still very much an open topic of research.  Detailed PIC simulations often show an energy distribution that is considerably more complex than the exponential in Eq.~(\ref{basic:injection_DF}).

The angular spread of the fast electrons has no simple or clean characterization either, all the more so since it is expected to depend on the electron energy.  Even if one neglects the energy dependence, it is virtually impossible to measure $\theta_{inj}$ directly, although experiments have {\em inferred} a wide range of characteristic angles in addtion to a range of simulation results.  Current research is tending to operate under the presumption that FI will have to contend with a scenario where the characteristic fast electron divergence half-angle is greater than 30$^\circ$, and possibly even exceeding 50--60$^\circ$.

The conversion efficiency from laser energy to fast electron energy, $\eta_L$, is  not well characterized either.  A wide range of experimental and theoretical results on this were compiled by Davies in \cite{jrd3}, who noted that the range of results spanned the range 10$\le \eta_L \le$90\%.  Detailed PIC simulations relevant to FI, amongst other results, indicate that achieving a $\eta_L$ in the range of 30--50\% is likely, thus making an `attractive' FI scheme still possible provided that efficient coupling to the hot spot can also be achieved.


\subsection{Effect of Macroscopic EM Fields}
\label{basic_macro}
\subsubsection{Return Current and Current Balance}
As the fast electron beam propagates through dense plasma, it will draw a {\it return current} that is both spatially coincident with the fast electron current density ${\bf j}_f$ and which nearly cancels the fast electron current to a good approximation \cite{bell-ppcf-2006}, i.e.\ if the return current density is ${\bf j}_b$ then,
\begin{equation}
\label{cbalance}
{\bf j}_f+ {\bf j}_b \approx 0.
\end{equation}
To see how this arises, one can consider the hypothetical case where there is no return current.  For a wide beam, one can estimate the electric field growth from $E \approx - j_ft/\varepsilon_0$.  Since the current densities in FI can easily reach 10$^{16}$ A m$^{2}$, one can see that the electric field can reach 10$^{12}$ V m$^{-1}$ in 1~fs, which is enough to stop MeV fast electrons on a few $\mu$m scale.  Thus it is clear that a return current will be drawn when the fast electrons propagate through dense plasmas.  In a fully 3D situation, one might imagine that the fast electron current is only globally balanced, but not locally balanced (as in Eq.~(\ref{cbalance})).  However, this will lead to the growth of magnetic fields that would destroy the beam, so the current neutralization must indeed be co-spatial.  The return current phenomenon is not particular to fast electron transport in the context of ultra-intense laser-plasma physics, and arises in a number of other contexts such as charged particle-beam dynamics \cite{mill82} and energetic electron transport in solar flares \cite{vandenoord1}.

\subsubsection{Resistive Inhibition and Ohmic Heating}
Current balance implies that $\mathbf{j}_b \approx - \mathbf{j}_f$, and on inserting this into a resistive Ohm's law, one obtains $\mathbf{E} = -\eta{\mathbf{j}_f}$.  The peak resisitivity in many conducting solids will be around 10$^{-6}\ \Omega$ m, and even low-Z plasmas at a temperature of a few hundred~eV will have resistivities of 10$^{-7}$--10$^{-8}\ \Omega$ m.  This means that the resistively generated electric field can be 10$^8$--10$^{10}$ V m$^{-1}$, which is sufficient to inhibit fast electron transport significantly.

The drawing of the return current also heats the background plasma via Ohmic heating with power density ${\bf j}_b.{\bf E} \approx \eta{j_f^2}$.  From the aforementioned typical values of resistivity and fast electron current density this means that the Ohmic heating can heat a solid-density target at a rate of 0.1--1~keV/ps.  Therefore at solid density the Ohmic heating must be included in the energy equation of the background plasma, as the heating and thus the effect on resistivity is strong.  However at very high density (e.g.\ DT fuel above 100~$\gcmcub$) this heating is very small, and thus Ohmic heating will not make any significant contribution to the generation of the hot spot.

\subsubsection{Resistive Magnetic Field Generation}
Current balance also has implications for the generation of magnetic field \cite{bell2}.  An improved estimate for the resistive electric field is,
\begin{equation}
{\bf E} = -\eta{\bf j}_f + \frac{\eta}{\mu_0}\nabla \times {\bf B}.
\end{equation}
Inserting this into Faraday's law $\partial_t{\bf B} = - \nabla \times \mathbf{E}$ yields
\begin{equation}
\label{maggen}
\frac{\partial{\bf B}}{\partial{t}} = \eta\nabla \times {\bf j}_f + \nabla\eta \times {\bf j}_f + \frac{\eta}{\mu_0}\nabla^2{\bf B} - \nabla\eta \times (\nabla \times {\bf B}).
\end{equation}
The last two terms correspond to resistive diffusion and resistive advection of magnetic field, and these are the normal terms that are found in the resistive MHD description of a static plasma.  The first two terms, on the other hand, correspond to resistive generation of magnetic field, and these are due to the presence of the fast electrons.  Davies noted that one can describe the first term as growing magnetic field which pushes fast electrons into regions of higher current density, whereas the second term grows magnetic field which pushes fast electrons into regions of higher resistivity.  These magnetic field growth rates are significant --- the magnitude of the growth rate is roughly $\dot{B} \sim \eta{j_f}/R$ (where $R$ is the fast electron beam radius).  Taking some typical figures ($j_f =$10$^{16}$ A m$^{-2}$, $\eta =$10$^{-7}\,\Omega$ m, $R =$5 $\mu$m), yields a growth rate of 2$\times$10$^{14}$ T s$^{-1}$, i.e. 200~T in 1~ps.  Magnetic fields on the order of 100--1000~T will have a significant effect on multi-MeV electrons if the fields extend over several microns, insofar as these fields can pinch or filament the beam.

\subsubsection{Self-Pinching of the Fast Electron Beam}
The magnetic field generated by the $\eta\nabla \times {\bf j}_f$ term in Eq.~(\ref{maggen}) grows in the sense which acts to pinch the fast electron beam \cite{jrd4,tatarakis1}.  The tendency of the beam to self-pinch can, in principle, be highly beneficial to FI.  Counteracting against this self-pinching is the angular divergence of the fast electron beam.  Bell and Kingham derived a condition for self-pinching or collimation for a plasma with Spitzer resistivity by noting that the self-pinching condition is $R/r_g > \theta_{1/2}^2$ with $r_g$ the electron gyroradius, i.e.\ the magnetic field can deflect a fast electron through the characteristic fast electron divergence half-angle, $\theta_{1/2}$, in the same distance that it takes the beam radius to double.  In the limit of strong heating ($T_b \gg T_{b,init}$), the Bell-Kingham condition \cite{bell-2003} for self-pinching is $\Gamma > 1$ where,

\begin{equation}
\label{bkcond}
\Gamma = \mbox{0.13}n_{23}^{3/5}Z^{2/5}(\log\Lambda)^{2/5}P_{TW}^{-1/5}T_{511}^{-3/10}(2 + T_{511})^{-1/2}R_{\mu{m}}^{2/5}t_{psec}^{2/5}\theta_{rad}^{-2},
\end{equation}
with $n_{23}$ being the background electron density in units of 10$^{23}$cm$^{-3}$, $T_{511}$ being the fast electron energy in units of the electron rest mass, $R_{\mu{m}}$ being the beam radius in microns, $t_{psec}$ is the fast electron pulse duration in ps, and $\theta_{rad}$ is $\theta_{1/2}$ in radians.  Eq. \ref{bkcond} shows that the self-pinching is most strongly dependent on the divergence angle of the fast electrons, with most other parameters exhibiting much weaker dependence.  For conditions relevant to FI, self-pinching is marginal and strongly dependent on $\theta_{1/2}$.
 
\subsubsection{Beam Hollowing}
Even in a homogeneous plasma, the $\nabla\eta \times {\bf j}_f$ term can still have a significant effect.  This occurs in the regime of strong heating as this will produce a significant $\nabla\eta$ as the Ohmic heating in the centre of the beam is much stronger than at the periphery of the beam ($\propto \eta{j_f^2}$).  This can lead to the sign of $\partial{E_x}/\partial{r}$ reversing ($\bf{x} || \bf{j_f}$), which leads to the generation of a de-collimating magnetic field in the beam centre.  In turn this will lead to the expulsion of fast electrons from the centre of the beam, and this effect is therefore referred to as {\it beam hollowing}.  Davies first identified this effect in \cite{jrd2}, where he analyzed heating and magnetic field generation in the case of a rigid beam model, and he considered different possible resistivity models via $\eta \propto T_b^\alpha$.   When $\alpha < 1$ beam hollowing will eventually occur, as all materials become Spitzer-like ($\alpha = -3/2$) at sufficiently high temperature. 

\subsection{Drag and Scattering of Individual Fast Electrons}
\label{drag+scattering}

Here we will consider the transport of individual fast electrons through plasmas and solids, in other words, we will not consider collective effects arising from the presence of more than one fast electron. 

Fast, in this context, refers to an electron traveling at speeds much greater than that of the electrons in the material.
In this case, the principal effects on the fast electron are energy loss and angular scattering.
We will present expressions for the rate of energy loss, or drag, and the rate of angular scattering and briefly outline their derivations and their implications for fast ignition.

This single particle model will be an adequate description of drag and scattering provided that the fast electron density is much less than the electron density of the material.  To determine exactly how much less requires an accurate calculation of collective effects, which we do not have. Work along these lines for the correlated stopping of $N$ fast electrons has been presented in \cite{deutsch1999,bret-stopping2008}.  In the case of a plasma, this effect should be negligible if the separation between fast electrons is greater than the screening length for the fast electron wake. This distance is the dynamical screening length $v/\omega_p$ for $v \gg (kT_e/m_e)^{1/2}$, and the plasmas Debye length $\lambda_{\rm D}=\sqrt{\varepsilon_0 kT/n_ee^2}$ in the opposite limit.  However, there could still be significant electromagnetic fields generated by the collective response of the material to the fast electrons as a whole that can then be considered independently from the drag and scattering. These effects, such as beam-plasma instabilities, are discussed elsewhere in this article.

We briefly note that drag and scattering have received much attention since the proposal of FI in 1994 \cite{atzeni-ppcf-2009}. However, calculations of drag date back to the 1930s, with the definitive reformulations of the basic theories being published in the 1950s  \cite{sternheimer1952,fano1956}. These frequently consider the more general problem of drag in matter with bound electrons. Free electrons (namely in conductors) are considered in these calculations, therefore they do apply to plasma. These results are embodied in \cite{icru37}, were recently summarized in \cite{solodov-stop-pop-2008,atzeni-ppcf-2009}, and are presented here. The latter two references differ slightly in their angular scattering formulas, and in some details of their logic. For a fully quantum-mechanical (but non-relativistic) treatment of drag due only to free electrons, including both binary collisions and interaction with the plasma medium (e.g.\ plasmon excitation), see \cite{ferrell1956}.

\subsubsection{Drag}

The standard expression for the drag on a fast electron in {\em all} matter (solid, liquid, gas or plasma, conductor or insulator) is \cite{icru37}
\begin{eqnarray}
\label{eq-drag}
\frac{dE}{dt} & = & - \frac{n_e e^4}{4\pi \varepsilon_0^2 m_e v}L_{\rm d}, \\
\label{eq-Ld}
L_{\rm d} & = & \ln \frac{pv}{\sqrt{\gamma+1}\hbar\omega_p} -\frac{\ln2}{2} + \frac{9}{16} + \frac{(1/2)\ln2+1/16}{\gamma^2} - \frac{\ln2 + 1/8}{\gamma},
\end{eqnarray}
where $E$, $p$, and $v$ are the kinetic energy, momentum and velocity of the fast electron, respectively. We have introduced the dimensionless parameter $L_{\rm d}$, which we call the drag number. In conventional plasma physics notation it would be called ``$\ln\Lambda$". 
Bremsstrahlung has been neglected.
As would be expected, fast electron drag does not depend on the velocity or the binding energy of the electrons in the material, since we are considering the limit in which these become negligible, it depends only on their density $n_e$, which here refers to total, not free, electron density and it is also total density that determines the plasma frequency $\omega_p$ in \eq{eq-Ld}. 
The value of $\hbar \omega_p/e$ at the typical solid density of $6\times10^{28}$ atoms m$^{-3}$ is $9.1\sqrt{Z}$ eV, where in this section $Z$ represents nuclear, not ionic, charge. In DT, $\hbar \omega_p/e=180\sqrt{\rho_{100}}$ eV, where $\rho_{100}=\rho/100\ \gcmcub$. This result has been extensively tested in cold matter, but not so extensively in plasma and never at the densities required for fast ignition, however, there is no reason to believe that this lies in a fundamentally different physical regime; drag due to degenerate, free electrons is present in metals.

What changes between materials when applying \eq{eq-Ld} is the implication of {\em fast}.
In plasma, it is sufficient for the fast electron to have a velocity a few times higher than the electron thermal velocity ($\sqrt{kT_e/m_e}$) or the Fermi velocity if the electrons are degenerate.
This will be true in the corona and in the core of ignition targets for all cases of interest.
In unionized matter, the fast electron must have an energy much greater than any binding energy, so in a cone we will have to consider the effect of electron binding.
Before we do this, we will outline the derivation of \eq{eq-Ld} as given in \cite{icru37}.

Fast electron energy loss $W$ above a cut-off $W_{\rm c}$ is calculated using a binary collision model and energy loss below $W_{\rm c}$ is calculated using a model for the collective response of the electrons in the material. It is assumed that $W_{\rm c}$ is much less than the fast electron energy yet much greater than the energy of any individual electron in the material. The cut-off $W_{\rm c}$ at which the two models are patched together cancels in the final result, giving some confidence that it is accurate, even though neither model is valid for intermediate energy losses, for which no analytical model is available. 

For the binary collision model the M{\o}ller cross section is used, an approximate solution to the Dirac equation to order $\alpha v/c$ (the first Born approximation), where $\alpha$ is the fine structure constant, so it includes relativistic effects and quantum spin and exchange effects.
Experimentally, deviations from this cross section have only been detected in close collisions at energies much higher than those of interest here, when radiation becomes important.
The target electron is assumed to be stationary, which requires its velocity to be much less than that of the fast electron, and any binding or potential energy is neglected, which requires this to be much less than the energy loss. 
It is also implicitly assumed that the energy loss occurs largely while the electrons are close together, because the cross section applies for isolated electrons coming in from infinity and being detected at infinity but is being applied to calculate fast electron energy loss to only one, immediately adjacent electron among many others.
Classically, it can be shown that this is an adequate approximation for sufficiently fast electrons by considering interaction over a limited distance \cite{ordonez1}, and this does not represent a significant additional restriction on the theory. 
We know of no rigorous demonstration that this carries over to the quantum case.
The calculation follows the familiar treatment of binary collisions, with a maximum energy loss of half the fast electron energy, since only the fastest electron is followed, giving
\begin{equation}
\label{eq-Ld_Moller}
\left. L_{\rm d} \right|_{W>W_{\rm c}} = \ln \sqrt{\frac{E}{W_{\rm c}}} + \frac{9}{16} -\ln2 + \frac{(\ln2)/2+1/16}{\gamma^2} - \frac{\ln2 + 1/8}{\gamma},
\end{equation}
neglecting terms in $W_{\rm c}/E$. 
The first term would be obtained using the Rutherford cross section, and the remaining terms represent small quantum corrections due to spin and exchange.

For the model of the collective response of the electrons in the material it is assumed that the fast electron moves at constant velocity and that its electric field causes a small perturbation of the electrons from their equilibrium positions, so a quantum harmonic oscillator model may be used. 
These are only adequate assumptions far from the fast electron and while its velocity changes on a time scale much slower than the collective response time of the electrons. 
The final result is
\begin{equation}
\label{eq-Ld_wave}
\left. L_d \right|_{W<W_c} =  \ln \frac{\sqrt{2m_ev^2W_c}}{\hbar \omega_p},
\end{equation}
which when added to \eq{eq-Ld_Moller} yields \eq{eq-Ld}.
This can be understood in terms of energy exchange to plasma waves (plasmons) in quanta of $\hbar \omega_p$; that this arises from a quantum treatment of electron oscillations in a plasma, which has been considered by a number of authors \cite{ferrell1956,bohm1953,pines1953}, is not surprising. What is remarkable is that this also arises as the limiting form for fast electrons from a general treatment, including electron binding. 

We will now consider the more general case where the binding energies of the electrons in the material cannot be neglected, since this will be the case in a cone.
This leads to the drag being reduced. 
For a combination of historical and mathematical reasons the energy loss due to the collective response of the material is artificially divided into two parts and written
\begin{equation}
\label{eq-Ld_I}
\left. L_d \right|_{W<W_c} =  \ln \frac{\sqrt{2p^2W_c/m_e}}{I_{\rm ex}} - \frac{1}{2}\frac{v^2}{c^2} - \frac{\delta}{2},
\end{equation}
giving
\begin{equation}
\label{eq-Ld_gen}
L_d = \ln \left( \sqrt{\gamma+1} \frac{E}{I_{\rm ex}} \right) -\frac{\ln2}{2} + \frac{1}{16} + \frac{(1/2)\ln2+9/16}{\gamma^2} - \frac{\ln2 + 1/8}{\gamma} -\frac{\delta}{2}.
\end{equation}
The first part is given by the first two terms of \eq{eq-Ld_I}, the basis for which was published by Bethe in 1930 (in German).
It gives the energy transferred to the excitation of electrons by the electric field of a charged particle moving at constant velocity.
The complexities of dealing with coupled, quantized oscillations of multiple bound electrons are hidden in $I_{\rm ex}$, known as the mean excitation potential, for which there exist a variety of theoretical models.
In very general terms it can be written as 
\begin{equation}
\label{eq-Iex}
\ln I_{\rm ex} = \sum_{i,j}f_{ij}\ln(E_j-E_i),
\end{equation} 
a weighted sum over all possible transitions of electrons in the material from initial energy $E_i$ to final energy $E_j$, $f_{ij}$ being the transition probability. 
In the simplest possible case of a single, undamped, harmonic oscillator of frequency $\omega$ it is $\hbar \omega$.
This is a good approximation for plasma, giving the mean excitation potential to be $\hbar \omega_p$.
The values normally used for unionized materials are determined by measurements of either ion or electron energy loss or of optical absorption, as drag can be treated in terms of the absorption of a virtual photon field.
Thus it becomes a free parameter used to fit experimental data.
The reference values are those published in \cite{icru37}, available online at \cite{estar}. For elements these can be adequately fitted by $9.43Z+26.1$ eV, except for hydrogen, where $I_{\rm ex}/e$ is $19.2$ eV. 
For compounds, the stopping due to its constituents can be added since chemical structure has been found to have only a small effect on the mean excitation potential. 
The second part is the $\delta$, first quantified by Fermi in 1940 \cite{fermi1} using a purely classical calculation representing the electron response with a single, harmonic oscillator.
It gives a reduction in the energy loss due to the electric field of the fast electron being shielded by the collective response of the electrons in the material, an effect neglected by previous treatments, hence the convention of a negative sign (the factor of $2$ is another, rather confusing, convention).
It is called the density effect correction because it increases with electron density.
The mathematical reason behind this division is the difficulty of giving a straightforward expression for $\delta$ in the general case of multiple bound electrons.
It can be obtained analytically in the limit of a strongly relativistic electron \cite{fermi1}
\begin{equation}
\label{eq-limit}
\delta \rightarrow 2\ln\left(\frac{\gamma\hbar\omega_{p}}{I_{\rm ex}}\right)-\frac{v^2}{c^2}  \qquad v \rightarrow c,
\end{equation}
giving the general result for fast electron drag that we started with.
For plasma, where $I_{\rm ex} = \hbar\omega_{p}$, this expression is valid for all cases of interest. 
For bound electrons, where typically $I_{\rm ex} \gg \hbar\omega_{p}$, this expression is only greater than $0$ for $\gamma > 1.65I_{\rm ex}/\hbar\omega_{p}$.
In practice, $\gamma \gg 1.65I_{\rm ex}/\hbar\omega_{p}$ is required for \eq{eq-limit} to be a good approximation for bound electrons. 
In solid gold, for example, this requires a fast electron energy much greater than 8 MeV.

Sternheimer has given a simple, approximate formulation of the density effect \cite{sternheimer1} that is used in \cite{icru37}, which we will write as
\begin{equation}
\label{eq-delta}
\delta = \sum_n  f_n \ln \left[ 1+\frac{E_l^2}{f_{\rm S}^2B_n^2+f_Bf_n(\hbar\omega_p)^2} \right] - \left( \frac{E_l}{\gamma\hbar\omega_p}\right)^2
\end{equation}
where $f_n$ is the fraction of electrons with binding energy $B_n$ (Sternheimer writes this as $h$ times the frequency of the absorption edge), $f_B$ is $1$ if $B_n=0$ (free electrons) and $2/3$ otherwise, $E_l$ is given implicitly by 
\begin{equation}
\label{eq-El}
\sum_n \frac{f_n(\hbar\omega_p)^2}{f_{\rm S}^2B_n^2+E_l^2}=\frac{1}{(p/m_ec)^2}
\end{equation}
and the Sternheimer factor $f_{\rm S}$ is given implicitly by
\begin{equation}
\label{eq-f_S}
\sum_n  f_n \ln \left[f_{\rm S}^2B_n^2+f_Bf_n(\hbar\omega_p)^2 \right] = \ln I_{\rm ex}^2,
\end{equation}
which ensures that \eq{eq-limit} is obeyed with experimental values of the mean excitation potential, a consistent problem with other formulations. 
The Sternheimer factor is typically between $1.5$ and $2.5$ \cite{sternheimer1}.

For plasma ($B=0$, $f_B=1$) \eq{eq-delta} and \eq{eq-El} give the density effect correction to be $2\ln(\gamma)-v^2/c^2$ and \eq{eq-f_S} simply gives the mean excitation potential to be $\hbar \omega_p$, reproducing results we have seen before.

To illustrate the result for bound electrons let us consider a single binding energy, for which it is straightforward to obtain
\begin{eqnarray}
\delta & = & \ln \left[ (\gamma^2-1/3) \left( \frac{\hbar\omega_p}{I_{\rm ex}}\right)^2 \right]  - \frac{v^2}{c^2} + \left( \frac{f_{\rm S}B}{\gamma\hbar\omega_p} \right)^2 \qquad \frac{p}{m_ec} > \frac{f_{\rm S}B}{\hbar\omega_p}, \\
 & = & 0  \phantom{\ln \left[ (\gamma^2-1/3) \left( \frac{\hbar\omega_p}{I_{\rm ex}}\right)^2 \right]  - \frac{v^2}{c^2} + \left( \frac{f_{\rm S}B}{\hbar\omega_p} \right)^2} \qquad \frac{p}{m_ec} \le \frac{f_{\rm S}B}{\hbar\omega_p}.
\end{eqnarray}
This shows that there is a threshold fast electron energy for the density effect to occur in insulators (conduction electrons are treated as free electrons) and that $p/m_ec$ should exceed $f_{\rm S}B/\hbar\omega_p$ for all electrons before \eq{eq-Ld} will be a good approximation, a similar constraint to that indicated by \eq{eq-limit}.  
For multiple binding energies a numerical solution is required.

Sternheimer  \cite{sternheimer1} gives a 5 parameter fit to the density effect correction for numerous elements and compounds.
We have found that for Cu and Mo the overall drag number from \cite{icru37} is reproduced to within 1\% by just using
\begin{equation}
\label{eq-delta_fit}
\frac{\delta}{2} = \ln \left[ 1 + \frac{E}{m_ec^2}\frac{\hbar\omega_p}{I_{\rm ex}}\exp(-0.5)\right],
\end{equation}
which reproduces the limiting forms of the density effect correction, but does not fit at intermediate energies. However, here the density effect correction makes a negligible contribution to the drag number.
This approach could be adapted for insulators with a threshold energy $E_0$ by using
\begin{equation}
\label{eq-delta_fit-ins}
\frac{\delta}{2} = \ln \left[ 1 + \frac{E-E_0}{m_ec^2}\frac{\hbar\omega_p}{I_{\rm ex}}\exp(-0.5)\right] \qquad E \ge E_0,
\end{equation}
but we have not verified the accuracy of this approach.

What is lacking are results for partially ionized matter when the electron is not fast enough for \eq{eq-Ld} to apply.
The only treatment we are aware of is an approximate model for the mean excitation potential of bound electrons in an ion, published in a difficult to obtain report by More \cite{more1}, the drag number of the free electrons being given by \eq{eq-Ld}.
For the mean excitation potential he used a simplified theoretical model known as the local plasma approximation
\begin{equation}
\label{eq-Bloch}
\ln I_{\rm ex} = \int f_e(\vec{r})\ln[\hbar \omega_p(\vec{r})] dV,
\end{equation}
where $f_e$ is the electron probability density function and $\omega_p(\vec{r})$ refers to the plasma frequency at the local mean electron density $N_ef_e$, $N_e$ being the number of electrons.
In this approximation the mean excitation potential of bound electrons is higher than that of free electrons because they are concentrated around the nucleus.
To obtain the electron distribution around an ion More used the Thomas-Fermi model and found that the result could be described by 
\begin{equation}
\label{eq-Iq}
I_{\rm ex}(q)=I_{\rm ex}(0)\frac{\exp(1.29(q/Z)^{0.72-0.18q/Z})}{\sqrt{1-q/Z}},
\end{equation}
where $q$ is the ionization state.
This should be an adequate description for weakly ionized many electron atoms when the interaction between electrons of neighbouring ions is negligible. 
It does not give the correct result for hydrogen-like ions, which would be expected to have a mean excitation potential of roughly $Z^2$ times the value for hydrogen; More argues that the contribution of this one electron will, in general, be negligible.
Equation (\ref{eq-Iq}) could also be used to estimate the density effect for the bound electrons by using \eq{eq-f_S} to calculate a new effective value of $f_{\rm S}$ from the new value of $I_{\rm ex}$, representing an average increase in binding energies, or by using our crude model. 
Binding energies of ions can be measured or calculated, but a limited number of results are available and no one appears to have made the effort to apply these to calculating fast electron drag.
The mean excitation potential and density effect correction will also have to be recalculated for compressed material, such as a cone tip in a compressed target.
The simplest approach would be to start from \eq{eq-f_S} to reevaluate the mean excitation potential.

\subsubsection{Scattering}

In solids where scattering is from atoms with a radius $a$ much less than the interatomic separation and the de Broglie wavelength of the fast electron is much less than $a$, it is clear that angular scattering can be adequately described in terms of binary collisions.
An approximate model for the average potential around an atom is the familiar exponentially screened potential with a screening distance $a$ \cite{motz1,joachain1}.
Measured and calculated values of atomic radii are readily available for all elements; the Thomas-Fermi model gives a simple, general result of $a_0/Z^{1/3}$, where $a_0$ is the Bohr radius ($5.3\times 10^{-11}$ m), although this is not accurate for all elements.
Using the scattering cross section for an exponentially screened potential obtained from the Dirac equation in the first Born approximation \cite{motz1,atzeni-ppcf-2009}, the familiar treatment of binary collisions, integrating over {\em all} scattering angles, $0$ to $\pi$, gives
\begin{eqnarray}
\label{eq-theta2}
\frac{d\mean{\theta^2}}{dt} & = & \frac{Zn_e e^4}{2\pi\varepsilon_0^2p^2v} L_{\rm s} \\
\label{eq-Lsa}
L_{\rm s-a} & \approx & \ln \frac{2ap}{\hbar}  - 0.234 - 0.659 \frac{v^2}{c^2} \qquad \frac{2ap}{\hbar} \gg 1,
\end{eqnarray}
where $\mean{\theta^2}$ is the mean square scattering angle with respect to the electrons instantaneous direction of motion, not its original direction of motion, and we have introduced the scattering number $L_{\rm s}$ with the -a indicating that it applies to atoms.
The last term, which is due to the electron spin, had to be evaluated numerically, so all terms have been expressed to the same accuracy of 3 significant figures.
This differs slightly from the expression in \cite{atzeni-ppcf-2009} because they calculated $\mean{\cos\theta}$ not $\mean{\theta^2}$. 
The integral does not diverge at zero scattering angle (infinite impact parameter) because in quantum mechanics any potential that falls faster than $1/r$ has a finite cross section for zero scattering angle \cite{joachain1}.
Here this cross section is approximately $\pi [2Z\alpha a/(v/c)]^2$ for $2ap/\hbar \gg 1$, which surprisingly gives an effective upper impact parameter that is smaller than the screening distance $a$.
Since close collisions are not modified, using a screened potential in place of the M{\o}ller formula to calculate the drag number would make no significant difference.

The accuracy of the first Born approximation for the exponentially screened potential has been carefully analyzed by Joachain \cite{joachain1}. 
He found that it is only accurate to order $\ln(ap/\hbar)Z\alpha/(v/c)$ and only converges for $ap/\hbar \gg 1$ (hence our use of this limit), which are quite severe limitations.
However, his comparison with more accurate solutions shows that what this approximation misses are oscillations in the cross section and that it is accurate for small angle scattering.
Since we are only interested in the mean scattering angle and the most important factor is the cross section for zero scattering, this approximation should not lead to significant errors, with the usual provisos that the fast electron energy is high enough for it not to have bound states and low enough that radiation is not important.

In plasmas, following the treatment used for the drag term, we should only use binary collisions above some scattering angle $\theta_c$ and a statistical treatment of the electric field due to random, thermal fluctuations from charge neutrality below $\theta_c$; hopefully $\theta_c$ will cancel out. 
However, there does not exist an adequate model for the effect of distant charge fluctuations; all existing models do not deal adequately with interparticle correlations due to the electrostatic field and do not include quantum effects, which we have seen to be important for distant interactions.
The best approach appears to be to use \eq{eq-Lsa} with the Debye length in place of the atomic radius, giving
\begin{equation}
\label{eq-Lsi}
L_{\rm s-i}  \approx  \ln \frac{2\lambda_{\rm D}p}{\hbar}  - 0.234 - 0.659 \frac{v^2}{c^2} \qquad \frac{2\lambda_{\rm D}p}{\hbar} \gg 1,
\end{equation}
where the -i indicates that it applies to ions.
We will now briefly review the theoretical models that lead us to this conclusion, in historical order.

Landau \cite{boyd1} used a series of coupled kinetic equations for joint probability densities and set the 3-body joint probability density to zero, because it cannot be solved, and obtained an approximate solution for the 2-body joint probability density in equilibrium, neglecting particle motion.
This showed that pairs of particles interact via the exponentially screened potential, but does not prove that interactions in a plasma can be reduced to sums over pairs of particles; rather it assumes this.
Pines and Bohm \cite{pines1} used Fourier transforms of individual particle positions, but used the random phase approximation, which is equivalent to assuming that the particles are uncorrelated.
Their treatment went beyond that of Landau by considering the effect of particle motion, showing that the exponentially screened potential is only accurate for particles with velocities below the thermal velocity.
Faster particles show reduced, asymmetric screening, for which an analytic solution cannot be obtained, but numerical solutions have been published by a number of authors \cite{wang1981,decyk1987,ellis2011}. 
Several authors have used the Holtsmark distribution for the distant interactions \cite{chandrasekhar1,gasiorowicz1}, which describes the electric field due to a completely random distribution of stationary point charges. This diverges, so an upper cut-off has to be introduced.
It shows that the net effect of a completely random distribution of charges is the same as summing the effect of individual binary collisions with each particle.
In practice, not all distributions are possible because some will have an electrostatic potential energy higher than the total energy of the system.
The Debye length, or something close to it, then appears as the natural cut-off because it gives the distance over which deviations from charge neutrality will give an electrostatic potential energy of the order of the thermal energy.
Spitzer used two different models \cite{cohen1}.
First, he calculated the random fluctuations in the electric field at a point by applying Poisson statistics to the charged particles in a sphere surrounding it. 
Like the Holtsmark distribution, this assumes a completely random distribution of point charges so diverges as the size of the sphere considered tends to infinity and again the Debye length appears as the natural order of magnitude for a cut-off to prevent this divergence. 
He then considered the autocorrelation function of the electric field for charged particles moving in straight lines, which yet again neglects interparticle correlations and again diverges, but this time a cut-off in correlation time is needed.
He used $1/\omega_p$, giving much the same result as a spatial cut-off at $\lambda_D$ in his first model.
This gives a slightly different physical picture, with distant interactions being curtailed due to the limited lifetime of fluctuations from charge neutrality.  
A treatment in terms of dipoles has also been tried, but not published, and also diverges, although this would not be the case if a quantum treatment had been used.

In summary, these models indicate two practical approaches: 
\begin{enumerate}
  \item Sum partial binary collisions over a distance of the order of the Debye length, in effect a $1/r$ potential cut at the Debye length.
  \item Sum full binary collisions with all particles using the screened potential.
\end{enumerate}
We used approach 2 principally because it is more elegant.  
It also seems reasonable to assume that ions move completely at random, which allows us to reduce the many-body problem to a sum of binary interactions when considering the mean effect of many interactions, because electrons will move to cancel any charge build up before the ions are significantly affected by their mutual electrostatic field. 
The imperfect nature of this neutralization due to the thermal motion of the electrons is accounted for by using the Debye screened potential, which will apply to the vast majority of the ions since they have velocities less than the electron thermal velocity.

These considerations also lead us to exclude ion shielding in the Debye length, which is sometimes included by using $\sqrt{\varepsilon_0kT/(Z+1)n_ee^2}$ in place of $\sqrt{\varepsilon_0kT/n_ee^2}$, a conclusion that has been supported by results from numerical modeling \cite{dimonte1}.
The Debye length will have to be modified for degenerate electrons.
A crude approximation is to replace the temperature with $\sqrt{kT^2+E^2_{\rm F}}$ where $E_{\rm F}$ is the Fermi energy \cite{lee-more-pof-1984}.

Unfortunately, it appears that approach 1 will give significantly greater scattering than approach 2 because the quantum mechanical result for the screened potential effectively cuts off the interaction at a distance significantly less than the Debye length (the finite cross section of $\pi [2Z\alpha a/(v/c)]^2$ for zero scattering).
A proper quantum treatment of approach 1 is really required, but we can resort to the uncertainty principle to iron out this difference; we are considering the interaction of the electron with particles within a region of size $\lambda_{\rm D}$ so they can be attributed a minimum momentum spread of order $\hbar/2\lambda_{\rm D}$, interpreting this as imposing a minimum scattering angle and using the small angle approximation gives $\theta_{\min} \sim \hbar/2p\lambda_{\rm D}$.
Using this cut-off with the scattering cross section for a $1/r$ potential obtained from the Dirac equation in the first Born approximation (the Mott formula \cite{mott1}) actually leads to a scattering number slightly smaller than approach 2, but the difference is not significant given the crude approximation being used.

We will now consider scattering from electrons, which is normally ignored because it is only significant in hydrogen.
It is not the same as scattering from ions because the maximum energy exchange of half the fast electron energy gives a maximum scattering angle of $\sin^{-1}\sqrt{2/(\gamma+3)}$ and the physical considerations that led us to sum binary collisions with all atoms and ions using a screened potential would not appear to apply to electrons.
For the case of atoms, it seems clear that the electrons do not share the same screened potential and that scattering from electrons will only occur while the fast electron is inside the atom; the mean effect of the electrons on the total potential has been included in the screened potential and we just need to add the effect of the irregularities in the potential apparent close to electrons.
For the case of a plasma, we cannot apply the same argument that electrons are free to move at random and the Debye (static) screened potential will not apply to most electrons. 
The contribution of the electrons to the effect of distant charge fluctuations would appear to have been included in the screened potential used for the ions, so we just need to include scattering due to the random thermal motion of nearby electrons.
This amounts to saying that approach 1 is more adequate for electrons, with the atomic radius replacing the Debye length for atoms.
However, we have already argued that both approaches should give comparable results, therefore an adequate approximation for electrons should be to account only for the reduced maximum scattering angle, giving
 \begin{equation}
\label{eq-Lse}
L_{s-e} \sim L_{\rm s} - \frac{1}{2}\ln \frac{\gamma+3}{2}.
\end{equation}
The final expression for scattering rate can be written 
\begin{equation}
\label{eq-theta2_2}
\frac{d\mean{\theta^2}}{dt}  \approx \frac{n_e e^4}{2\pi\varepsilon_0^2p^2v} \left[ (Z+1)L_{\rm s}  - \frac{1}{2}\ln \frac{\gamma+3}{2} \right],
\end{equation}
with $L_{\rm s}$ given by \eq{eq-Lsa} for unionized material and by \eq{eq-Lsi} for fully ionized material.
In \cite{atzeni-ppcf-2009} scattering from electrons was dealt with using an exponentially screened potential and the result does not differ significantly. 

As with the drag term, we lack results for partially ionized material.
In the absence of a better treatment, we suggest summing scattering by the ion charge $q$ with a screening distance given by the Debye length for the free electrons $\lambda_{\rm D}(q)$ and scattering by the full nuclear charge $Z$ with a screening distance given by the ion radius $a(q)$.
This amounts to replacing the screening distance in either  \eq{eq-Lsa} or \eq{eq-Lsi} with $\lambda_{\rm D}(q)^{q/Z}a(q)$, so as we are only modifying the argument of a logarithm the approximation does not have to be particularly good.
Ion radii for low values of $q$ are available and for hydrogen-like ions it is $a_0/Z$, but values for intermediate ionization states are not readily available.

\subsubsection{Implications of Drag and Scattering for Fast Ignition}

We are interested in fast electron transport in compressed DT plasma and, for cone-in-shell fast ignition, in the cone material, for which gold has been the preferred candidate.
When the ignition laser is fired the cone tip will have been heated and shock compressed, so it is not entirely accurate to treat it as a cold solid, but we will use these values as an estimate.

The quantity of principle interest arising from the drag is the stopping distance.
Assuming the drag number is a constant we can obtain this analytically;
\begin{equation}
\label{eq-s}
s = \frac{4\pi\varepsilon_0^2}{n_ee^4L_{\rm d}} \frac{E^2}{\gamma}.
\end{equation}
It is tempting to use the relativistic limit $s \propto E$, but for this to be accurate to within 10\% requires energies greater than $4.6$ MeV, so for most cases of interest the full expression should be used.

Numerical calculations of stopping distances in cold matter are tabulated in \cite{icru37} and are available online \cite{estar}. 
As an example, a 1 MeV electron can penetrate up to 400 $\mu$m of gold ($\rho s=$ 0.77 g cm$^{-2}$), so stopping should not be an issue in the cone. 
These calculations also include bremsstrahlung, which allows us to determine when this is indeed negligible.
In hydrogen energy loss to bremsstrahlung (radiation yield) only exceeds $0.1E$ for $E>$ 100 MeV, while in gold this is reached for $E>$ 2 MeV. It is undesirable for an ignition design to have a significant number of electrons above 1 MeV stopping in a cone, so bremsstrahlung should never be a significant energy loss mechanism.

Fast electron stopping in compressed DT plasma has been considered using \eq{eq-Ld} in \cite{atzeni-ppcf-2009}. 
They give an approximate expression for the stopping distance:
\begin{equation}
\label{eq-ASD}
\rho s \approx 1.94 \frac{E_{\rm MeV}^2}{1+1.96E_{\rm MeV}} \rho_{100}^{0.066} \mbox{ g cm}^{-2},
\end{equation}
which was found to be accurate to within 10\% for energies from 1 to 10 MeV and DT mass densities from 300 to 1000 g cm$^{-3}$.
As an example, at 400 g cm$^{-3}$ a stopping distance less than $1.2$ g cm$^{-2}$ requires an energy less than $1.5$ MeV.

Scattering leads to an undesirable increase in the angular spread of electrons, which could be quite serious in the tip of a high-Z cone.
While energy loss remains negligible, the accumulated root mean square scattering angle over a path length $s$ is given by 
\begin{equation}
\label{eq-theta_rms}
\mean{\theta^2}^{1/2} \approx \frac{Ze^2}{\varepsilon_0pv} \sqrt{\frac{n_{\rm a}sL_{\rm s}}{2\pi}},
\end{equation}
where $n_{\rm a}$ is atom number density. 
If we wish to maintain this below, say, $45^{\circ}$ ($\pi/4$) for a 1 MeV electron then for solid gold, tip thickness should be less than 13 $\mu$m ($\rho s =$0.025 g cm$^{-2}$). 
Even if the {\em spatial} spread of electrons is reduced by a collimating magnetic field or vacuum gaps this will not reduce the angular spread, so as soon as the collimating effect ends the electrons will diverge.
This indicates that a lower $Z$ cone tip would be desirable, because even taking into account that thickness should then be increased to avoid shock break out roughly as $Z^{-1/2}$, the net angular scattering will still vary as $Z^{3/4}$. 

The effect of scattering on ignition requirements for an initially parallel beam of electrons entering a uniform sphere of compressed DT plasma has been considered in \cite{atzeni-ppcf-2009} using a Monte Carlo model.
They found that it led to a 10 to 20\% increase in the energy requirement.



\subsection{Beam-plasma instabilities}
\label{bpinstab}
\subsubsection{Motivation}

Electron beam-plasma instabilities are a long-standing field of plasma physics \cite{davi84}. It was early understood that, for a broad parameter range, the beam-driven
excitation of plasma waves can lead to energy and momentum transfer rates between the incident beam and the ambient plasma largely exceeding classical (collisional)
values \cite{bune59,robe67,mors69,fain70,onei71,davi72,lee73,thod76,okad80a}. Such ``anomalous'' relaxation or scattering processes underlie many scenarios of
intense electron beam transport in laboratory \cite{malk02b,sento03} or space  \cite{musc90,medv99,acht07b} plasmas. For instance, they were at the basis of the
pioneering concept of electron beam-driven fusion explored in the 1970s and 1980s \cite{mosher1975,lamp75,thod75,thod76,mill82,suda84,hump90}. Because it relies upon the
propagation and dissipation of an intense electron current into a large-scale plasma, the fast ignition scheme (FIS) has spurred renewed interest in this topic.

The influence of microscopic beam-plasma instabilities in the FIS could be twofold. First, the magnetic turbulence generated by a Weibel-like instability \cite{weib59,frie59}
in the laser absorption region tends to isotropize the fast electrons through random deflections \cite{adam06}. As a result, the electrons are injected into the target with a
large angular spread, which severely constrains the beam energy required for ignition: according to Atzeni \emph{et al.} \cite{atze08}, the ignition energy increases from
$\sim 25\,\mathrm{kJ}$ to $\sim 50\,\mathrm{kJ}$ when the half-angle divergence of the electron source increases from $20^\circ$ to $40^\circ$. Second, the variety of
instabilities arising during the beam transport could entail an enhanced stopping power which could relax the ignition requirements (e.g.\ \cite{yabuuchi2009}). Assuming the beam electrons' mean
energy, $\langle E_b \rangle$, obeys the ponderomotive scaling \cite{wilks1}, the laser ignition energy, $E_L$, is predicted to vary as \cite{atzeni-fastig-pop-2007}
\begin{equation}
  E_{L} \ge 93\left ( \frac{\rho }{300\,\gcmcub} \right)^{-0.9} \left( \frac{f_\mathcal{R}\lambda_0}{0.5 \,\mu\mathrm{m}} \frac{0.25}{\eta_L}Ê\right)^2 \, \mathrm{kJ} \, , 
\end{equation} 
where $\rho$ is the DT core density,  $\lambda_0$ the laser wavelength, $\eta_L$ the laser-to-electron coupling efficiency and $f_\mathcal{R}$ a parameter (close to unity in
the collisional regime) quantifying the effective beam range:
\begin{equation}
  \rho \mathcal{R} = 0.6f_\mathcal{R} \langle E_b \rangle\, \gcmsqu \,.
\end{equation}
The question therefore arises as to whether the excitation of beam-plasma instabilities may entail $f_\mathcal{R} \ll 1$ so as to significantly decrease $E_L$. This could proceed
either directly, through the unstable wave-beam interaction \cite{malk02b}, or indirectly, through an instability-induced increased plasma resistivity \cite{sento03}. 

In contrast to past studies, which mostly focused on electrostatic beam-aligned instabilities, recent FIS-related theoretical works have considered the whole unstable
$\mathbf{k}$-spectrum \cite{bret04,cali06,diec06,grem07,bret07a,bret08,cott08,karm09,bret10a,bret10b}, paying particular attention to the quasi-magnetic filamentation
modes developing normal to the beam direction \cite{pego96,cali98,sento00,hond00,tagu01,silv02,hill05,kato05,tzou06,adam06,scha06,mart08,polo08,karm08a,shve09,khud12}. 
Being all the stronger when the beam and plasma densities are comparable \cite{bret08}, the collisionless instabilities are most likely to disrupt the early propagation of the
beam into the ``low''-density regions of the target. Note, however, that from the optimal beam intensity found in Ref.\ \cite{atzeni-fi-pop-1999}, the beam density is expected to be
\begin{equation}
  n_b \sim 8 \times 10^{21} \, \left(\frac{\rho}{100\, \gcmcub} \right) \left( \frac{\langle E_b \rangle}{1\,\mathrm{MeV}} \right)^{-1} \, \mathrm{cm}^{-3} \,.
\end{equation}
Given such extreme values, collisionless instabilities may arise up to solid densities, which encompasses the laser absorption region, the cone tip (if any) and
part of the DT plasma.

\subsubsection{Main instability classes and their related properties} \label{subsec:linear_instabilities}

Unless otherwise noted, we shall restrict our review to uniform, infinite and initially field-free 2-D beam-plasma systems. The most general (kinetic) description
is afforded by the relativistic Vlasov-Maxwell equations, whose linearization yields the following dispersion relation for electromagnetic perturbations
$\propto e^{i(\mathbf{k}.\mathbf{x}-\omega t)}$
\cite{ichi73}
\begin{equation}\label{eq:dispersion}
  \left(\omega^2\epsilon_{zz} - k_x^2  c^2\right)\left(\omega^2\epsilon_{xx} - k_z^2  c^2\right)
  - \left(\omega^2\epsilon_{xz} + k_xk_z c^2\right)^2 = 0 \,, 
\end{equation}
where the dielectric tensor elements read
\begin{eqnarray}\label{eq:epsi_general}
    \epsilon_{\alpha \beta }(\mathbf{k},\omega) = \delta _{\alpha \beta }
    +\sum_j\frac{\omega_{pj}^2}{\omega^2}\int\int\int d^3p \, \frac{p_{\alpha }}{\gamma(\mathbf{p}) }\frac{\partial f_j^{(0)}}{\partial p_{\beta }} \nonumber \\
+\sum_j\frac{\omega_{pj}^2}{\omega^2}\int\int\int d^3p\, \frac{p_{\alpha }p_{\beta }}{\gamma(\mathbf{p})^2 }
\frac{\mathbf{k}\cdot \left(\partial f_j^{(0)}/\partial \mathbf{p}\right)}{m_j\omega -\mathbf{k}\cdot \mathbf{p}/\gamma(\mathbf{p}) } \,.
\end{eqnarray}
Here, $\mathbf{k} = (k_x,k_z)$ is the real wave number, $\omega$ is the complex frequency, $\omega_{pj} = (n_je_j^2/m_j \epsilon_0)^{1/2}$ is the plasma frequency
of species $j$ and $\gamma (\mathbf{p}) = [1+(p/m_jc)^2]^{1/2}$ is the Lorentz factor. In the following, the index $j = (b,p)$ stands for the electron beam and plasma
components. Collisional effects are neglected at this stage and will be discussed in Sec.\ \ref{subsec:collisions}. The main ingredient in \eref{eq:epsi_general} is the unperturbed
distribution function $f_j^{(0)}(\mathbf{p})$.  In the context of the FIS, there is no obvious physical reason supporting a particular model distribution for the beam electrons.
A variety of descriptions can be found in the literature, ranging from monokinetic \cite{blud60,pego96} to Maxwellian-like \cite{yoon89,yoon07,taut05,taut06} through waterbag \cite{yoon87,silv02,bret05,grem07,cott08} and Kappa \cite{laza08} distributions. However, in order to address potentially large (relativistic) thermal spreads, it
appears convenient to model the beam-plasma system by means of drifting Maxwell-J\"uttner distribution functions \cite{jutt11,wrig75}
\begin{equation} \label{eq:fMJ}
  f_j^{(0)}(\mathbf{p}) = \frac{\mu_j}{4 \pi m_j c^3 \gamma_j^2 K_2 (\mu_j/\gamma_j)} \exp \left[-\mu_j \left(\gamma - \beta_j \frac{p_z}{m_j c}\right) \right] \, ,
\end{equation}
where $\beta_j = \langle p_z/m_j \gamma c \rangle $ is the $z$-aligned mean drift velocity, $\gamma_j = (1-\beta_j^2)^{-1/2}$, $\mu_j = m_j c^2/T_j$ is the
normalized inverse temperature and $K_2$ is a modified Bessel function. Two arguments can be made for this model distribution. First, it permits an exact
resolution of the 2-D fully relativistic spectrum at an affordable numerical cost \cite{bret08}. Second, it has been shown, under certain conditions, to model
with some accuracy the relativistic electron phase space observed in laser-plasma simulations \cite{cott08}. Care must be taken, though, in the numerical
evaluation of Eqs.\ (\ref{eq:epsi_general}-\ref{eq:fMJ}) in the complex $\omega$-plane as detailed in Ref.\ \cite{bret10a}.

\begin{figure}[tbp]
\begin{center}
\includegraphics[width=0.45\textwidth]{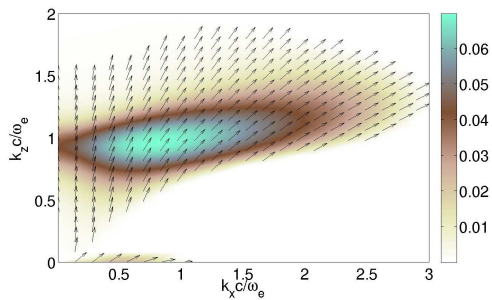}
\end{center}
\caption{Overlay of the normalized growth rate $\delta=\Im\omega/\omega_e$ (shaded colors) and of the electric field orientation (arrows) in the $(k_x,k_z)$ space.
Maxwell-J\"uttner distribution functions are considered with $n_b/n_p=0.1$, $\gamma_b=4$, $T_b=50$ keV, $T_p=5$ keV and the beam drifting along the $z$ axis.}
\label{fig:exsez_wi_gb_4_alpha_1e-1_Tb_50_Tp_5}
\end{figure}

Three instability classes can be identified according to their wave vector's orientation and electromagnetic properties. This is exemplified in
Fig.\ \ref{fig:exsez_wi_gb_4_alpha_1e-1_Tb_50_Tp_5} which displays the $\mathbf{k}$-dependence of the normalized growth rate 
\begin{equation}
\delta = \Im \frac{\omega}{\omega_e},
\end{equation}
where  $\omega_e=[n_ee^2/\epsilon_0m_e]^{1/2}$ is the nonrelativistic total plasma frequency ($n_e=n_b+n_p$) for a dilute-beam configuration: $n_b/n_p = 0.1$, $\gamma_b=4$, $T_b=50$ keV and
$T_p=5$ keV. The plasma drift velocity follows from the current neutrality condition $\beta_p = - \beta_b n_b/n_p$. 

The well-known two-stream modes \cite{bohm49} are located along the beam direction ($k_x =0$), with a peak growth rate $\delta_\mathrm{max} \sim 0.04$ at
$k_{z, \mathrm{max}}c/\omega_e \sim 1/\beta_b$. These are purely electrostatic plasma waves propagating at the phase velocity $\Re \omega/k \sim \beta_b$.
Their maximum growth rate is given by the approximate analytical expressions (in the weak $n_b/n_p$ limit) 
\begin{equation}
  \delta_\mathrm{max}^\mathrm{TS} \approx \left\{
\begin{array}{ll}
  \frac{\sqrt{3}}{2^{4/3}} \frac{1}{\gamma_b}\left(\frac{n_b}{n_p}\right)^{1/3} &\mathrm{if} \quad \frac{T_b}{m_e c^2} \le \beta_b^2 \gamma_b \left( \frac{n_b}{n_p}\right)^{2/3} \,,\\
  \beta_b^2 \frac{m_e c^2}{T_b} \frac{n_b}{n_p} & \mathrm{otherwise} \, ,
\end{array}
  \right.
\end{equation}  
in the hydrodynamic (cold) and kinetic regimes, respectively \cite{suda84}.  The orientation of the associated electric perturbation can be evaluated from the linear
relation $E_x(\omega,\mathbf{k})/E_z(\omega,\mathbf{k}) = (k_x^2c^2/\omega_e^2-\omega^2\epsilon_{zz})/(k_xk_zc^2/\omega_e^2+\epsilon_{xz})$ \cite{bret04}.
As expected, Fig.\ \ref{fig:exsez_wi_gb_4_alpha_1e-1_Tb_50_Tp_5} shows that the two-stream modes fulfill $\mathbf{k}\times \mathbf{E} =0$.

\begin{figure}[tbp]
\begin{center}
\includegraphics[width=0.4\textwidth]{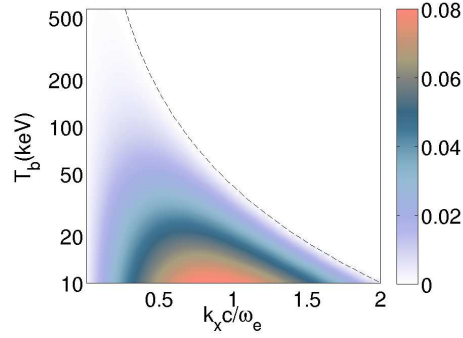}
\end{center}
\caption{Filamentation growth rate $\delta$ as a function of the tranverse wave vector $k_x$ and the beam temperature $T_b$.
Maxwell-J\"uttner distribution functions are considered with  $\gamma_b = 2$, $n_b/n_p = 0.1$ and $T_p = 5$ keV. The cut-off wave vector $k_\mathrm{lim}$ [Eq.\ (\ref{eq:klim})] is plotted in dashed line.}
\label{fig:wi_k_Tb_gb_2_Tp_5}
\end{figure}

\begin{figure*}
\begin{tabular}{c}
\includegraphics[width=0.8\textwidth]{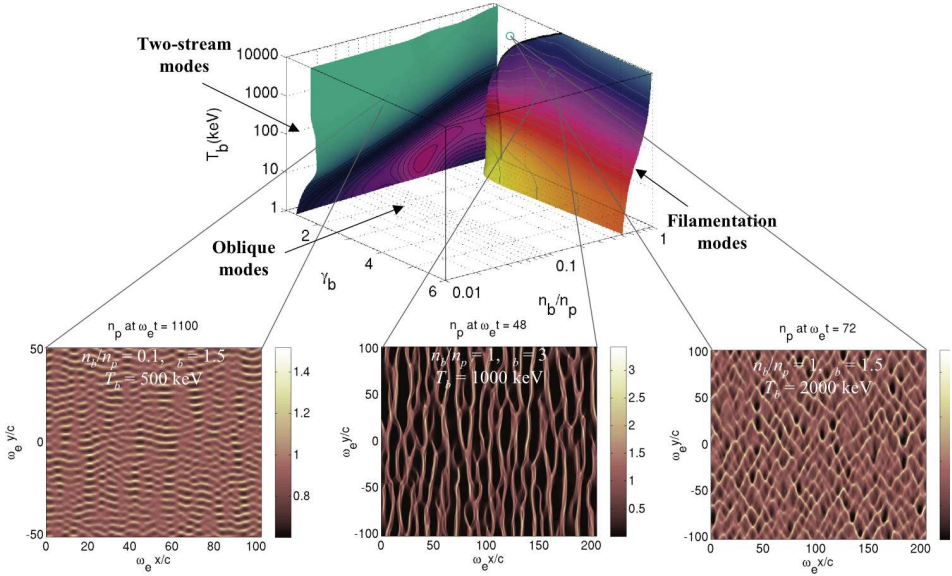}
\end{tabular}
\caption{(top) Hierarchy of the two-stream, oblique and filamentation modes in the $(n_b/n_p, \gamma_b, T_b)$ parameter space for
Maxwell-J\"uttner distribution functions. (bottom) Plasma density profiles at the end of the linear phase as predicted by 2-D PIC simulations, each
ruled by a specific instability class. The plasma temperature is $T_p = 5$ keV and the beam flows along the $y$-axis.}
\label{fig:hierarchie}
\end{figure*}

The filamentation instability, which arises in systems composed of counterstreaming species, belongs to the family of anisotropy-driven instabilities typified by the Weibel
instability \cite{weib59,frie59}.  Hence, the two designations are often used interchangeably in the literature. As the classical Weibel instability, the filamentation modes
develop preferentially normal to the ``hot'' (beam) direction $k_z \approx 0$. They correspond to aperiodic ($\Re \omega =0$), mostly magnetic fluctuations amplified
by the repulsive force between electron currents of opposite polarity. In Fig.\ \ref{fig:exsez_wi_gb_4_alpha_1e-1_Tb_50_Tp_5}, the filamentation growth rate  is seen
to maximize at $k_xc/\omega_e \sim 0.5$ with $\delta_\mathrm{max} \sim 0.02$. An analytical estimate can be derived in the cold limit ($T_b = T_p = 0$), which reads
\cite{pego96}.
\begin{equation} \label{eq:filam_cold}
  \delta_\mathrm{max}^\mathrm{F} \approx \beta_b\sqrt{\frac{n_b}{\gamma_b n_p}}
\end{equation}
for $1\le k_xc/\omega_e <\infty$. In the kinetic (hot) regime associated to Fig.\ \ref{fig:exsez_wi_gb_4_alpha_1e-1_Tb_50_Tp_5}, by contrast, the unstable domain
is restricted to $0\le k_x \le k_\mathrm{lim}$, with the cut-off wave vector \cite{bret10b}
\begin{equation} \label{eq:klim}
  k_\mathrm{lim}^2 = \frac{1}{2}
  \left[ \mathcal{F}_2 - \mathcal{F}_0 + \sqrt{(\mathcal{F}_2-\mathcal{F}_0)^2+4(\mathcal{F}_0\mathcal{F}_2-\mathcal{F}_1^2)} \right] \frac{\omega_e^2}{c^2} \, ,
\end{equation}
where $\mathcal{F}_n = \sum_j (n_j/n_e) \mu_j \beta_j^n$. Assuming $n_b/n_p \ll 1$, we have $k_\mathrm{lim} \sim \beta_b \sqrt{(n_b/n_p)(m_ec^2/T_b)}$. Interestingly,
the fastest-growing filamentation wave vector has the same scalings as $k_\mathrm{lim}$ \cite{bret10a}.  The shrinking of the unstable domain for increasing beam
temperatures is illustrated in Fig.\ \ref{fig:wi_k_Tb_gb_2_Tp_5} for $\gamma_b = 2$, $n_b/n_p = 0.1$ and $T_p = 5$ keV.  Along with the decrease in $k_\mathrm{lim}$,
the peak growth rate is found to drop as $\delta_\mathrm{max}\propto T_b^{-3/2}$ \cite{bret10a}. Further analysis shows that, similarly to the cold-fluid scaling
(\ref{eq:filam_cold}), the instability is also quenched in the high-$\gamma_b$ limit as $\delta_\mathrm{max} \propto \gamma_b^{-1/2}$ due to the  beam's increasing
inertia \cite{bret10a}. Note that any combination of Maxwell-J\"{u}ttner functions with non-vanishing $\beta_j$'s is filamentation unstable (\emph{i.e.}, $k_\mathrm{lim} > 0$)
due to a finite anisotropy. In practice, though, \eref{eq:klim} sets an effective stabilization threshold when $k_\mathrm{lim} \le 1/L_x$, where $L_x$ is the
transverse size of the beam-plasma system (typically of the order of the laser spot). This incomplete stabilization contrasts with the total suppression occurring for
model distributions allowing for independent longitudinal and transverse thermal spreads \cite{silv02,bret05,bret10b,taut08}. In fact, filamentation proves mostly
vulnerable to the transverse temperature $T_{bx} = \langle p_x^2/\gamma \rangle$, causing a pressure force counteracting the magnetic pinching force. In the
simplified waterbag case with weak beam density and temperature, stabilization is thus predicted for \cite{silv02}
\begin{equation} \label{eq:stabil_waterbag}
  \frac{\Delta \beta_b^2}{\beta_b^2} \ge \frac{n_b}{\gamma_b n_p} \,,
\end{equation}
where the beam's transverse velocity spread, $\Delta \beta_b$, is related to the transverse temperature, $T_{b\perp} \equiv T_{bx}$, through \cite{silv02}
\begin{equation}
   T_{bx} = \frac{m_e\gamma_bc^2}{2}\left[ 1 + \frac{1-\Delta \beta_{b}^2}{2\Delta \beta_{b}}
  \ln \left(\frac{1-\Delta \beta_{b}}{1+\Delta \beta_{b}}\right) \right] \,,
\end{equation}
which simplifies to $T_{bx} \sim m_e\gamma_b c^2 \Delta \beta_b^2/3$ in the limit $\Delta \beta_b \ll 1$. 

Although the filamentation modes are essentially magnetic, Fig.\ \ref{fig:exsez_wi_gb_4_alpha_1e-1_Tb_50_Tp_5} demonstrates that their electric-field component is not
purely inductive ($\mathbf{k}.\mathbf{E} \neq 0$). This follows from the fact that, except for perfectly symmetric systems (\emph{i.e}, with $n_b=n_p$, $\beta_b = -\beta_p$
and $T_b=T_p$), the off-diagonal term $\epsilon_{xz}$ in \eref{eq:dispersion} is generally nonzero \cite{bret07a}. The ion response to the resulting space-charge force
should therefore be taken into account in the weaky-unstable regime \cite{tzou06,ren06}.

The spectrum in Fig.\ \ref{fig:exsez_wi_gb_4_alpha_1e-1_Tb_50_Tp_5} turns out to be governed by off-axis modes, thus propagating obliquely to the beam. The
fastest-growing oblique mode with $\delta_\mathrm{max} = 0.07$ is located at $(k_x,k_z)= (0.8,0.95)$. As shown in Ref.\ \cite{bret10b}, these modes are quasi-electrostatic
in a broad system-parameter range including the configuration of Fig.\ \ref{fig:exsez_wi_gb_4_alpha_1e-1_Tb_50_Tp_5}. For $n_b/n_p \ll 1$, their maximum growth
rate can be estimated to be \cite{ruda71,suda84,bret10b}
\begin{equation}
  \delta_\mathrm{max}^\mathrm{0} \approx \frac{\sqrt{3}}{2^{4/3}}\left(\frac{n_b}{\gamma_b n_p}\right)^{1/3} \, ,
\end{equation}
in the hydrodynamic regime defined by
\begin{equation}
  \frac{T_b}{m_e c^2} < \frac{3}{2^{10/3}} \left(\frac{n_b}{n_p}\right)^{2/3}\gamma_b^{1/3} \frac{\left(1+\gamma_b^{-2} \right)^{2/3}}{\left( 1+\gamma_b^{-1} \right)^2}  \,. \\
\end{equation}
In the opposite kinetic limit, one has approximately
\begin{equation}
  \delta_\mathrm{max}^\mathrm{0} \approx \beta_b^2 \frac{m_e c^2}{T_b} \frac{n_b}{n_p} \, .
\end{equation}
In both regimes, the longitudinal wave vector of the dominant oblique mode is correlated to the dominant two-stream mode ($k_zc/\omega_e \sim 1/\beta_b$),
whereas the transverse component $k_xc/\omega_e$ decreases below unity when moving into the kinetic regime \cite{bret10a}. 

The domain of preponderance of each instability class has been computed in the ($n_b/n_p$,$\gamma_b$,$T_b$) parameter space for a fixed plasma temperature
$T_p=5\,\mathrm{keV}$ \cite{bret08,bret10a}. The surfaces that delimit regions governed by different instability classes are displayed in Fig.\ \ref{fig:hierarchie} and
colored according to the local maximum (in $\mathbf{k}$-space) growth rate. The two-stream instability prevails for non-relativistic beam drift energies ($\gamma_b -1 \ll 1$),
as well as in weakly relativistic systems with hot enough beams. This follows from the quenching of the filamentation and oblique instabilities with decreasing $\beta_b$
and increasing $T_b$, respectively. Filamentation modes govern systems where the beam and plasma densities are similar (in the FIS, this mostly concerns the laser
absorption region), whereas oblique modes are dominant for dilute relativistic beams. The  filamentation-to-oblique transition is mostly determined by $\gamma_b$ for dense,
cold and weakly relativistic beams, and by $n_b$ in the relativistic and ultra-relativistic regimes. Note that oblique modes always dominate for hot enough relativistic
beams. These results are illustrated by the lower panels of Fig.\ \ref{fig:hierarchie} showing the plasma density profiles observed in three 2-D PIC simulations, each
ruled by a distinct instability class. The spectral characteristics of each modulated pattern have been checked to perfectly agree with linear theory.

\subsubsection{Collisional effects}
\label{subsec:collisions}

Collisions are expected to influence the development of the instabilities in the high-density, low-temperature regions penetrated by the electron beam, that is, at a
distance from the laser absorption region. As a consequence, most of the studies performed in this respect have considered dilute collisionless beams interacting
with dense, nonrelativistic collisional plasmas \cite{cott08,hao08,karm08a,fior10,hao12}. Collisional effects are frequently described by simplified Krook-like models,
which consist in introducing phenomelogical relaxation terms in the Vlasov equation \cite{ophe02}. The most accurate approach of this kind is the particle-number-conserving
BGK model \cite{bhat54}
\begin{eqnarray} 
\label{eq:bgk}
  \frac{\partial f_p}{\partial t} + \mathbf{v}.\nabla_\mathbf{x} f_p -e \left(\mathbf{E} + \mathbf{v}\times \mathbf{B}\right).\nabla_\mathbf{v} f_p = C(f_p) \nonumber \\
   = -\nu \left(\delta f_p - \delta n_p f_p^{(0)} / n_p^{(0)} \right) \, ,
\end{eqnarray}
where $\delta f_p = f_p-f_p^{(0)}$ and $\delta n_p  = \int d^3v \delta f_p$. The BGK model can be generalized to conserve momentum and energy as well. A more rigorous approach makes use of the Landau collision operators \cite{shka66,bran06}.
In the case of a large ion charge ($Z \gg 1$), electron-ion collisions prevail over electron-electron collisions and are described by the operator 
\begin{equation} \label{eq:landau}
  C_{ei}(f_p) =\frac{Z n_p e^4 \ln \Lambda}{8\pi\epsilon_0^2m_e^2}  \nabla_{\mathbf{v}}.\frac{1}{v}\left(I-\frac{\mathbf{v}\mathbf{v}}{v^2}\right). \nabla_{\mathbf{v}}f_p(\mathbf{v}) \, ,
\end{equation}
where $I$ denotes the identity operator and $\ln \Lambda$ is the Coulomb logarithm. In principle, the BGK collision frequency $\nu$ should be adjusted so as to
reproduce the plasma susceptibility obtained from the Landau operator in the collisional limit, which yields $\nu = \nu_{ei}$, where $\nu_{ei}$ is the usual collision
frequency \cite{bend93}.

\begin{figure}[tbp]
\begin{center}
\includegraphics[width=0.35\textwidth]{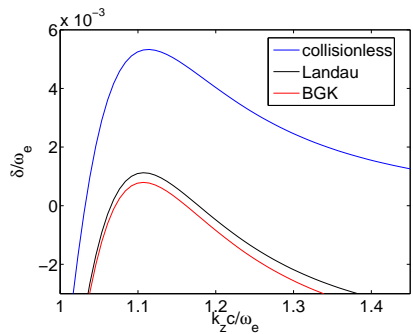}
\end{center}
\caption{Growth rate of the two-stream instability with (red, black) and without (blue) $e-i$ collisions for $n_e=10^{23}\textrm{cm}^{-3}$, $n_b/n_p=0.01$, $T_b=100$ keV, $\gamma_b=3$,
$T_p=1$ keV, $Z=10$ and $\ln\Lambda=2$. The BGK curve (red) is found to understimate the exact Landau curve (black).}
\label{fig:Comparaison4cwi}
\end{figure}

An exact evaluation of the collisional two-stream instability using a Maxwell-J\"uttner-distributed beam and the electron-ion Landau operator has been recently carried
out \cite{verm12}. Figure \ref{fig:Comparaison4cwi} plots the $k_z$-dependence of the growth rate for the parameters $n_e=10^{23}\textrm{cm}^{-3}$, $n_b/n_p=0.01$,
$T_b=100$ keV, $\gamma_b=3$, $T_p=1$ keV, $Z=10$ and $\ln\Lambda=2$. There follows a collision frequency $\nu_{ei}/\omega_e =0.01$, that is, approximately twice
the maximum collisionless  growth rate (blue curve). In the presence of collisions, the peak growth rate drops from $\delta_\mathrm{max} = 5.3\times 10^{-3}$ to
$1.1\times 10^{-3}$, while the dominant wave number only slightly decreases. Note that the BGK model (red) yields a peak growth rate about 30\% lower than the Landau
value (black).

\begin{figure}[tbp]
\begin{center}
\includegraphics[width=0.35\textwidth]{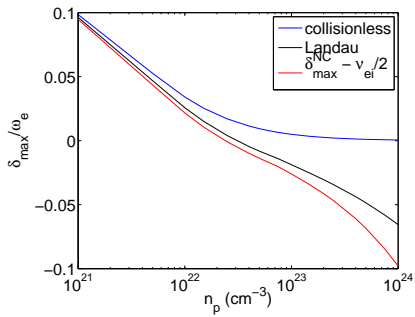}
\end{center}
\caption{Maximum growth rate of the two-stream instability for $n_b=10^{21}\,\textrm{cm}^{-3}$, $T_b=100$ keV, $\gamma_b=2$, $T_p=300$ eV, $Z=10$ and $\ln\Lambda=2$.}
\label{fig:Gamma300ev}
\end{figure}

If strong enough, collisions are able to completely stabilize the two-stream modes \cite{sing64,hao12}. This is illustrated in Fig.\ \ref{fig:Gamma300ev}, where the
maximum growth rate is plotted as a function of the plasma density, the beam density being fixed at $n_b = 10^{21}\,\textrm{cm}^{-3}$. The other parameters are
$\gamma_b=2$, $T_b=100\,\mathrm{keV}$, $T_p=300\,\mathrm{eV}$, $Z=10$ and $\ln\Lambda=2$. The exact collisional curve (black) is fairly approximated by
the expression
\begin{equation}
\label{eq:deltaapp}
  \delta_\mathrm{max} \approx  \delta_\mathrm{max}^\mathrm{NC} - \nu_{ei}/2
\end{equation}
where $ \delta_\mathrm{max}^\mathrm{NC}$ is the maximum collisionless growth rate (blue curve). The relative error between \eref{eq:deltaapp} and the exact values is
found to increase as the instability weakens. Complete stabilization ($\delta_\mathrm{max}\le 0$) is achieved here for $n_p \ge 3.2\times 10^{22}\,\textrm{cm}^{-3}$, which
corresponds to $\nu_{ei}/\omega_e \ge 0.04$. By contrast, \eref{eq:deltaapp} yields a somewhat underestimated stabilization threshold ($n_p \ge 2.5\times 10^{22}\,\textrm{cm}^{-3}$). 

Because of the close connection between two-stream modes and oblique modes in a broad parameter range \cite{bret10a}, the latter are affected by collisions in a similar fashion,
exhibiting, in particular, complete stabilization in the strong collisional limit \cite{hao12}.

\begin{figure}[tbp]
\begin{center}
\includegraphics[width=0.35\textwidth]{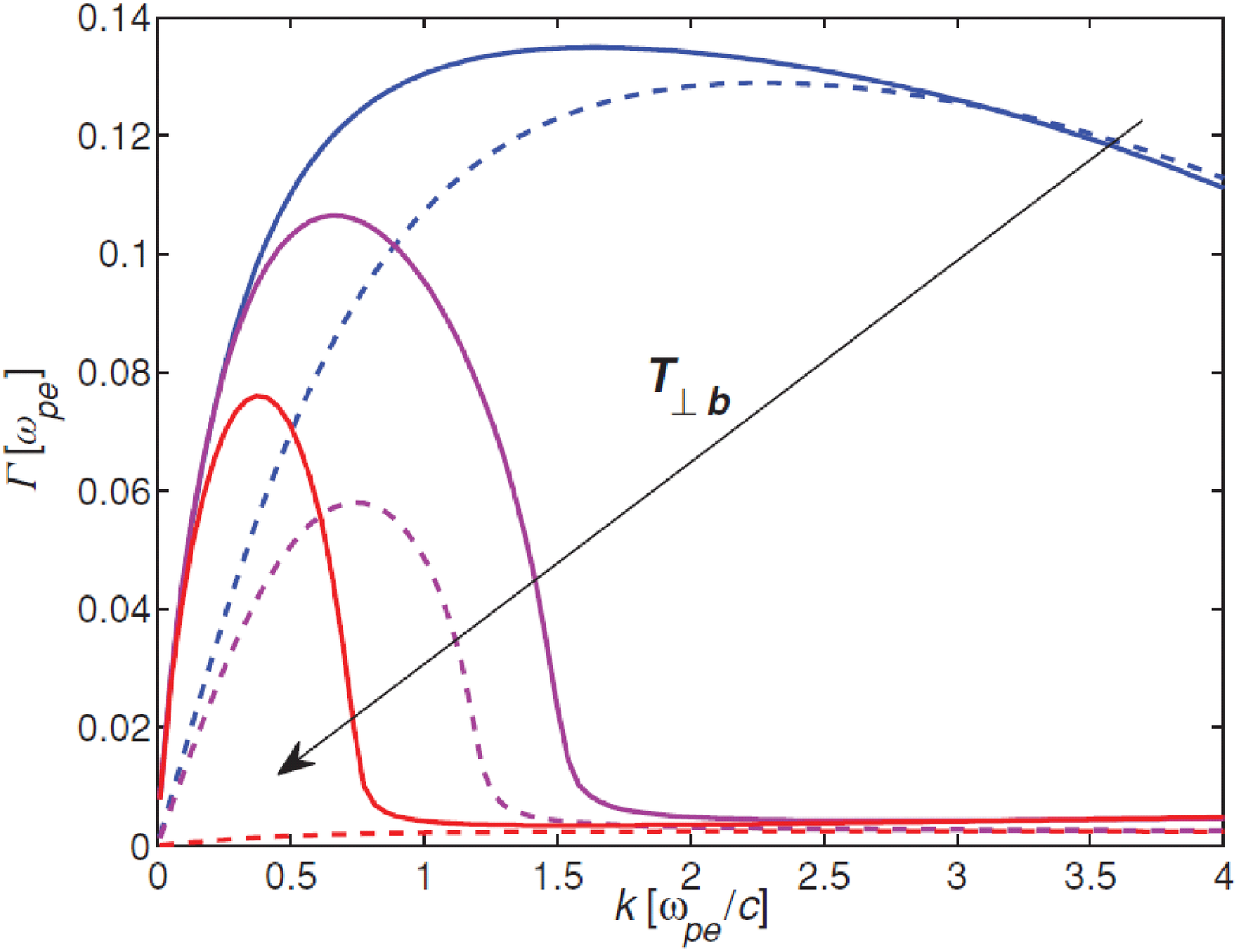}
\end{center}
\caption{Filamentation growth rate as a function of the wave vector in collisionless (dashed lines) and collisional (solid lines)
configurations for $\gamma_b = 5$, $T_p = 10\,\mathrm{keV}$, $n_b/n_p = 0.1$, $\nu/\omega_e = 0.5$ and increasing beam transverse temperatures:
$T_{b\perp} =0.5\,\mathrm{keV}$ (blue), $T_{b\perp} =9\,\mathrm{keV}$ (magenta) and $T_{b\perp} =34\,\mathrm{keV}$ (red) (after Fiore \emph{et al.} \cite{fior10}).}
\label{fig:Fiore}
\end{figure}

As first demonstrated by Molvig \cite{molv75} and further investigated in Refs.\ \cite{hao08,karm08a,fior10}, a starkly different scenario takes place for the filamentation
instability. The main result is that for dilute and energetic enough beams, collisions keep the system unstable regardless of the transverse beam temperature. Moreover,
collisions shrink the unstable domain towards small $k$'s. Figure \ref{fig:Fiore}, which is extracted from Ref.\ \cite{fior10}, illustrates these effects by comparing the
$k$-variations of the collisionless and collisional filamentation growth rates for waterbag distribution functions with $\gamma_b = 5$, $T_p = 10\,\mathrm{keV}$, $n_b/n_p = 0.1$.
A BGK collision model is employed with $\nu/\omega_e = 0.5$. As expected, the instability is weakened and confined to decreasing wave numbers as the beam transverse temperature
is raised. The instability is enhanced in the presence of collisions, especially in the large-temperature limit ($T_{b\perp} = 34\,\mathrm{keV}$) where, according to
\Eref{eq:stabil_waterbag}, it should be stabilized in the collisionless regime. PIC simulations confirm the predicted robustness of the collisional filamentation
and the generation of filamentary structures larger than in the collisionless regime \cite{karm08b,fior10}. Note that the highly-collisional filamentation instability 
corresponds to the so-called resistive filamentation instability seen in hybrid simulations \cite{grem02,honr06,solo08,solo09}, which is derived assuming the return
current obeys Ohm's law $\mathbf{E} = \eta \mathbf{j}_p$, where $\eta$ is the electrical resistivity \cite{hump90,grem02}.
 
\subsubsection{Weibel/filamentation instability in fast electron generation and transport}

Multidimensional PIC simulations of the fast electron generation in overcritical plasmas have shown that the filamentation instability
plays a major role in the laser-absorption region \cite{pukh96,sento00,sento03,adam06,ren06,okad07,debayle1}. This is so because,
for a large enough laser spot ($\gg \lambda_0$) and normal incidence, the electron acceleration initially takes place within an essentially
1-D geometry. As a result of this plane-wave approximation, the transverse canonical momentum is conserved: $p_\perp(x(t),t) -eA_\perp(x(t),t) = p_{\perp0}$.
As the vector potential $A_\perp$ vanishes over a few plasma skin depths, the fast electrons quickly recover their initial (thermal) transverse
momentum $\vert p_{\perp0}\vert$ ($\ll \vert p_x \vert$) when penetrating into the target. There follows an input electron  distribution strongly
elongated along the longitudinal direction, hence prone to the Weibel/filamentation instability. Magnetic fluctuations are then spontaneously
generated along the target surface, leading to a fragmentation of the fast electron profile into small-scale filaments. 

\begin{figure}[tbp]
\begin{center}
\includegraphics[width=0.45\textwidth]{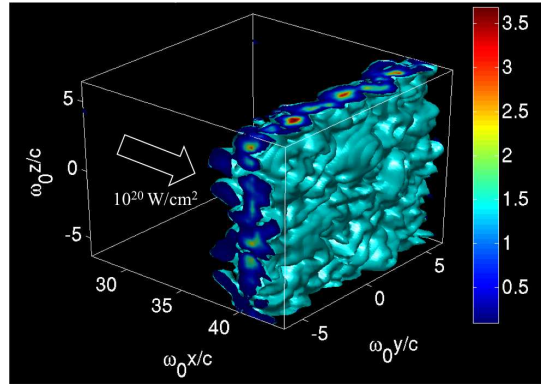}
\end{center}
\caption{Isosurface of the magnetic field $\vert \mathbf{B} \vert $ associated to Weibel-generated fluctuations (averaged over a laser cycle) in a 3-D PIC simulation
of a $10^{20}\,\Wcmsq$ laser plane wave interacting with a $50n_c$ plasma. The arrow points along the incoming laser direction and the
magnetic field is normalized to $m_e \omega_0/e$.}
\label{fig:weibel_ilp}
\end{figure}

This process is illustrated in Fig.\ \ref{fig:weibel_ilp} in the case of a 3-D PIC simulation of a  $10^{20}\,\Wcmsq$ laser plane wave impinging
onto a $50n_c$ plasma. At $t = 100\omega_0^{-1}$ (where $\omega_0$ is the laser frequency), magnetic fluctuations have grown in the interaction
region to an amplitude  $\vert \mathbf{B} \vert \sim m_e \omega_0/e$ with a transverse size $\lambda \sim c/\omega_0$. The underlying physics
can be understood as follows.  Let us assume for simplicity that the plane wave approximation initially holds and that the hot electron distribution
is cold in the transverse direction, $f_b(\mathbf{p}) \propto \exp [\mu \sqrt{1+p_x^2/(m_ec)^2}]\delta(p_y)$. In the high-energy limit
($\mu < 1$), $\mu$ is related to the mean Lorentz factor as $\langle \gamma \rangle =  K_2(\mu)/K_1(\mu) \sim 2/\mu$.  Substitution of the above
distribution into Eqs. \ref{eq:dispersion}-\ref{eq:epsi_general} (noting that the ``hot'' axis is now the $x$-axis) readily yields the maximum
Weibel growth rate  $\delta_\mathrm{max} = \omega_{ph} \sqrt{K_0(\mu)/K_1(\mu)} \sim \omega_{pb}\sqrt{-\mu \ln \mu}$, where $\omega_{pb}$
is the hot electron plasma frequency. Because of the vanishing dispersion in the transverse momentum, the growth rate saturates to
$\delta_\mathrm{\max}$ for wave vectors $k_y \ge \sqrt{\mu}$.  Assuming that the hot electron density is equal to the critical density
($\omega_{pb} = \omega_0$) and that the normalized mean electron energy scales as $\langle \gamma \rangle \sim a_0^\alpha$, where
$\alpha \sim 2/3-1$ \cite{wilks1,beg1,ping08}, one gets $\delta_\mathrm{max} \sim \omega_0 \sqrt{\ln a_0 /a_0^{\alpha} }$. As a result, the
growth rate is comparable to the laser frequency for $a_0 > 1$. 

The saturated level of the magnetic fluctuations, $B_\mathrm{sat}$, can be estimated from the widely used trapping criterion \cite{davi72,yang94,silv02,acht07b,okad07} which expresses the fact that the exponential growth phase comes to an end when the electron bouncing frequency inside
a magnetized filament of period $2\pi/k_y$, $\omega_\mathrm{B} \sim \omega_0 \sqrt{\langle 1/\gamma \rangle (k_yc/\omega_0)(eB_\mathrm{sat}/m_e \omega_0)}$, 
becomes of the order of the growth rate $\delta (k_y)$. Using the above estimates with $\langle 1/\gamma \rangle = K_0(\mu)/K_1(\mu)$ and $k_y = \mu^{1/2}$,
one finds the saturated magnetic amplitude $eB_\mathrm{sat}/m_e \omega_0 \sim \mu^{-1/2} = a_0^{\alpha/2}$. The maximum quiver momentum being
$p_y = m_e \langle \gamma \rangle \omega_\mathrm{B}/k_y$, the approximate divergence is $p_y/p_x \sim \omega_\mathrm{B}/k_y c \sim \sqrt{\ln a_0}$.
Magnetic deflections within the self-generated magnetized filaments then rapidly cause the hot electrons to acquire a divergence of the order
unity. The Weibel/filamentation instability therefore appears as the mechanism mainly responsible for the large angular spread seen in simulations
\cite{adam06,debayle1} and experiments \cite{green1}.  

The assumption of a zero transverse temperature for the hot electrons actually holds only a few skin depths away from the laser absorption region.
In reality, however, the instability develops within the skin layer, where the electron distribution has a finite anisotropy, thus yielding a weaker
growth rate. This process has been addressed in Ref.\ \cite{okad07} by fitting the simulated hot electron distribution to a semi-relativistic,
two-temperature Maxwellian \cite{okad80a}. Defining $A = T_x/T_\perp-1 > 0$ and applying the same reasoning as above, the saturated magnetic
field is expected to scale as
\begin{eqnarray}
B_\mathrm{sat} &\approx& 0.16 a_0^{2\alpha} \sqrt{\frac{n_b}{n_c}} \frac{A^{5/2}}{(A+1)^3} \frac{m_e \omega_0}{e} \nonumber \\
&\le& 0.04 a_0^{2\alpha}\sqrt{\frac{n_b}{n_c}}  \frac{m_e \omega_0}{e} \, .
\end{eqnarray}
3-D PIC simulations performed with  a laser amplitude $a_0 = 3$ and plasma densities $n_e = (1-2) n_c$ predict a maximum anisotropy $A \sim 2-10$
and a saturated magnetic amplitude in reasonable agreement with the above estimate. 

\begin{figure}[tbp]
\begin{center}
\includegraphics[width=0.48\textwidth]{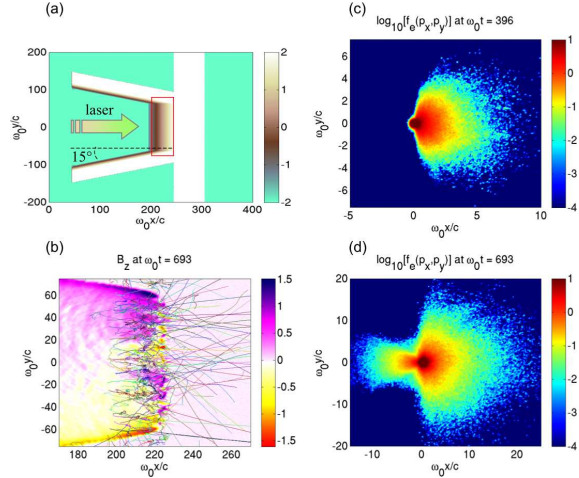}
\end{center}
\caption{2-D PIC simulation of the interaction of a $10^{19}\,\Wcmsq$ laser pulse with a cone-attached target: initial density profile (a);
typical electron trajectories within the box shown in (a) and map of the quasistatic magnetic field $B_z$ (normalized to $m_e /\omega_0e$) at the time
of the on-target laser peak (b); $p_x-p_y$ electron phase space around the absorption region at $t=300\omega_0^{-1}$ before (c) and at the time (d)
of the laser peak. See text and Ref.\ \cite{bato08} for further details.}
\label{fig:cone_divergence}
\end{figure}

The 2-D PIC results displayed in Figs. \ref{fig:cone_divergence}(a-d) further depict this self-generated magnetic scattering effect
in the case of a $10^{19}\,\Wcmsq$ ($a_0 = 3$) laser pulse injected into a $100n_c$ cone-guided target \cite{bato08}. The pulse has a
$500\omega_0^{-1}$ duration and a $16\lambda_0$ width.  A $1\,\mu\mathrm{m}$ scale-length exponential preplasma is added on the inner target walls
[panel (a)]. A set of typical electron trajectories inside the absorption region are plotted in panel (b). Beside being reflected by the laser field
in the low-density region ($x \sim 200c/\omega_0$), the fast electrons undergo strong deflections across the skin layer ($x\sim 220c/\omega_0$) due
to magnetic modulations of amplitude $B_z \sim 1.5 m_e \omega_0/e$. The resulting electron momentum distribution is shown at $t=300\omega_0^{-1}$
before  the laser maximum [panel (c)] and at the time of the laser maximum [panel (d)]. The root-mean-square angle of the fast ($> 1\,\mathrm{MeV}$)
electrons is found to increase during this time interval from $\langle \theta^2 \rangle ^{1/2} \sim 34^\circ$ to $\langle \theta^2 \rangle ^{1/2} \sim 48^\circ$.

In the plane-wave case, the rapid magnetic build-up breaks the invariance along the transverse directions, and hence the transverse canonical momentum
is no longer conserved. The electron acceleration is modified due to the coupling between the laser field and the quasistatic magnetic fluctuations. This
multidimentional effect causes the transverse velocity of the electrons injected into the target to oscillate at the laser frequency. More precisely,
it has been found in Ref.\ \cite{adam06} using a quasilinear analysis that the averaged transverse electron velocity behaves as
\begin{eqnarray}
  \langle v_y(x,t) \rangle &\approx \gamma^{-2} \sum_k \int_0^{x} dx' \int_0^{x'} dx'' \int_0^{x''} dx''' k c_k(x',x''')  \nonumber \\
  &\times \sin \left( k \gamma^{-1} \int _{x'''}^{x'} d\xi A_y(\xi,t) \right) \, , \label{eq:adam}
\end{eqnarray}
where $A_y(t,x)$ is the laser vector potential, $c_k(x_1,x_2) = \langle B_k(x_1) B_{-k}(x_2)\rangle $ is the spectral density of the perturbative
Weibel-generated field $B(x,y) = \sum_k B_k(x) e^{iky}$. The above equation shows that $ \langle v_y \rangle$ changes sign with the laser field in
accordance with PIC simulations \cite{adam06}. 

The late-time dynamics of the magnetized filaments has been frequently investigated by means of simulations resolving only the plane orthogonal
to the beam's axis \cite{lee73,hond00,saka02,medv05,diec09}. In the case where $n_b \ll n_p$, the beam electrons are strongly pinched by the
magnetic field, which expels the plasma electrons from the filament's interior. Further magnetic pinching of the beam electrons generates a
strong space-charge field accelerating the ions in the radial direction \cite{hond00,saka02}. The nonlinear stage is dominated by the merging
of magnetically-interacting neighboring filaments (due to incomplete current shielding by the plasma electrons), leading to increasingly large 
filaments. While this merging process is accompanied by a steadily-declining total beam current,  Polomarov \emph{et al.} \cite{polo08} have
demonstrated that during its earliest phase, the fusion of sub-Alfv\'enic filaments (\emph{i.e.}, carrying current $I < I_A = \gamma_b\beta_b 4\pi m_e c/\mu_0e \approx 17.05 \gamma_b\beta_b$ kA)
entails an increase in the magnetic energy and, consequently, a decreasing kinetic energy, whereas the coalescence of super-Alfv\'enic filaments
($I > I_A$) occurring at later times gives rise to a slowly-decaying magnetic energy. In the case of comparable beam and plasma densities, simulations
indicate that the typical filament size increases roughly linearly with time as a result of successive coalescences \cite{medv05,diec09}. 

\begin{figure}
\centering
\begin{tabular}{c}
(a) \\
\includegraphics[width=0.36\textwidth]{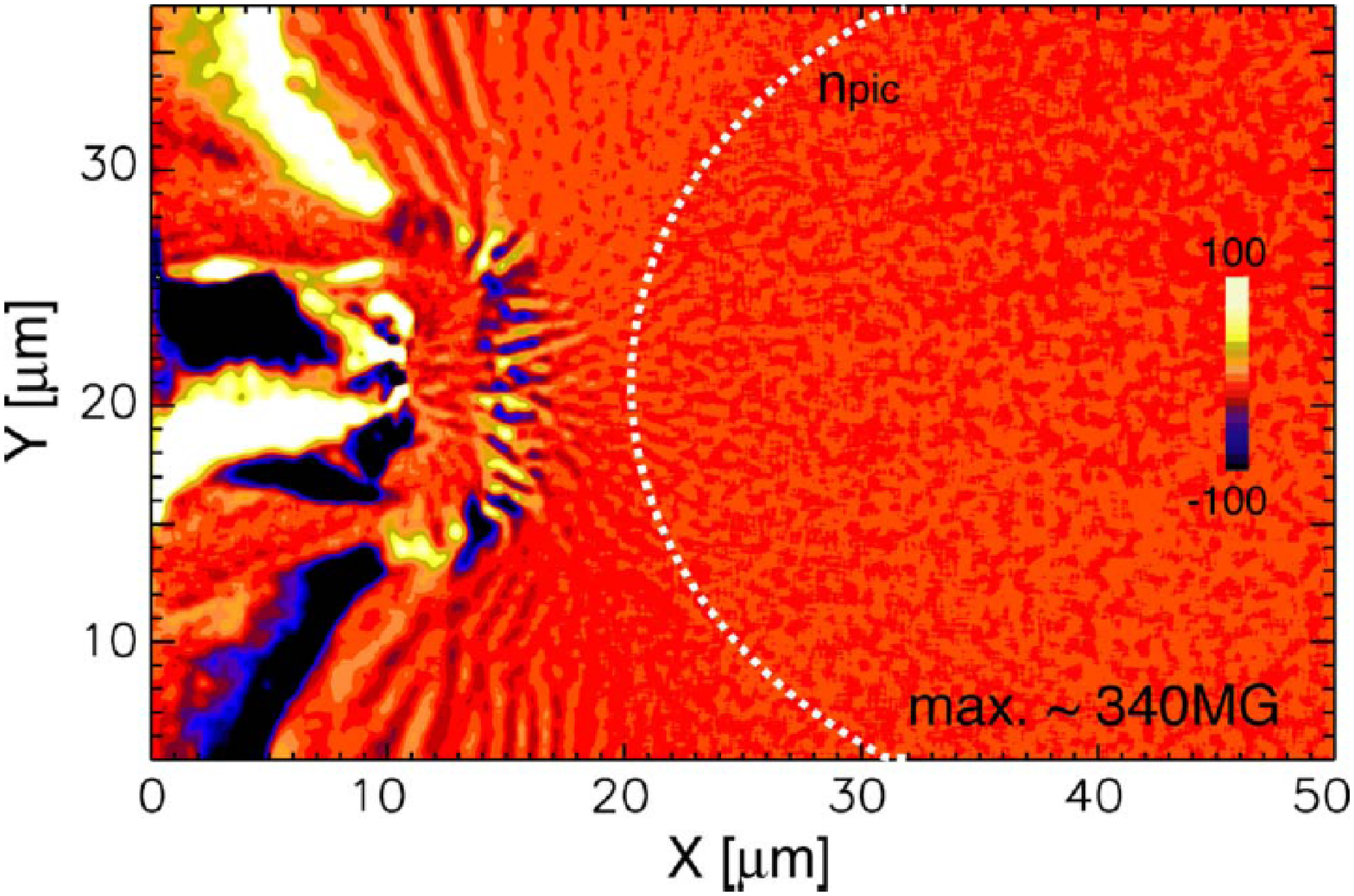} \\
(b)\\
\includegraphics[width=0.4\textwidth]{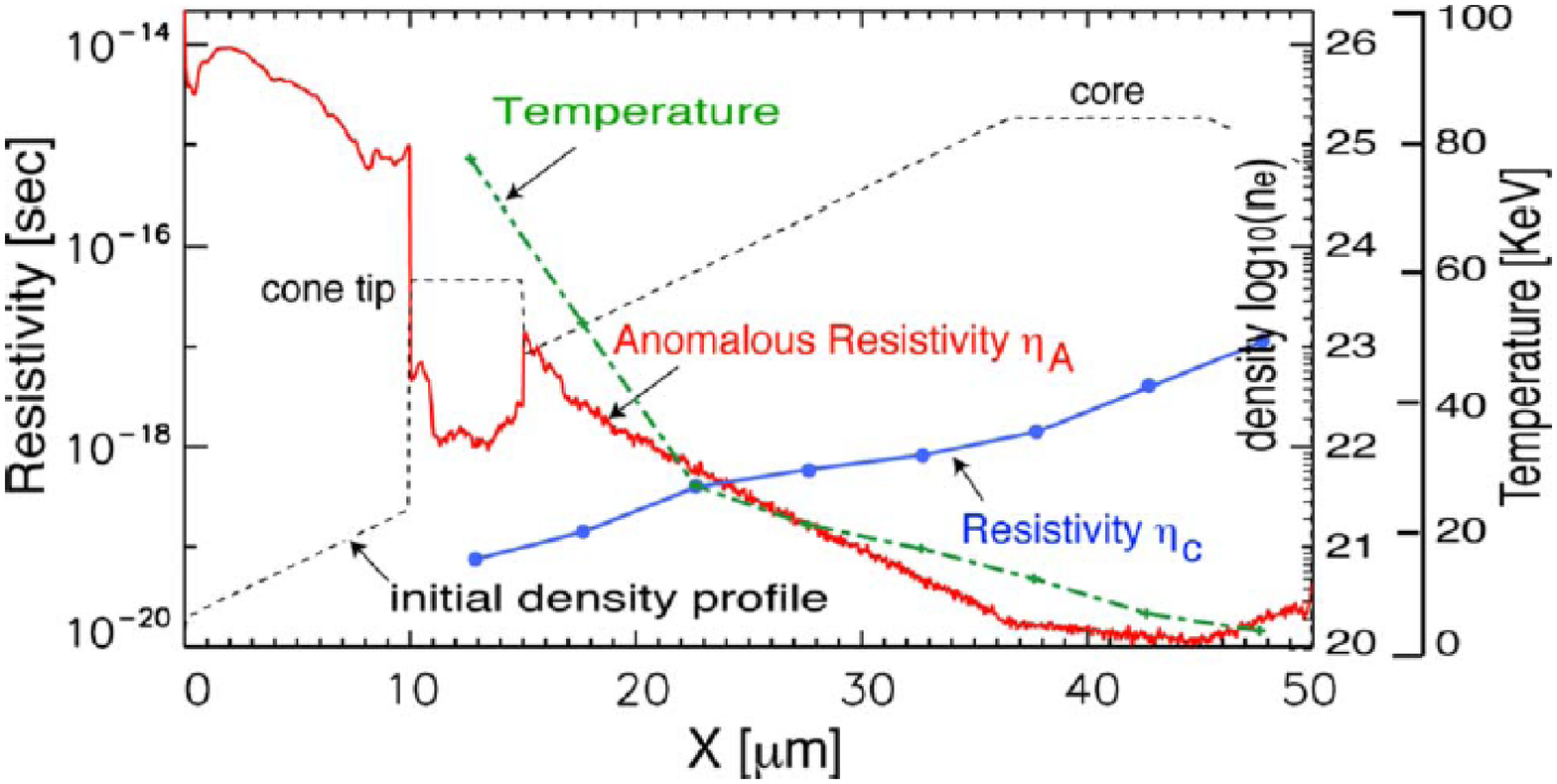}
\end{tabular}
\caption{Collisional 2-D PIC simulation of the interaction of a $10^{20}\,\Wcmsq$ laser pulse with a cone-guided compressed target: (a) quasistatic
magnetic field (in MG) at 860 fs; (b) $x$-profiles of the anomalous resistivity, collisional resistivity, density and bulk electron temperature at $y = 21\,\mu\mathrm{m}$
(after Chrisman \emph{et al.} \cite{chri08}).}
\label{fig:Christman}
\end{figure}

Provisos, however, must be made concerning the practical relevance for the FIS of the aforementioned simulations. First, all of them consider beam electrons with
weak thermal spreads in contradistinction with the momentum distributions of laser-accelerated electrons seen in PIC simulations [Figs. \ref{fig:cone_divergence}(c,d)].
Second, by describing the electron dynamics in the plane perpendicular to the beam's flow only, they do not capture the parallel or oblique unstable modes
described in Sec.\ \ref{subsec:linear_instabilities}. As shown in Ref.\ \cite{silv06} through comparisons with 3-D simulations, the overall influence of these
multidimensional processes on the beam transport is best reproduced by 2-D calculations performed in the plane of the beam's flow. Finally, the simulations
carried out in Refs.\ \cite{lee73,hond00,saka02,medv05,polo08,diec09,shve09,khud12} employ initially uniform beam profiles with periodic boundary conditions,
thus neglecting the stabilization provided by the dilution of the diverging beam as it propagates away from the injection region. It is then no surprise that 
a somewhat different picture emerges from more realistic simulations of the fast electron generation and transport. In particular, for laser intensities
$10^{20-21}\,\Wcmsq$, filamentation is found to be confined to the vicinity of the laser-irradiated zone \cite{adam06,ren06,tong09,debayle1}.
While this region remains (weakly) Weibel-unstable in the nonlinear stage due to the destabilizing effect of the ion motion, the interior region becomes
stable owing to the important dilution of the fast electrons \cite{ren06}. Importantly, the filamentation-driven rippling of the target surface triggers additional
laser heating mechanisms such as the Brunel effect \cite{ren06,bato08}. Furthermore, the surface ions may be accelerated by the laser radiative pressure to
velocities high enough to trigger the ion-Weibel instability. The magnetic turbulence thus generated may give rise to a collisionless shock of astrophysical
interest \cite{fiuz12}. 

In Ref.\ \cite{sento03}, the magnetized beam filaments have been shown to act as random scattering sources for the return current electrons, yielding an
effective electrical resistivity of the order of $\nu_\mathrm{A} = \omega_c/\omega_p^2 \epsilon_0 $, where $\omega_c$ is the electron cyclotron frequency
in the average magnetic field amplitude $\langle \vert B \vert \rangle$. Yet, large-scale laser-plasma simulations indicate that this ``anomalous'' effect
only arises within a few microns of the irradiated surface, where the backgound temperature is high enough to quench collisional processes \cite{chri08}.

\subsubsection{Electrostatic instabilities in fast electron transport}

Few studies have addressed the influence of the electrostatic (two-stream or oblique) instabilities in the FIS context. One notable exception is the simulation work of
Kemp \emph{et al.} \cite{kemp06} who showed that, in a 1-D geometry and for a laser intensity of $10^{19}\,\Wcmsq$, two-stream kinetic modes govern
the energy transfer  from hot to thermal electrons in plasma densities $< 10^{23}\,\mathrm{cm}^{-3}$, whereas they prove strongly inhibited by Coulomb collisions
at higher densities. 

\begin{figure}[tbp]
\begin{center}
\begin{tabular}{cc}
(a) & (b) \\
\includegraphics[width=0.242\textwidth]{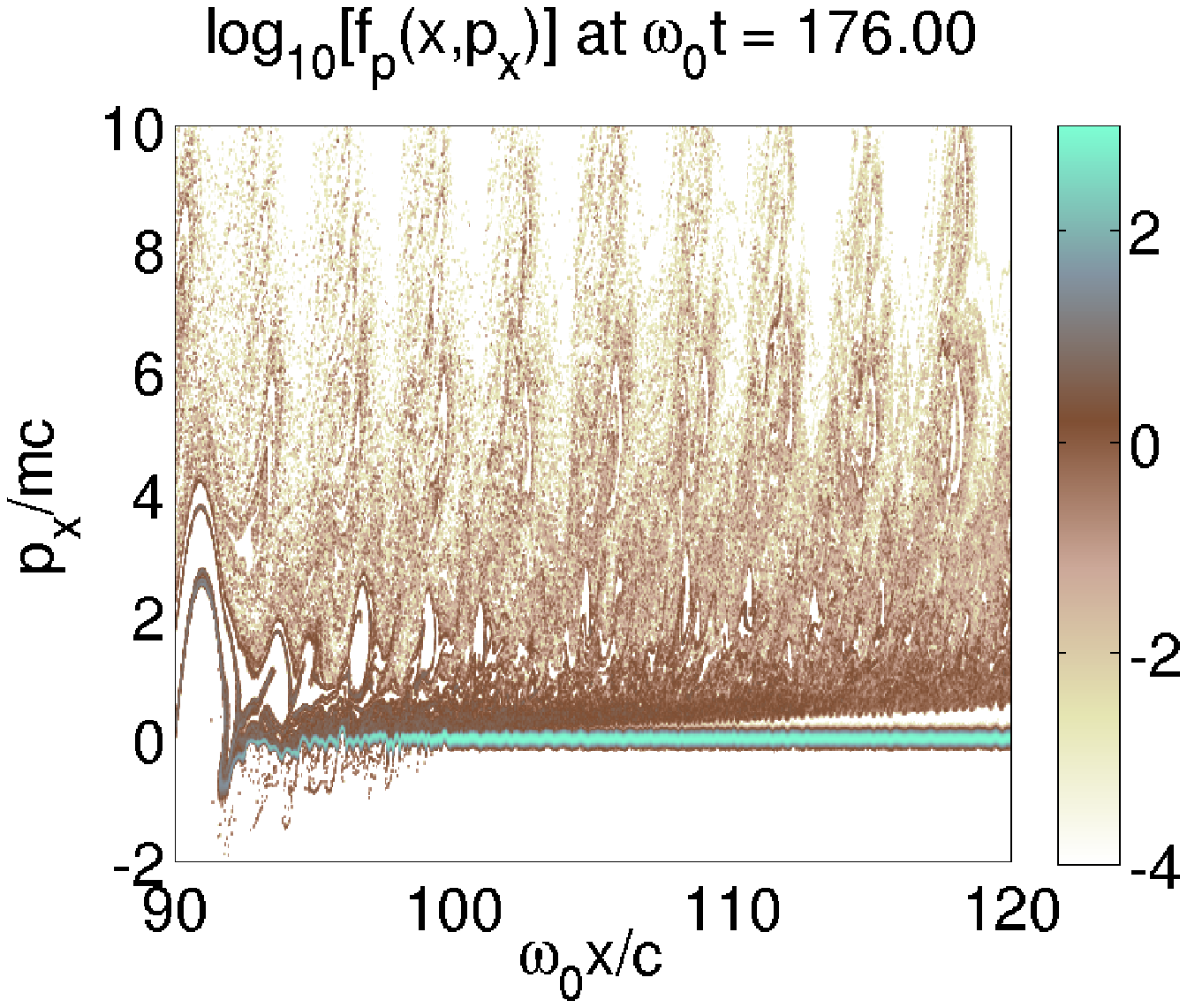}
&\includegraphics[width=0.242\textwidth]{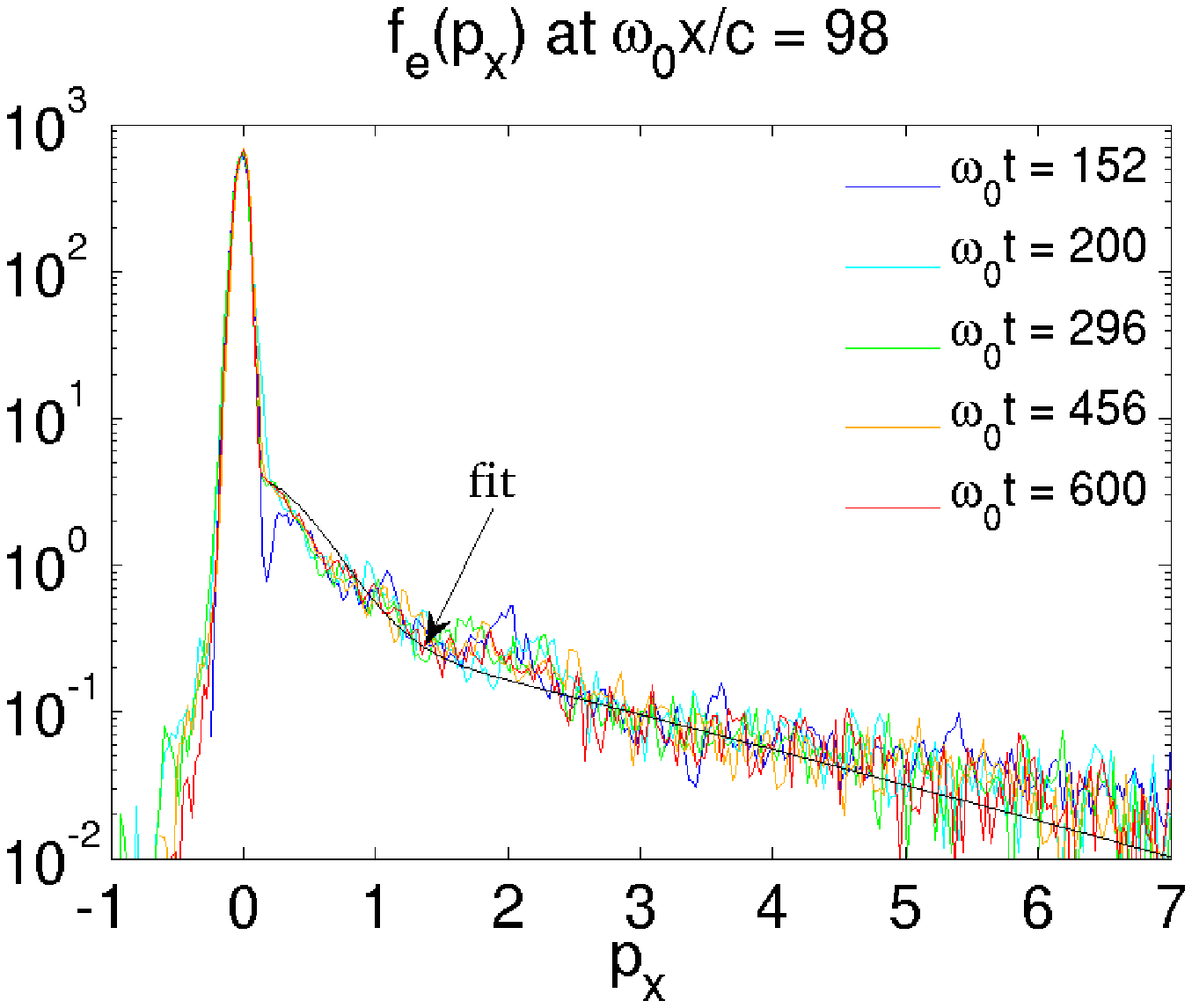}
\end{tabular}
\end{center}
\caption{1-D PIC simulation of the interaction of a $10^{19}\,\Wcmsq$ laser wave with a $100n_c$, $1\,\mathrm{keV}$  plasma: (a) $x-p_x$ electron phase space
of the interaction region at $t = 176\omega_0^{-1}$; (b) momentum distribution at various times at $x = 98c/\omega_0$ and averaged over half a laser wavelength.
The solid line plots the high-momentum fit \Eref{eq:gremfit}.}
\label{fig:ilp_quasilineaire1}
\end{figure}

It is well-known that the nonlinear evolution of the two-stream instability depends on the monochromatic or broadband character of the unstable spectrum \cite{suda84}.
The latter case corresponds to the weakly-unstable, kinetic limit and, to first order, is amenable to quasilinear theory \cite{davi72b}. Through resonant wave-particle interaction
(\emph{i.e.},  involving waves satisfying $\omega = \mathbf{k}.\mathbf{v}$), the beam distribution tends to flatten down to the plasma thermal velocity. This weak-turbulence
problem has been tackled in Refs.\ \cite{fain70,ruda71,suda84} where the plateau formation was found to be disturbed by secondary, nonlinear ion-induced scattering
and parametric processes. If not collisionally suppressed, this kinetic regime seems to prevail in the FIS context due to the broadly-spread and monotonically-decreasing
momentum distribution of the hot electron source. 

These mechanims are illustrated here by 1-D PIC simulations of the interaction of a $3\times 10^{19}\,\Wcmsq$ laser pulse with a $100n_c$ plasma
\cite{grem12}. The initial temperature is 1 keV and a $1\lambda_0$ scale-length exponential preplasma is added on the target surface. In order to obtain a
quasi-stationary kinetic energy flux into the plasma and, therefore, help identify the unstable beam-plasma processes, the ions are kept fixed in a first stage.
As a result, the instantaneous laser-to-plasma absorption rate has an approximately constant value of $\sim 12\%$. Beyond the laser-irradiated surface
($x \sim 90c/\omega_0$), the $x-p_x $ electron phase space displayed in Fig.\ \ref{fig:ilp_quasilineaire1}(a) exhibits $2\omega_0$ high-energy jets ($p_x/m_ec \ge 4$)
typical of the $J\times B$ acceleration mechanism \cite{krue85b}. The electron vortices centered on $p_x/m_ec \sim 1$ point to the beam-driven excitation
of a strongly nonlinear wave close to the absorption region ($x<100 c/\omega_0$). This wave, however, rapidly damps out due to bulk electron trapping, hence yielding
a monotonically-decreasing average momentum distribution, as plotted, at various times, in Fig.\ \ref{fig:ilp_quasilineaire1}(b). The high-momentum part ($p_x>0.2m_e c$)
of this distribution carries a density $n_b/n_c \sim 0.2$ and, to a good approximation, can be fitted to \begin{equation} \label{eq:gremfit}
f_b(p_x) \approx 1.67 \times 10^3 e^{-5.6 \gamma_x} + 0.68e^{-0.57 \gamma_x}
\end{equation}
where $\gamma_x = \sqrt{1+p_x^2/(m_e c)^2}$. Because of its decreasing shape, the source distribution is locally stable with respect to electrostatic fluctuations,
as also observed by Tonge \emph{et al.} \cite{tong09}. Deeper into the target ($x > 105 c/\omega_0$), though, time-of-flight differences generate a transient
positive gradient  destabilizing the hot electron distribution. 

\begin{figure}[tbp]
\begin{center}
\begin{tabular}{cc}
(a) & (b) \\
\includegraphics[width=0.23\textwidth]{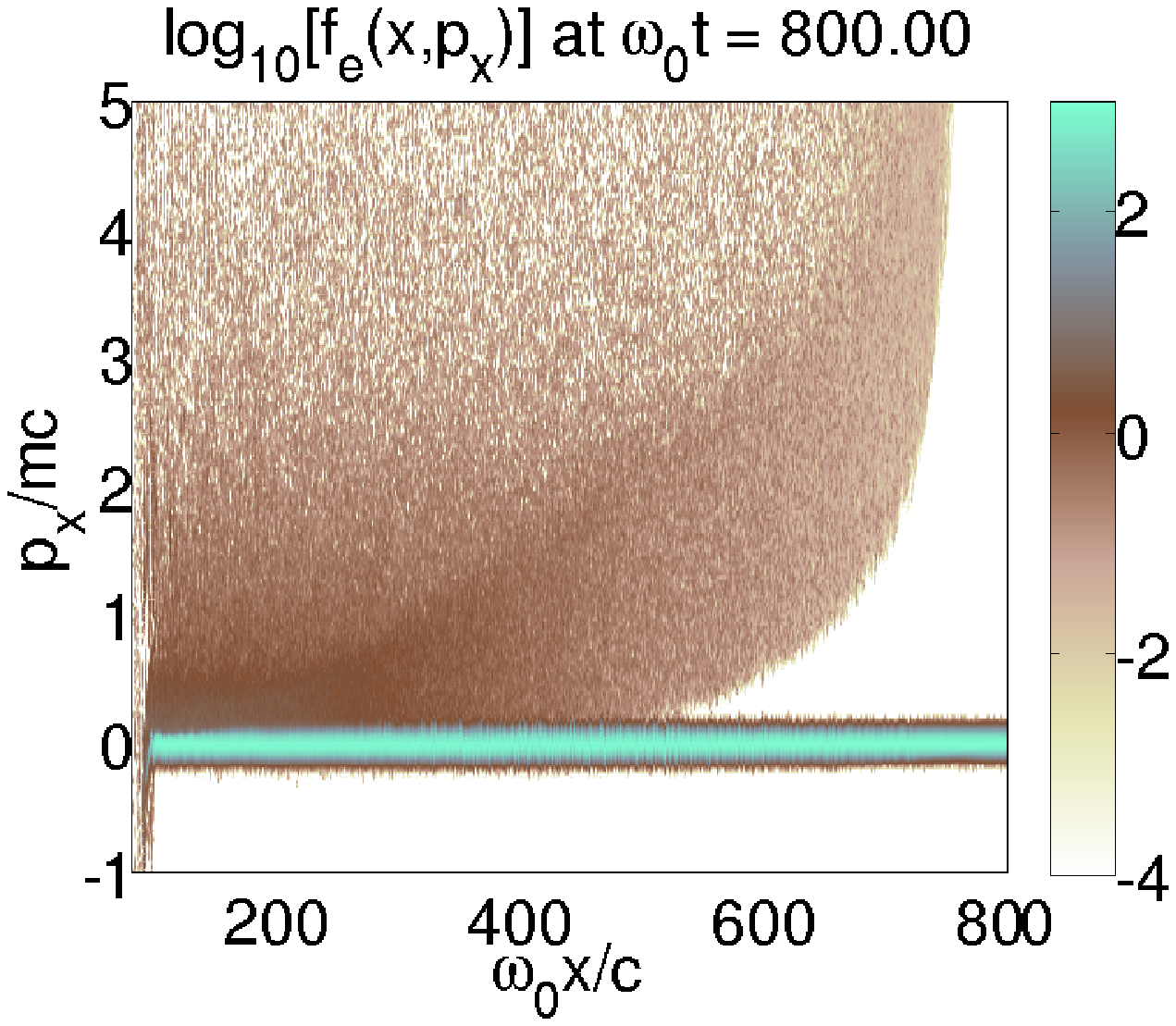}
&\includegraphics[width=0.24\textwidth]{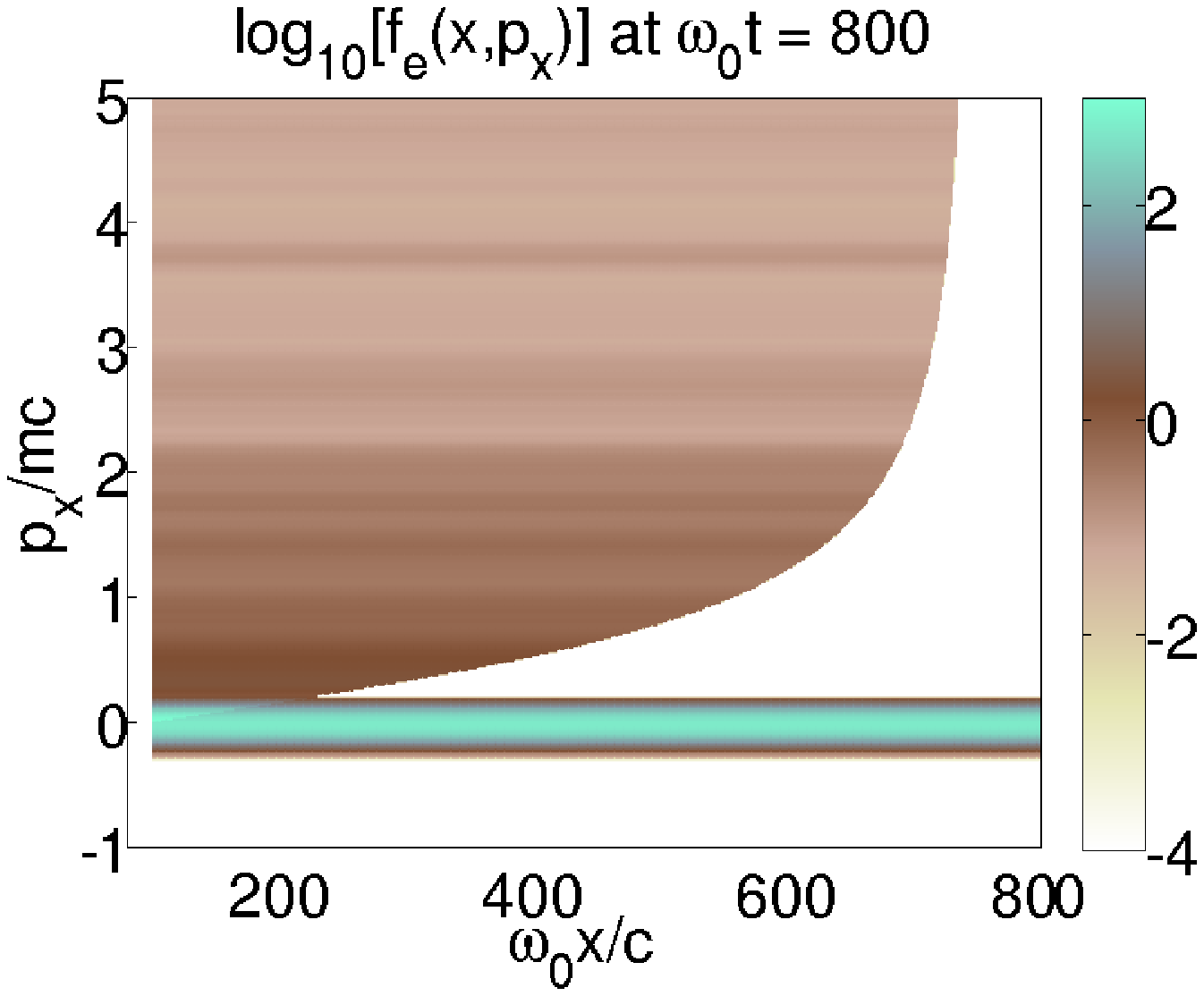}
\end{tabular}
\end{center}
\caption{Electron $x-p_x$ phase space at $t = 800 \omega_0^{-1}$ as predicted by a self-consistent PIC simulation (a) and the ballistic propagation of the hot electron
source (b). Same parameters as in Fig.\ \ref{fig:ilp_quasilineaire1}.}
\label{fig:ilp_quasilineaire2}
\end{figure}

\begin{figure}[tbp]
\begin{center}
\begin{tabular}{cc}
(a) & (b) \\
\includegraphics[width=0.242\textwidth]{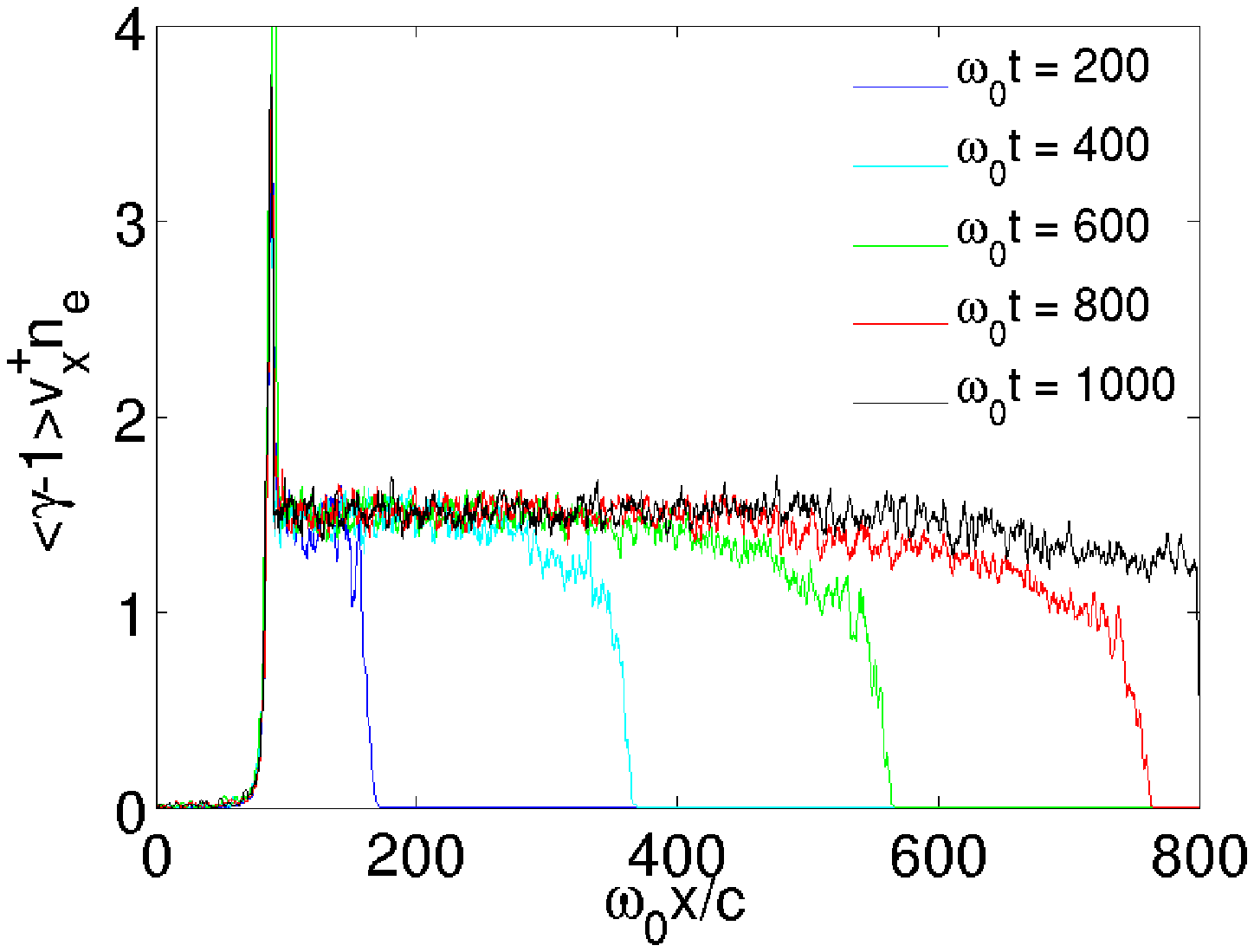}
&\includegraphics[width=0.242\textwidth]{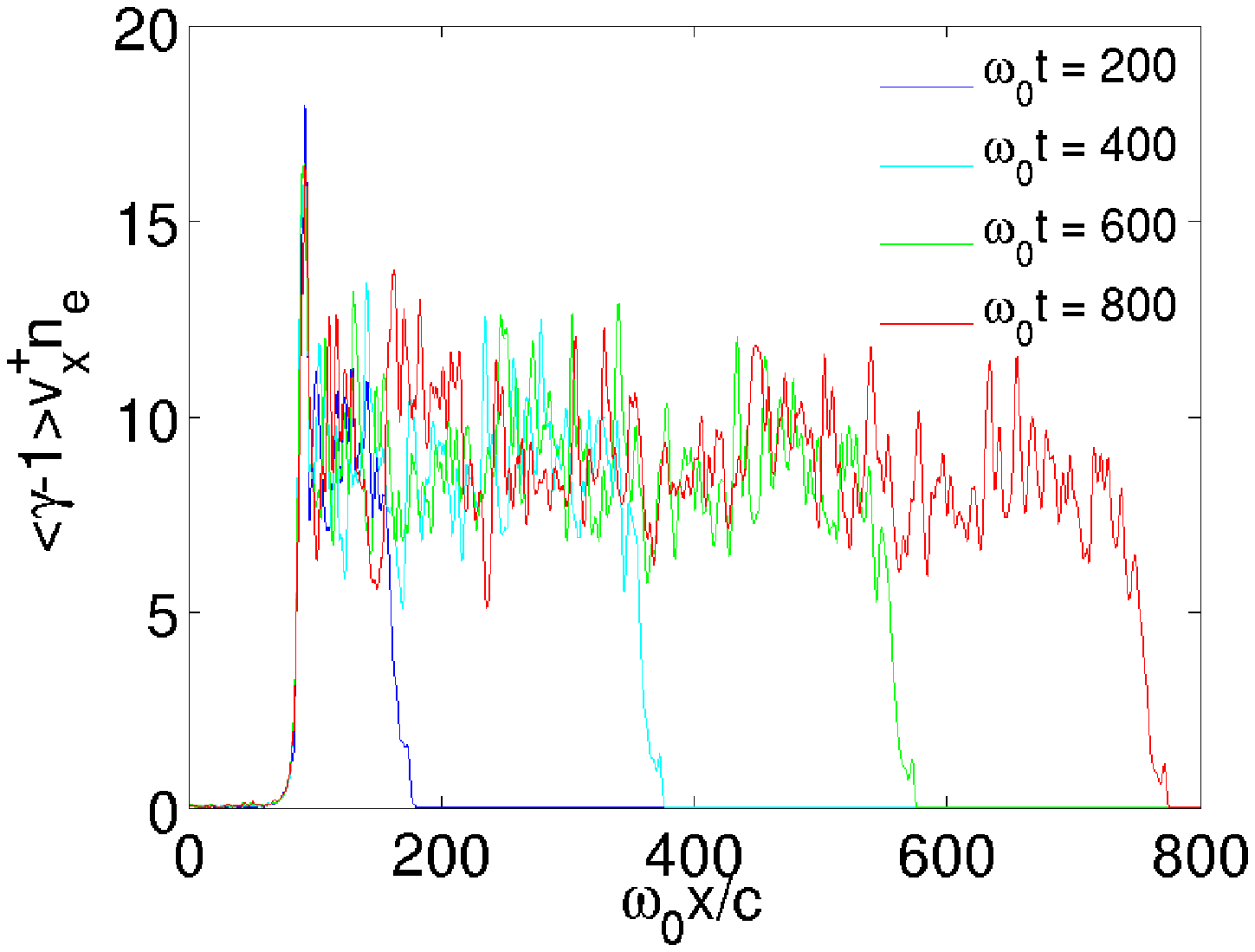}
\end{tabular}
\end{center}
\caption{Energy flux density of the forward-going electrons (in units of $m_e c^3n_c$) as a function of $x$ for a laser intensity of $3\times 10^{19}\,\Wcmsq$ (a)
and $10^{20}\,\Wcmsq$ (b).}
\label{fig:ilp_quasilineaire3}
\end{figure}

The quasilinear relaxation induced by the two-stream modes developing in the $\partial f_b /\partial p_x > 0$ region is evidenced in Fig.\ \ref{fig:ilp_quasilineaire2},
where the PIC-simulated electron phase space at $t = 800 \omega_0^{-1}$ is compared to that obtained by ballistically evolving the source distribution \eref{eq:gremfit}. A plateau
clearly forms in the gap region separating the hot and thermal electrons, with a width increasing with the distance from the injection surface. This proves
that the advection time $\tau_\mathrm{adv} = x(v_\mathrm{min}^{-1}-c^{-1})$ (where $v_\mathrm{min}$ is the minimum velocity of the hot electrons) is much larger than
the characteristic growth time $\tau_\mathrm{TS} \sim (n_e/n_b)(\Delta p_x^2/p_x^3)\omega_p^{-1}$. This is indeed expected in the present case where $n_b/n_e \sim 10^{-3}$,
$p_x/m_ec \sim 2-4$, $\Delta p_x/m_ec \sim 1$, $v_\mathrm{min} \sim 0.2c$, $x \sim 1000c/\omega_p$, and hence $\tau_\mathrm{adv} \sim 1000\omega_p^{-1} \gg \tau_\mathrm{TS} \sim 100 \omega_p^{-1}$.
Note that the plateau formation is sped up at higher laser intensities due to increased beam density. The quasilinear equations describing the space-time evolution
of the averaged beam distribution function and the spectral density of the beam-resonant waves can be analytically solved along the lines of Ref.\ \cite{zait74},
by assuming instantaneous plateau formation and using \Eref{eq:gremfit} for the source distribution \cite{grem12}. In agreement with the simulation results,
this model predicts that, for a $10^{19}\,\Wcmsq$ laser intensity, a maximum of $\sim 6\%$ of the beam energy is converted to resonant waves. Owing to
the stable distribution source, these waves are subsequently reabsorbed by slower electrons arriving at later times. Overall, the wave energy is too weak to affect the
beam energy flux. This is demonstrated in \mbox{Fig.\ \ref{fig:ilp_quasilineaire3}(a)}, where the spatial profile of the energy flux carried by forward-going ($p_x > 0$) electrons
is plotted at various times. Energy is seen to propagate at a velocity $\sim c$ with negligible dissipation over $\sim 800c/\omega_0$. The spatial variations near the right-hand
edges of the profiles stem from time-of-flight differences.  Figure \ref{fig:ilp_quasilineaire3}(b) corresponds to a $10^{20}\,\Wcmsq$ laser intensity: albeit more
strongly modulated than in panel (a), the energy flux profiles do not reveal significant dissipation either. These findings contrast with the fast relaxation found in the 2-D
simulation study of Tonge \emph{et al.} \cite{tong09}. The origin of this discrepancy is not as yet clearly understood: it may stem from 2-D physical effects or from the
artificial collisionality caused by insufficient numerical resolution. 

\begin{figure}[tbp]
\begin{center}
\begin{tabular}{cc}
(a) & (b) \\
\includegraphics[width=0.242\textwidth]{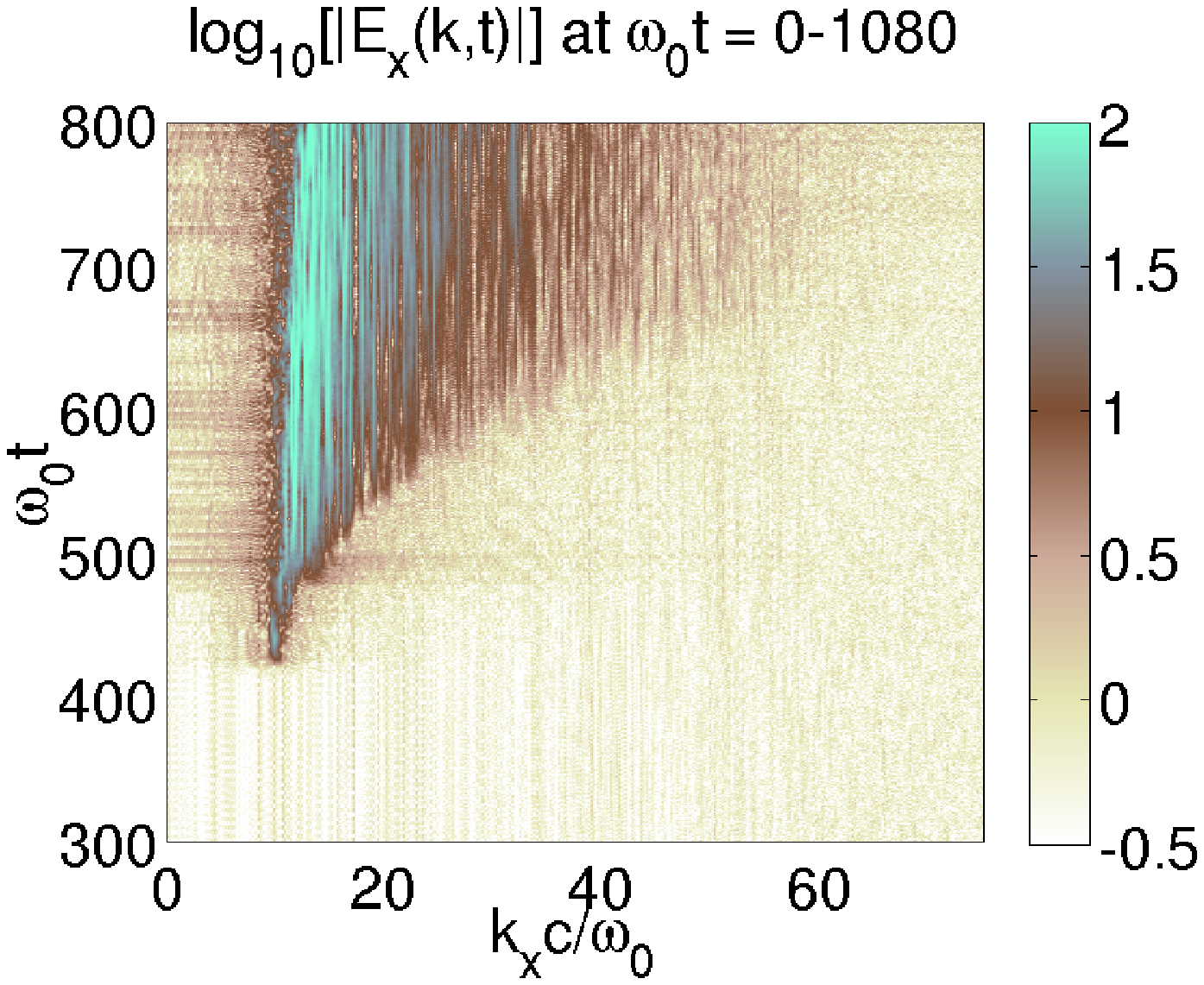}
&\includegraphics[width=0.242\textwidth]{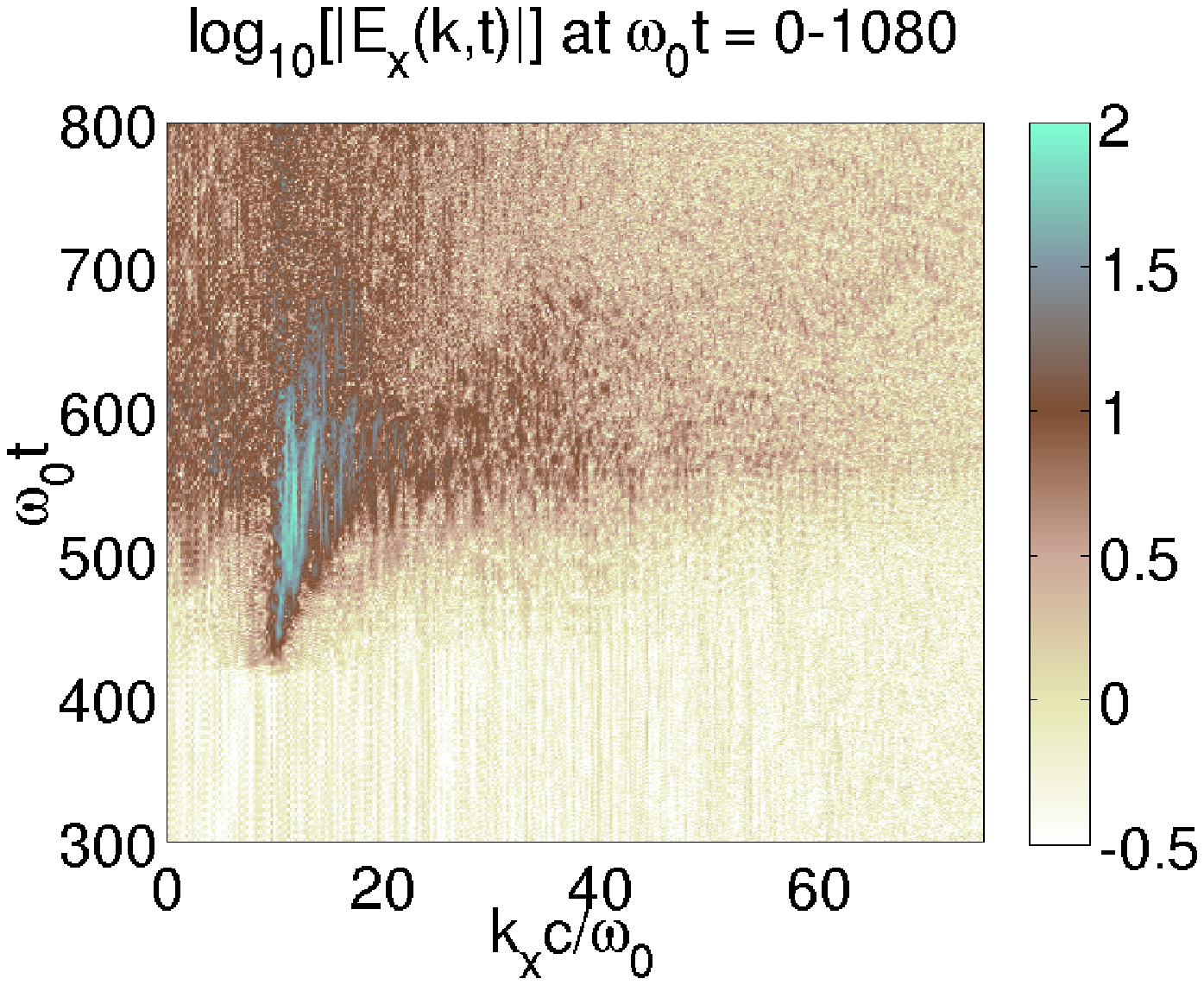}
\end{tabular}
\begin{tabular}{c}
(c) \\
\includegraphics[width=0.242\textwidth]{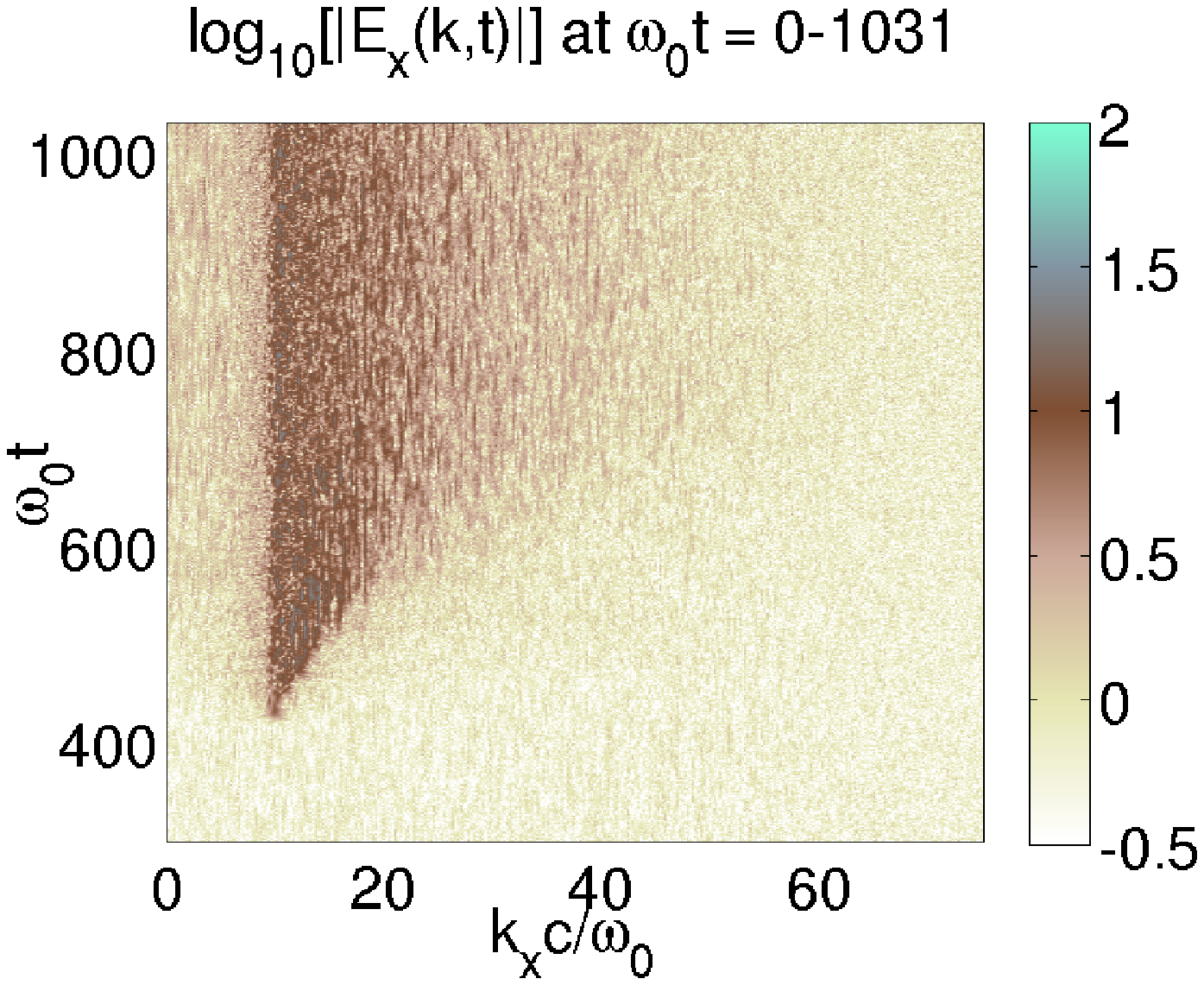}
\end{tabular}
\end{center}
\caption{Electric field spectrum $\vert E_x(k,t)\vert$ in the space region $375 < \omega_0x/c < 400$. Panels (a) and (b) correspond to collisionless
cases with immobile (a) and mobile (b) ions, while $e-i$ and $e-e$ Coulomb collisions are described in panel (c). The ion charge and temperature
are $Z = 13$ and $T_i = 0.2\,\mathrm{keV}$, respectively (other parameters identical to those of Fig.\ \ref{fig:ilp_quasilineaire1}).}
\label{fig:ilp_quasilineaire4}
\end{figure}

The time evolution of the wave spectrum in the space region $375 < \omega_0x/c < 400$ is displayed in Fig.\ \ref{fig:ilp_quasilineaire4}(a). As slower and slower electrons
reach the detection region, waves of decreasing phase velocity are emitted, hence the observed spectral broadening towards high $k$'s. The case of mobile aluminum ions with
charge $Z=13$ and temperature $T_i = 0.2\,\mathrm{keV}$ is treated in Fig.\ \ref{fig:ilp_quasilineaire4}(b). Weaker and shorter-lived electric fluctuations are then generated,
as a result of a modulational instability which efficiently scatters the beam-excited waves outside the beam-resonant region \cite{thod75,frie76,mima84}.
This parametric process can be modeled assuming the primary waves behave as an monochromatic pump wave $(\omega_1,k_1)$ decaying into an ion wave
$(\omega,k)$ and Langmuir waves $(\omega_1 \pm \omega, k_1 \pm k)$. The corresponding dispersion relation is \cite{mima84}
\begin{equation} \label{eq:modulationnel}
  1 + \frac{\omega_p^2(k\lambda_D)^2}{4} \frac{W_E}{n_eT_e} \frac{(1+\chi_i) \chi_e}{1 + \chi_e + \chi_i} \,
  \Bigg( \frac{1}{D_{-}} + \frac{1}{D_{+}} \Bigg) \,  = 0
\end{equation}
where $D_{\pm} = (\omega \pm \omega_1)^2 -\omega_p^2 - 3v_e^2(k \pm k_1)^2$, $\lambda_D$ is the Debye length, $\chi_j$ is the $j$th component susceptibility and
$W_E$ is the wave energy density. In the present case, one has $\omega_1/\omega_p = 0.98$, $k_1 \lambda_D = 0.53$ and $W_E/n_e T_e \sim 0.13$. Numerical resolution
of \eref{eq:modulationnel} then yields a peak modulational growth rate $\delta_\mathrm{max} = 2.5\times 10^{-3} \omega_p$ for the wave number
$k_\mathrm{max} = 0.13\lambda_D^{-1}$ and the real frequency $\omega_\mathrm{max} = 2\times 10^{-4} \omega_p$. We have checked that these values  closely reproduce
the simulation results. The high-$k$ secondary waves generated by this instability are strongly Landau-damped by the bulk plasma electrons, which, as observed in \cite{kemp06},
gives rise to suprathermal tails but negligible ion heating (not shown). When Coulomb collisions are switched on, Fig.\ \ref{fig:ilp_quasilineaire4}(c) shows that the beam-plasma
instability is strongly weakened. This is expected since, for the parameters under consideration ($n_e = 100n_c$, $T_e = 1 \,\mathrm{keV}$ and $T_i = 0.2\,\mathrm{keV}$),
the collision frequency ($\nu_{ei} \sim 0.03\omega_p$) is comparable to the collisionless two-stream growth rate. The primary waves are then too weak to trigger the modulational
instability and the beam-to-plasma energy transfer essentially proceeds through the resistive electric field.

In summary, 1-D kinetic simulations indicate that electrostatic instabilities play only a minor role in the energy relaxation of fast electrons generated by $10^{19-20}\,\Wcmsq$
laser pulses into $100 n_c$ plasmas due to the decreasing shape of the electron source distribution.  Generalization of these results to more realistic 2-D geometries, as attempted
in Refs.\ \cite{tong09,schm12}, requires further investigation. In particular, the influence of the oblique modes remains to be clearly demonstrated in a FIS-relevant laser-plasma setup.



\subsection{Background plasma physics}
\label{bkgdphysics}

This section discusses the physics of the background medium (i.e.\ excluding fast electrons), that is relevant to fast electron transport and fast ignition.  This broadly falls in the realm of radiation-hydrodynamics, which we will not review in detail.  Instead, we focus on aspects of special interest to fast electron transport, which are frequently not emphasized in traditional rad-hydro models. These in particular are fluid models that include fast electrons, incorporate electromagnetic fields, and account for Fermi-Dirac (F-D) statistics (namely background electron degeneracy) in transport coefficients like electrical resistivity.

It is extremely productive to separate transport problems into a background medium and fast electrons.  This requires distinguishing between fast and background electrons, which is generally done based on an intermediate electron energy well below that of most fast electrons yet well above that of most background electrons.  This becomes invalid if the background temperature is comparable to typical fast electron energies, either because the background is strongly heated or the fast electrons have slowed down significantly.  We shall assume that the distinction can be validly made.

Fast electrons and the background interact via collisions and macroscopic electromagnetic fields.  Fast electron collisions are discussed in detail in Sec.\ 3.3 of the present article.  The e/m fields evolve according to the Maxwell equations, which contain the charge and current densities, $\rho$ and $\mathbf{J}$, carried by the fast electrons and background.  Since the fast electrons are not atomically bound, it is trivial to find their $\rho$ and $\mathbf{J}$.  The background can be much more complicated, depending on whether it is neutral matter, a conductor, or a partially or fully ionized plasma.  We assume the background can be described by a fluid model, meaning it is not a collisionless plasma requiring a fully kinetic description.  A fluid model applies to neutral matter (with appropriate and perhaps difficult models for material properties like equation of state), and for plasmas that are sufficiently collisional that the background distribution functions are close to equilibrium (e.g.\ Maxwellian or Fermi-Dirac).

We focus on the background electron momentum and energy equations, which we write in a form following ``notation II'' of Ref.\ \cite{epperlein-xport-pof-1986}:
\begin{eqnarray}
m_e\frac{\partial\mathbf{v}_e}{\partial t} = -e\left( \mathbf{E} - \mathbf{E}_C - \mathbf{E}_{NC} \right), \\
\mathbf{E}_C = \mat{\eta} \cdot \mathbf{J}_e - e^{-1}\mat{\beta}\cdot\nabla T_e, \\
\mathbf{E}_{NC} = -\frac{\nabla p_e}{en_e} - \mathbf{v}_e\times\mathbf{B}, \\
\left[ \frac{\partial}{\partial t} + \mathbf{v}_e\cdot\nabla \right] (\rho c_V T_{e}) + p_e \nabla\cdot\mathbf{v}_e &= \nabla\cdot\left[ \mat{\kappa}\cdot\nabla T_e + e^{-1}T_e\mat{\beta}\cdot\mathbf{J}_e \right] \\
&+ \nu_{ei,T}n_e(T_i-T_e) + \mathbf{J}_e\cdot \mathbf{E}_C. \nonumber
\end{eqnarray}
The Maxwell equations including fast electrons are
\begin{eqnarray}
\frac{\partial \mathbf{B}}{\partial t} = -\nabla\times\mathbf{E}, \\
\epsilon_0\mu_0 \frac{\partial \mathbf{E}}{\partial t} = \nabla\times\mathbf{B} - \mu_0(\mathbf{J}_e+\mathbf{J}_i+\mathbf{J}_f), \label{eq:djsamp} \\
\nabla\cdot\mathbf{E} = e \epsilon_0^{-1}\left( -n_e-n_f + \sum_iZ_in_i \right).
\end{eqnarray}
This section uses SI units and expresses temperature in energy units. Subscripts $e,f$ refer to (background, fast) electron quantities. The electron momentum and energy equations are written in the rest frame of the ions, so that for instance $v_e$ in $\eta J_e$ should be replaced by $v_e-v_i$ in a frame where the ions move. $c_V$ is the specific electron heat capacity, which differs from the Maxwellian ideal-gas result due to e.g.\ Fermi-Dirac statistics. Fluid equations of this type go back at least to Braginskii \cite{braginskii-xport-1965}, and require departures from collisional equilibrium to be small. This breaks down, for example, when $\mathbf{E}$ is large enough that a significant portion of the background electrons become runaways, or when $v_e$ exceeds the ion thermal speed and triggers the ion acoustic drift instability.

We have expressed the forces as equivalent electric fields $(\mathbf{E}_C, \mathbf{E}_{NC})$, which arise from (collisional, collisionless) effects, respectively.  The specific $\mathbf{E}_C$ and $\mathbf{E}_{NC}$ given above are those currently implemented in the Zuma code \cite{larson-zuma-dpp-2010,strozzi-fastig-pop-2012}, and neglect certain effects.  Namely, $\mathbf{E}_{NC}$ lacks the advective term $\mathbf{v}_e\cdot\nabla \mathbf{v}_e$ and off-diagonal components of the pressure tensor, and $\vec E_C$ neglects collisions of fast with background electrons. $\mat{\eta}$ (resistivity), $\mat{\beta}$ (thermal force), and $\mat{\kappa}$ (thermal conductivity) arise from collisional or other dissipative effects, which in a weakly-coupled plasma are mainly electron-ion (e-i) collisions. They are tensors due to magnetic fields or anisotropic distributions, and reduce to scalars for $\mathbf{B}=0$ and isotropic distributions.

Transport problems frequently consider situations where background electron inertia can be neglected.  In plasmas this typically applies for time scales much longer than the period of Langmuir waves. Dropping $\partial\mathbf{v}_e/\partial t$ from the momentum equation gives an algebraic equation for $\mathbf{E}$ in terms of other quantities: $\mathbf{E}=\mathbf{E}_C + \mathbf{E}_{NC}$. We call this an Ohm's law.  Since this approximation gives $\mathbf{E}$, we cannot treat Amp\`ere's law \eref{eq:djsamp} as a time evolution equation for $\mathbf{E}$.  Generally the displacement current $\partial\mathbf{E}/\partial t$ is dropped from \eref{eq:djsamp}, although it may be fruitful to include it.  Amp\`ere's law instead gives $\mathbf{J}_e$ and thus $\mathbf{v}_e$, which is no longer specified by the inertialess momentum equation. We call dropping both background electron inertia and displacement current the Ohmic approximation.  Langmuir and light waves are excluded by construction. In addition, quasi-neutrality is commonly assumed, which entails dropping $\nabla\cdot\mathbf{E}$ from Gauss's law and is valid on length scales much longer than the Debye length.  One should remember that these approximations are independent, even though they are often generically called ``hybrid models.''

The Ohmic and quasi-neutral approximations currently made in Zuma are
\begin{eqnarray}
\mathbf{E} = \mathbf{E}_C + \mathbf{E}_{NC}, \\
\mathbf{J}_e = -\mathbf{J}_f  + \mu_0^{-1}\nabla\times \mathbf{B}, \\
n_e = \sum_iZ_in_i.
\end{eqnarray}
Zuma itself does not handle ion motion, but has been coupled to the rad-hydro code Hydra \cite{marinak-hydra-pop-2001} as detailed in \cite{strozzi-fastig-pop-2012}.

Regardless of whether an Ohmic approximation is made, a key ingredient is specifying the transport coefficients and ionization state. These are aspects where ideal (fully-ionized, weakly-coupled, non-degenerate) plasma physics is insufficient in transport problems.  At high density, it is important that the free electrons obey F-D statistics, so that their equilibrium distribution is not the classical Maxwellian. The exclusion principle causes the electrons to have random momentum with respect to the ions even at zero temperature. This becomes significant when $T_e <$ the Fermi energy $E_F \equiv (\hbar^2/2m_e)(3\pi^2n_e)^{2/3}$. For a sense of scale, fully-ionized solid beryllium ($\rho=$ 1.84 $\gcmcub$) has $E_F=$ 22.7 eV. Below we sometimes combine $T_e$ with $E_F$ in a qualitatively correct way, although exact expressions involve F-D integrals.

A general-purpose framework for transport coefficients in dense plasmas is Lee and More's model \cite{lee-more-pof-1984}, which connects the plasma and non-plasma (solid, liquid, neutral gas) states. Desjarlais \cite{desjarlais-metal-cpp-2001} provides improvements to their model, as well as to the Thomas-Fermi ionization model based on the Saha equation and particularly relevant near the metal-insulator transition.  Zuma currently employs an extended Desjarlais model for ionization and transport coefficients. Whether a material at room temperature is a conductor or insulator can be important for experiments, and difficult for models geared toward plasmas to capture correctly. 

The main result of the Lee-More model is the electron relaxation time $\tau$.  From this follows the various transport coefficients, including off-diagonal components due to magnetic fields. We present the Lee-More model as embodied in Zuma. It has been extended to include electron-electron (e-e) collisions along the lines of Refs.\ \cite{braginskii-xport-1965,epperlein-xport-pof-1986}. We cast our results in terms of the resistivity $\eta$, which has direct physical meaning via the electron momentum slowing-down rate $\nu_m$. For no magnetic field (or the component of $\mat{\eta}$ along $\mathbf{B}$), $\eta=(m_e/n_ee^2)\nu_m$. $\tau$ and $\nu_m$ are related by $1/\nu_m = \tau A^\alpha$. $A^\alpha$ accounts for electron F-D statistics and involves F-D integrals $F_n$:
\begin{eqnarray}
A^\alpha = \frac{4}{3} \frac{F_2(\hat\mu) }{ \left[ 1+\exp(-\hat\mu)\right]F_{1/2}(\hat\mu)^2}, \\
F_n(x) \equiv \int_0^\infty dt\ t^n \left[ e^{t-x}+1 \right]^{-1}.
\end{eqnarray}
$A^\alpha$ is given in Eq.\ (25a) of Ref.\ \cite{lee-more-pof-1984}, except with a typographical error that $F_2$ incorrectly reads $F_3$ there. $\hat\mu \equiv \mu/T_e$ where $\mu$ is the electron chemical potential, defined implicitly by $F_{1/2}(\hat\mu)=(2/3)\theta^{-3/2}$ with $\theta\equiv T_e/E_F$. Antia \cite{antia-fd-apj-1993} provides rational function approximations to $F_n$ for several half-integer orders and their inverses, which we use to directly find $\hat\mu(\theta)$.

A formula for $\tau$ that spans the plasma, neutral-gas, and condensed regimes is
\begin{equation}
\tau = \mathrm{max}(\tau_{ec}, \taumelt, \taumin).
\end{equation}
$\tau_{ec}^{-1} = \tau_{ei}^{-1}+\tau_{en}^{-1}$ defines the electron collision time off both charged ions ($\tau_{ei}$) and neutral atoms ($\tau_{en}$), with rates added. $\taumelt$ and $\taumin$ stem from a Bloch-Gr\"{u}neisen melting model \cite{ziman-philmag-1961}, and a minimum time based on inter-atom spacing $R_i\equiv(3/4\pi n_i)^{1/3}$. $\tau_{en}^{-1}=n_n\sigma_n\bar v$ where $n_n$ is the number density of neutral atoms, $\sigma_n$ the cross-section, and $\bar v \equiv 3\cdot2^{-1/2}(T_e/m_e)^{1/2}\theta^{3/2}F_1(\hat\mu)$ the average electron speed. The limiting values of $\bar v^2$ are $(8/\pi)T_e/m_e$ for $\theta\gg1$ and $(9/8)\epsilon_F/m_e$ for $\theta\ll1$. Approximately, $\taumin \approx R_i / \bar v$, and Desjarlais has discussed refinements to this \cite{desjarlais-metal-cpp-2001}. The melt model gives $\taumelt \approx 50 (\Tmelt/T_e)\taumin$, with the material-dependent constant 50 decreasing somewhat for $T_e>\Tmelt$ (see Lee and More for details). The melt model applies to conductors with strong ion correlations, such as a periodic lattice, and not to insulators or gases.  In the periodic case, the electron wavefunction becomes a Bloch wave in a periodic potential, and essentially does not undergo Coulomb collisions off single ions.  Instead, electrons slow down due to interactions with phonons.  As temperature increases and the ions become uncorrelated, Coulomb collisions with ions dominate, and $\tau_{ec}$ applies. Taking the maximum of the three $\tau$'s is a crude way of capturing the real, more complicated physics.

We now discuss $\tau_{ei}$, which falls closest to the realm of traditional plasma physics. We consider one electron species colliding with one ion species; for multiple ion species the collision rates and therefore the $\eta$'s add.  Lee and More find
\begin{equation}
\frac{\eta}{\eta_0} = \frac{1}{3\theta^3F_2(\hat\mu)}, \qquad \eta_0 \equiv \pi\sqrt2 \frac{e^2m_e^{1/2}}{E_F^{3/2}} \delta_{ee} \frac{n_i}{n_e}Z_i^2\ln\Lambda_{ei}.
\end{equation}
$\eta_0$ is the fully-degenerate result, and $\delta_{ee}$ accounts for e-e collisions.  In the non-degenerate (Spitzer) limit $\eta=\eta_S$ with $\eta_S/\eta_0 = (\pi^{1/2}/8) \theta^{-3/2}$, or $\eta_S \propto T_e^{-3/2}$ (neglecting temperature dependence of $\eta_0$ via $Z_i$ or $\ln\Lambda_{ei}$). An approximate form with the correct small and large $\theta$ limits is
\begin{equation}
\frac{\eta}{\eta_0} \approx \left[ 1 + (4\pi^{-1/3}\theta)^p\right]^{-3/2p}.
\end{equation}
For $p=1.72$ the relative error in $\eta/\eta_0$ is at most 2\% for all $\theta$.  Some workers approximately include Fermi degeneracy by capping $\eta$ at the value at some temperature.  The Spitzer $\eta_S$ equals the fully degenerate $\eta_0$ for $\theta=(\pi^{1/2}/8)^{2/3}=0.366$.

We use the Lee-More Coulomb logarithm:
\begin{equation}
\ln\Lambda_{ei} = \mathrm{max}\left( 2, \frac{1}{2}\ln\left[1+\Lambda^2 \right]  \right),
\end{equation}
with $\Lambda \equiv b_{max}/b_{min}$ the coupling parameter; $\Lambda\gg 1$ for a weakly-coupled plasma.  $b_{max}=[\lambda_{DH}^2+R_i^2]^{1/2}$ is the overall screening length. The Debye-H\"{u}ckel screening length is given by 
\begin{equation}
\lambda_{DH}^{-2} = \frac{e^2}{\epsilon_0} \left( \frac{n_e}{\bar T} + \sum_i \frac{n_iZ_i^2}{T_i} \right)
\end{equation}
with $\bar T \equiv (T_e^2+(4/9)E_F^2)^{1/2}$. The minimum impact parameter is 
\begin{eqnarray}
b_{min}^2 = b_{min,Q}^2 + b_{min,C}^{2}, \\
b_{min,Q} \equiv \frac{\hbar}{(12m_e\bar T)^{1/2}}, \\
b_{min,C} \equiv \frac{e^2}{12\epsilon_0\bar T} \frac{\sum_in_iZ_i^2}{\sum_in_iZ_i}.
\end{eqnarray}
$b_{min,Q}$ is the de Broglie wavelength, and $b_{min,C}$ is the classical distance of closest approach.

The Lee-More model has proven successful at capturing the results of experiments or more detailed models.  Most concrete implementations of Lee-More involve several material-dependent adjustable parameters.  They can frequently be chosen to replicate more correct results.  Like any semi-analytic model, Lee-More can be applied over wide parameter ranges and usually gives smooth results.  Tabulated output from more detailed models can include more physics.  The typical drawbacks of tables include the limited parameter range over which they were generated, and the difficulty of tabulating a high-dimension domain (for instance, a dopant ion species of variable concentration increases the table dimensionality).

More sophisticated models than Lee-More exist, and are particularly necessary in the non-plasma regime. One is the Purgatorio code developed at LLNL \cite{wilson-purg-jqsrt-2006,sterne-purg-hedp-2007}.  It solves the Dirac equation for bound and continuum electron states surrounding a single ion. Transport coefficients like resistivity are found using an extended Ziman formulation \cite{ziman-philmag-1961,evans-ziman-1973,rinker-xport-pra-1988}, and require specification of the ion correlation function, or equivalently the ion structure factor. See Ref.\ \cite{hansen-purg-nedpc-2005} for details.  A similar approach to transport coefficients is presented in Ref.\ \cite{rozsnyai-xport-hedp-2008}.

\begin{figure}
\center \includegraphics[height=6cm]{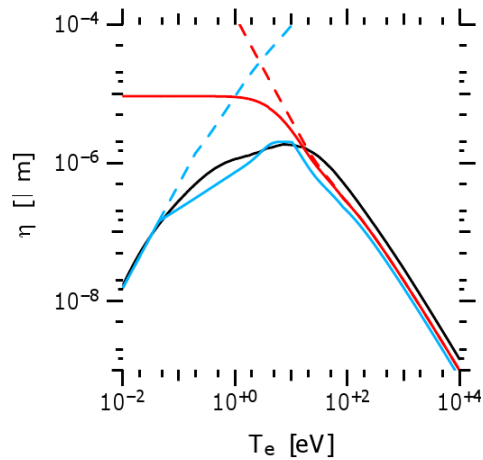}
\caption{Resistivity for beryllium at density 1.84 $\gcmcub$ from various models.  See text for details.}
\label{fig:etaBe}
\end{figure}

Figure \ref{fig:etaBe} plots $\eta$ vs.\ $T_e$ for 1.84 $\gcmcub$ beryllium from various models. The solid black curve is from Purgatorio \footnote{Data kindly provided by P.\ Sterne, LLNL}. The ion structure factor used was a combination of the results of Baiko et al.\ \cite{baiko-ionstruct-prl-1998}, the one component plasma model, and Debye-H\"{u}ckel theory.  Purgatorio calculates the charge state as well, which increases smoothly with $T_e$, from 1.5 at room temperature, to 3.2 at 100 eV, and asymptotically approaching 4 for higher $T_e$. We use Purgatorio's $Z_i$ in the other calculations. Since $Z_i>1$ for all $T_e$ we neglect electron-neutral collisions ($\tau_{en}\rightarrow\infty$).  The solid red curve comes from Lee-More's $\tau_{ei}$ for e-i collisions, modified to include e-e collisions ($\delta_{ee}\neq1$).  The dashed red curve is the Spitzer $\eta_S$ ($E_F/T_e\rightarrow0$).  The solid blue curve is the full Lee-More model, with numerical parameters chosen to give a decent agreement with Purgatorio at low $T_e$, and with e-e collisions neglected ($\delta_{ee}=1$). This last choice gives a slight difference between the red and blue curves at high $T_e$.  It is easy to include $\delta_{ee}$, but we omit it to demonstrate its magnitude. The dashed blue curve comes from just the melting model $\tau=\taumelt$.  Although we have found parameters that bring the Lee-More model into decent agreement with the more complete Purgatorio results for the chosen density, those values are likely not optimal for all densities.

\begin{figure}
\begin{center}
(a)\includegraphics[height=6cm]{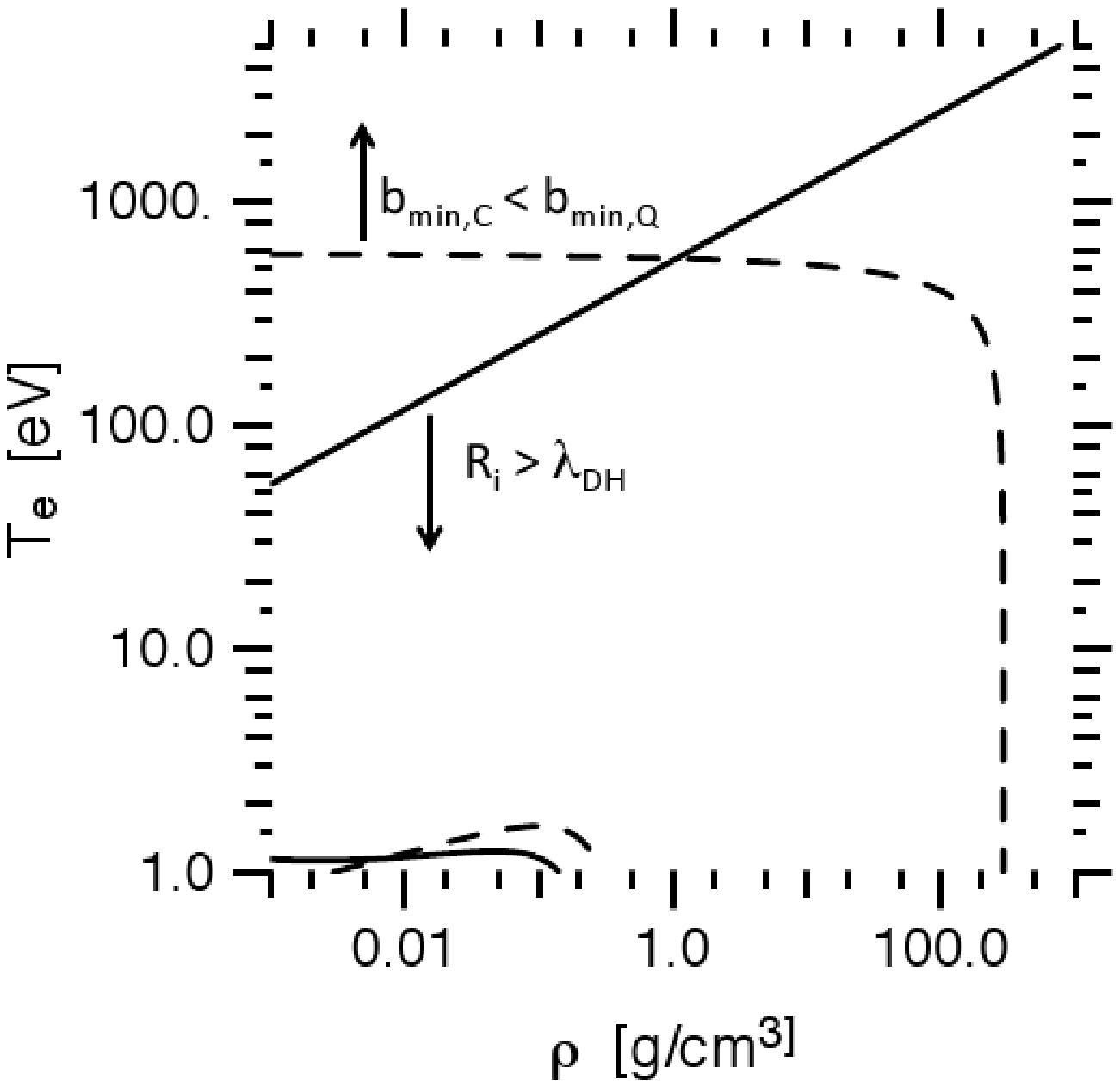} (b)\includegraphics[height=6cm]{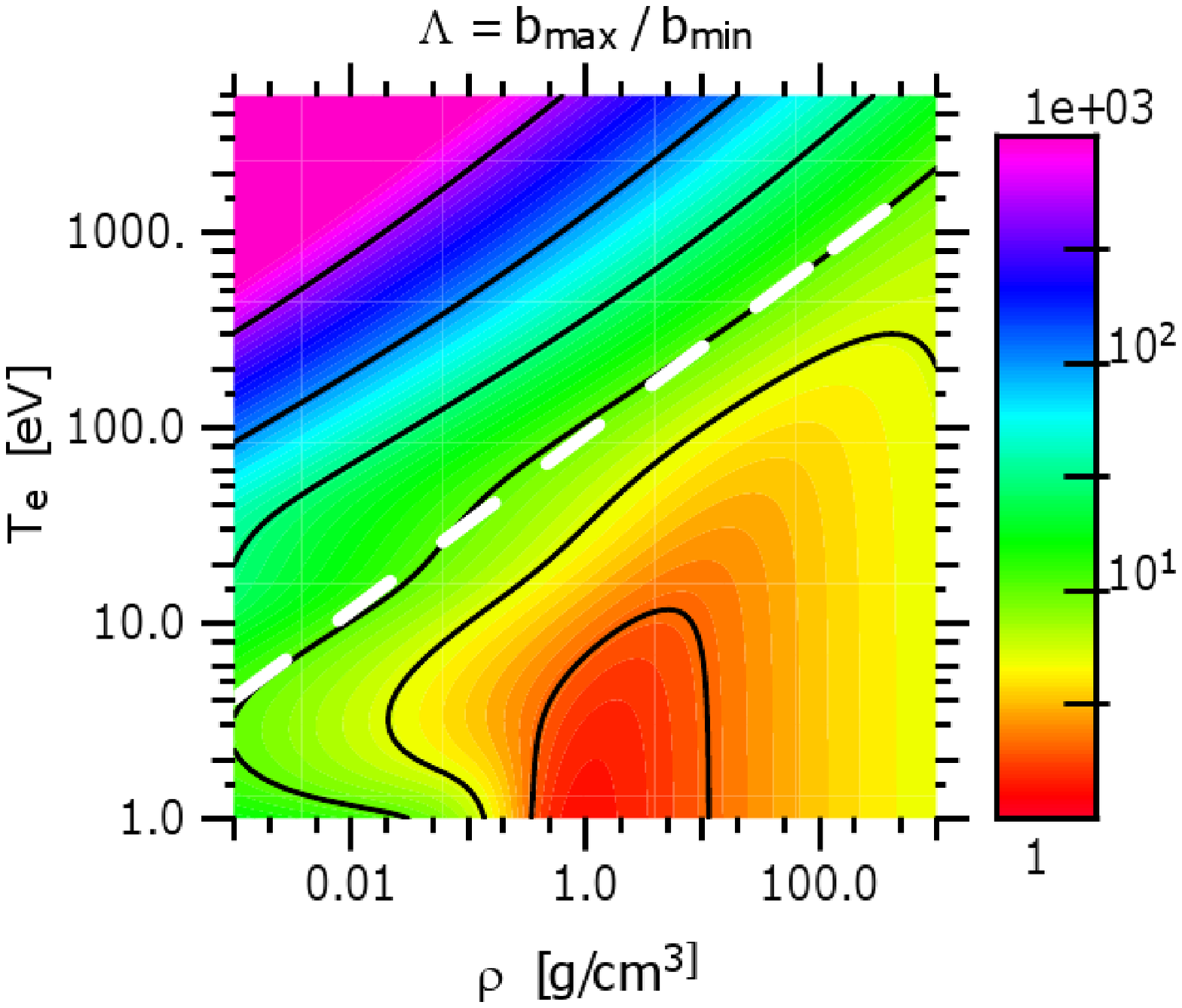}
\caption{For beryllium: (a) Where different terms dominate $b_{min}$ and $b_{max}$. (b) Coupling parameter $\Lambda$ with black curves for $\Lambda=2, 5, 10, 30, 100, 500$. White dashed curve is fit to $\Lambda=10$ contour: $T_e = 1.5$ keV$(\rho/500\, \gcmcub)^{0.45}$.}
\end{center}
\label{fig:coupBe}
\end{figure}

Some recent attention has been paid to transport coefficients for strongly coupled plasmas, i.e.\ $\Lambda\sim1$ \cite{baalrud-xport-pop-2012,glosli-md-pre-2008,daligault-temprel-pre-2009}.  These usually do not consider F-D statistics and connection with the non-plasma state, so that a strongly-coupled generalization of Lee-More is not yet at hand.  Following Baalrud \cite{baalrud-xport-pop-2012}, strong coupling significantly affects the Coulomb logarithm when $\Lambda < 10$. We assess the potential importance of strong coupling for beryllium, using the Purgatorio $Z_i$, in Fig.\ \ref{fig:coupBe}. Panel (a) indicates where different terms dominate $b_{min}$ and $b_{max}$, and panel (b) plots $\Lambda$ including a fit to the $\Lambda=10$ contour. Strong coupling is estimated to be significant below this contour, which includes regimes of interest to fast ignition and high energy density physics more generally.


\section{Simulation Methods}
\label{simmeth}
In this section we will review the simulation methods that have been developed for studying fast electron transport, and particularly those that are used
in the Fast Ignition context.

\subsection{Vlasov-Fokker-Planck Codes}
\label{vfpcodes}
The Vlasov-Fokker-Planck (VFP) equation for electrons describes their
motion through phase space under the action of the local average Lorentz
force and the microscopic field fluctuations that give rise to small-angle
collisions with other electrons and ions. It is usually expressed
in Cartesian geometry without giving the details of the collisional
term \cite{thomas12}:
\begin{equation}
\frac{\partial f}{\partial t}+\mathbf{v\cdot}\frac{\partial f}{\partial\mathbf{x}}-e\mathbf{\left(E+v\times B\right)\cdot}\frac{\partial f}{\partial\mathbf{p}}=\left(\frac{\partial f}{\partial t}\right)_{c},
\label{FPcart}
\end{equation}
where $\mathbf{x}$ and $\mathbf{p}$ are the phase-space position
and momentum coordinates (respectively), $f=f\left(\mathbf{x},\mathbf{p},t\right)$
is the electron distribution function, $\mathbf{E}+\mathbf{v}\times\mathbf{B}$
is the Lorentz force and the term on the RHS accounts for the scattering
in momentum-space due to collisions. When this equation is solved
by the use of computational particles it is known as {}``collisional
Particle-In-Cell'' (collisional-PIC). The collisional-PIC technique
has many advantages: robustness; good momentum and energy conservation;
no stability or magnitude restrictions in momentum-space; very accurate
advection in momentum-space; and it naturally concentrates computational
effort in well-populated regions of phase-space. The collisional-PIC
technique also has the great benefit of over 40 years of
research experience behind it. However, it does suffer one major drawback
in that it also introduces noise into the numerical result. The effect
of this noise is arguably not yet well-explored in the regime of the
dense plasmas that arise in fast-ignition research. For this reason
a small number of codes based on finite-difference techniques \cite{bell-ppcf-2006,robinson-switchyard-pop-2007,sherlock:103101}
have arisen in recent years. While techniques based on finite-difference in phase space eliminate
noise, they do not currently possess most of the aforementioned advantages
inherent in the particle approach (and in fact are often deleteriously
affected by their converses). Numerical diffusion also occurs (although
modern techniques do minimize it) and this may adversely affect the
physics. Nevertheless the finite-difference approaches have much merit
and also serve as a useful {}``reality-check'' on the results gained
with the particle techniques. 

Although the VFP equation is largely valid in FI-relevant plasmas,
it should be kept in mind that the collision term requires corrections
of order $1/\ln\Lambda$ (the inverse Coulomb logarithm), which should
be necessary when the plasma is initially cold and dense. Since the
VFP equation is valid over all of momentum space, it is possible to
solve it for the distribution function of the energetic particles
only, and this is the basis of the VFP-hybrid technique (as used in
e.g. \cite{robinson-switchyard-pop-2007}), where the background electrons are
treated as a simple fluid (see the section on hybrid methods). In
fact due to the heavy computational demand of solving the VFP equation,
the hybrid technique is by far the most common approach.

Since angular-scattering is important for electrons, 
it is advantageous to use a spherical coordinate system in momentum-space,
as angular scattering can be easily expressed as diffusion of the
distribution function in the angular coordinates. Recasting \eref{FPcart} in spherical-coordinates in momentum-space ($p,\theta,\phi$)
and introducing diffusive and drag-like (i.e. in $p$) collision terms gives:
\begin{eqnarray}
\fl \frac{\partial f}{\partial t}+v\cos\theta \sin\phi\frac{\partial f}{\partial
  x}+v\sin\theta \sin\phi\frac{\partial f}{\partial y} \nonumber \\
+F_{x}\left\{ \cos\theta \sin\phi\frac{\partial f}{\partial
    p}-\frac{\sin\theta}{p\sin\phi}\frac{\partial
    f}{\partial\theta}+\frac{\cos\theta \cos\phi}{p}\frac{\partial
    f}{\partial\phi}\right\} \nonumber  \\
+F_{y}\left\{ \sin\theta \sin\phi\frac{\partial f}{\partial
    p}+\frac{\cos\theta}{p\sin\phi}\frac{\partial
    f}{\partial\theta}+\frac{\sin\theta \cos\phi}{p}\frac{\partial
    f}{\partial\phi}\right\} \nonumber  \\
=\left(\frac{Y_{ee}n_{e}}{m_{e}}+\frac{Y_{i}n_{i}}{m_{i}}\right)\frac{m_{e}^{2}}{p^{2}}\frac{\partial}{\partial
  p}\left(\gamma^{2}f\right) \nonumber  \\
+\frac{1}{2}\left(Y_{ee}n_{e}+Y_{i}n_{i}\right)\frac{m_{e}}{p^{3}}\left\{
  \frac{1}{\sin^{2}\phi}\frac{\partial^{2}f}{\partial\theta^{2}}+\frac{1}{\sin\phi}\frac{\partial}{\partial\phi}\left(\sin\phi\frac{\partial
      f}{\partial\phi}\right)\right\},
\label{VFP}
\end{eqnarray}
where 
\begin{eqnarray}
F_{x}=-e\left(E_{x}-vB_{z}\sin\theta \sin\phi\right), \\
F_{y}=-e\left(E_{y}+vB_{z}\cos\theta \sin\phi\right)
\end{eqnarray}
are the components of the Lorentz force and $v=p/\gamma m$, $\gamma=\sqrt{1+p^{2}/m^{2}c^{2}}$, $Y_{ee}=4\pi\left(e^{2}/4\pi\epsilon_{0}\right)^{2}\ln\Lambda_{ee}$
and $Y_{i}=4\pi\left(Ze^{2}/4\pi\epsilon_{0}\right)^{2}\ln\Lambda_{ei}$.
The collision terms on the RHS are taken from \cite{johzaki:062706}
and are valid for hybrid VFP simulations only: the first term accounts
for the dynamic friction of fast electrons with the cold electrons
and ions, while the second term accounts for the angular scattering
of fast electrons off cold electrons and ions.

\Eref{VFP} can be readily solved in flux-conservative form,
as is done in the FIDO simulation code \cite{sherlock:103101}, which
uses Piecewise-Parabolic-Interpolation to compute the fluxes combined
with the Finite-Difference Time-Domain (FDTD) scheme for Maxwell's
Equations to compute the fields (see e.g. \cite{villasenor1992}).
This treatment is particularly advantageous when the field (acceleration)
terms dominate, as is the case when absorption in strong laser fields
is modelled.

An alternative form of \Eref{VFP} is possible by expanding
the distribution function in momentum-space in a spherical-harmonic
basis:
\begin{equation}
f\left(t,{\bf
    x},\mathbf{v}\right)=\sum_{n=0}^{N}\sum_{m=-n}^{n}f_{n}^{m}\left(t,{\bf x},v\right)P_{n}^{m}\left(\cos\theta\right)e^{im\phi}
\end{equation}
where $\theta$ is the angle between the velocity vector and the spatial
coordinate and the $P_{n}^{m}$ are the associated Legendre functions
\cite{bell-ppcf-2006}. This gives rise to a large set of coupled
partial differential equations for the coefficients:
\begin{eqnarray}
\fl \frac{\partial f_{n}^{m}}{\partial
  t}+\frac{n\left(n+1\right)}{2}\frac{1}{2}\left(Y_{ee}n_{e}+Y_{i}n_{i}\right)\frac{m_{e}}{p^{3}}f_{n}^{m}-C_{ee}
= \\
\frac{eB_{z}}{2m}\left\{ \left(n-m\right)\left(n+m+1\right)f_{n}^{m+1}-f_{n}^{m-1}\right\} \\
-\left(\frac{n-m}{2n-1}\right)v\frac{\partial f_{n-1}^{m}}{\partial x}-\left(\frac{n+m+1}{2n+3}\right)v\frac{\partial f_{n+1}^{m}}{\partial x}\\
-eE_{x}\left\{ \frac{n-m}{2n-1}G_{n-1}^{m}+\frac{n+m+1}{2n+3}H_{n+1}^{m}\right\} \\
-\frac{eE_{y}}{m}\left\{ \frac{1}{2n-1}\left[G_{n-1}^{m-1}-\left(n-m\right)\left(n-m-1\right)G_{n-1}^{m+1}\right]\right\} \\
-\frac{eE_{y}}{m}\left\{ \frac{1}{2n+3}\left[-H_{n+1}^{m-1}+\left(n+m+1\right)\left(n+m+2\right)H_{n+1}^{m+1}\right]\right\}, 
\label{FPharmonic}
\end{eqnarray}
where $G_{n}^{m}=\partial f_{n}^{m}/\partial p-nf_{n}^{m}/p$ , $H_{n}^{m}=\partial f_{n}^{m}/\partial p+\left(n+1\right)f_{n}^{m}/p$
and $C_{ee}$ accounts for the friction between fast electrons and
the background plasma. Equation \ref{FPharmonic} is for the 1D case
only (see \cite{bell-ppcf-2006} for the full 2D equations). This
form of the VFP equation has the advantage that it can be solved with
relatively fast and simple (for example Runge-Kutta) integration
schemes, provided the driving fields are small (in comparison to the
laser fields).  Numerical schemes that can handle large perturbations to the distribution function
are complex and slow.  It also allows for a particularly accurate treatment
of the magnetic field terms, which are reduced to algebraic form when
differenced. This form for the VFP equation was used in, e.g. \cite{bell-2003}.


\subsection{Hybrid Ohmic Codes}
\label{hybridcodes}
It is apparent from the disparity between the cold and fast electron characteristics, as described in Section \ref{basic} that the problem of fast electron transport is computationally `stiff' (disparate length and time scales), and that this also allows a natural separation of the problem into two interlinked models.  It is this observation that has motivated the development of the `hybrid' code.  The term `hybrid code' appears in many places in plasma physics, and the term often denotes very different things.  In the case of fast electron transport, the term denotes a code in which a kinetic treatment is applied to a distinct fast electron population, and a fluid treatment is applied to a distinct background plasma.  It is frequently assumed that the background plasma will respond instantaneously to the fast electrons to ensure quasineutrality, which is valid on length scales larger than the Debye (or other screening) length.  An independent assumption, which we call the Ohmic approximation, is that the electric field can be determined from a suitable generalized Ohm's Law, with displacement current $\partial_t{\bf E}$ dropped from Amp\`ere's law. In all cases the magnetic field is evolved from Faraday's Law.

This splitting of the populations will only be a good approximation when $n_f \ll n_b$, and when the fast electron mean energy is very much greater than the mean thermal energy of the background electrons.  In the case of the ultra-intense laser-generated fast electron transport problem these conditions will be quite easily satisfied at material densities above 1~g~cm$^{-3}$.  Although the fast electron population is very much less dense than the background electrons, the fast electron current density is still sufficiently large to generate electric and magnetic fields with energy densities comparable to the fast electron energy density {\it unless} there is a compensating return current carried by the background electrons (see \S \ref{basic_macro}).  This leads to the conclusion that, to a good approximation,  ${\bf j}_f + {\bf j}_b \approx 0$.  A more accurate approximation is ${\bf j}_f + {\bf j}_b \approx \mu_0^{-1}\nabla \times {\bf B}$.  The key equations for {\bf E} and {\bf B} in the hybrid approximation then become,
\begin{equation}
\label{ohms}
{\bf E} = -\eta{\bf j}_f + \frac{\eta}{\mu_0}\nabla \times {\bf B},
\end{equation}
and
\begin{equation}
\label{induction}
\frac{\partial{\bf B}}{\partial{t}} = \eta\nabla \times {\bf j}_f + \nabla\eta \times {\bf j}_f +\frac{\eta}{\mu_0}\nabla^2{\bf B}-\frac{1}{\mu_0}\nabla{\eta} \times {\bf B}. 
\end{equation}
The Ohm's Law shown in \Eref{ohms} can be easily extended to include a number of additional terms (see Sec.\ \ref{bkgdphysics} of this review) --- here it is just given in one of the simplest forms.  The kinetic treatment of the fast electrons is mathematically described by a suitable kinetic equation, i. e. \eref{FPcart}.  Usually this is solved by using the standard Particle-in-Cell methods, but with collision operators for the angular scattering from background ions and electrons, and drag from background electrons, included via a Monte Carlo method.  The background plasma is described, in general, by a set of hydrodynamic equations, although for some problems it is reasonable to treat the background plasma as essentially static.  Even if the background plasma is static, its temperature must evolve due to both Ohmic heating of the background electrons, and collisional drag on the fast electrons. These effects must be incorporated into the background electron energy equation.  

This must also be accompanied by a prescription for the resistivity.  This can be from a theoretical model (Spitzer resistivity or Lee-More), an empirical model, or even a heuristic model.  Although the background resistivity is not calculated self-consistently (unlike a purely kinetic model), one advantage of the hybrid approach is that it is relatively easy to use a resistivity model that better treats the `warm dense matter' regime. This regime is unavoidable both in solid density interactions and Fast Ignition, as the resistivity will only be {\it very} well approximated by the Spitzer resistivity well above the Fermi energy.  For DT at 1~g~cm$^{-3}$ this would be temperatures above 14~eV.

Although the hybrid approach makes a number of approximations it also has a number of powerful advantages.  Firstly the fluid treatment of the background means that the very small length and time scales of the background plasma can be ignored, and thus much larger time-steps can be used which allows large problems to be run quickly.  Secondly the model is very computationally robust.  Thirdly hybrid codes are easy to write and maintain.  Fourthly, hybrid codes allow a lot of physics to be included easily.  There are a number of such hybrid codes both in current use and reported in the literature.  This includes the unnamed code of Davies \cite{jrd1}, {\sc PETRA}\cite{honr06}, {\sc LEDA} \cite{robinson-switchyard-pop-2007}, and {\sc ZEPHYROS} \cite{robinson2}.
\subsection{Hybrid Implicit PIC Codes}
\label{hybridpiccodes}
LSP (Large Scale Plasma) is a one-, two-, and three-dimensional particle-in-cell (PIC) code developed by Mission Research Corporation \cite{welch1} and currently maintained by Voss Scientific \cite{welch2,thoma1}. LSP has a hybrid mode, as well as several types of electromagnetic field solution available: the standard (explicit) leapfrog algorithm and implicit algorithms using iterative Alternate-Direction Implicit (ADI), a two-step ADI, and matrix inversion. The implicit algorithms, particularly the two-step ADI, are useful in relaxing the Courant limit on the time step. An iterative electrostatic algorithm is also available for simulations in which fields are slowly varying. For short-pulse LPI and fast electron transport problems, the relativistic implicit-PIC code ELIXIRS \cite{drouin10} employs a similar approach.

LSP has several options for advancing particles: the standard momentum-conserving and energy-conserving PIC algorithms, cloud-in-cell (CIC) algorithm, and direct implicit particle/field algorithm \cite{hewett-imppic-jcp-1987} which can be used in either the PIC or CIC models. The direct implicit algorithm is used most often. The benefits of the direct implicit algorithm are that the usual charged particle limitations on the time step, namely the need to resolve the cyclotron and plasma frequencies, are relaxed although both frequencies cannot be under resolved at the same time and position. The implicit algorithm is useful for very dense plasmas (that occur in fast ignition studies) where the details of electron plasma oscillations can be ignored. By damping unresolved high-frequency Langmuir modes the direct implicit algorithm allows stable modeling of dense plasmas without needing to resolve these fast modes.

The energy conservation as well as the speed of the direct implicit calculation is further improved by including a nonrelativistic inertial fluid model for the electrons in which the directed and thermal energy of the electrons are treated separately. The equation of motion for the fluid electrons is of standard Braginskii type \cite{braginskii-xport-1965}. It includes a frictional force to model collisions with other particle species. For temperature, the new energy equation for an ideal gas is added including the $pdV$ work, energy exchange between species, thermal conduction, and Ohmic heating rate. Inelastic losses with neutrals can also be included. In some circumstances, kinetic effects become important. Examples include runaway, where a hot electron population coexists with a thermal one, or acceleration of less dense electrons or ions from a biased plasma. In the hybrid mode, LSP permits dynamic reallocation of particles between the fluid and kinetic description. Fluid particles with directed energy much greater than thermal energy transition to kinetic particles. Kinetic particles with energy less than the ambient fluid thermal energy transition to fluid particles. These transitions result in energetic electrons treated kinetically and dense thermal plasma electrons as a fluid.

The LSP simulations include an algorithm to model electron-electron, ion-ion and electron-ion collisions. For kinetic particles, this involves first constructing drifting Maxwellian distributions at each grid cell. A particle of a given species is then elastically scattered isotropically in the center-of-mass frame off a particle obtained by sampling this distribution. Collisions between different species (both kinetic and fluid) are separated into an energy push and a frictional momentum push. The energy and momentum transfers from one species to another are accomplished by summing the changes from each interaction on the grid. The collision frequencies are determined from the Spitzer formulation. Optionally, the more accurate Lee-More \cite{lee-more-pof-1984} model with Desjarlais corrections \cite{desjarlais-metal-cpp-2001} is available for collisions of fluid electrons and background ions with the ion charge state calculated with a Thomas-Fermi equation-of-state with pressure ionization corrections. Monte-Carlo type scattering model with the drag and scattering formulas of \cite{solodov-stop-pop-2008,atzeni-ppcf-2009} is also available for collisions of kinetic electrons and background plasma electrons and ions.

The LSP hybrid implicit approach to dense plasma modeling is alternative to that of hybrid Ohmic codes, described in the previous subsection. In those codes, the background plasma is modeled as a collisional fluid and charge neutrality is assumed. The electric field is found from Ohm's law and the background return current is found from Amp\`ere's law without displacement current. This reduced-model approach is inapplicable to laser-plasma interactions, or low-density regions with, e.g., Debye sheaths. The LSP model solves full Maxwell equations with displacement current and is applicable in the laser-plasma interaction and low-density regions, provided a kinetic description for plasma electrons is used there (while a fluid description can be used elsewhere in the same run). Time steps and cell sizes can be chosen that resolve the laser-plasma interaction (LPI) near the critical surface. These time steps, while explicit in the LPI region (electron density, $n_e=1.1\times 10^{21}$ cm$^{-3}$ for 1~$\mu$m laser light), can be still highly implicit to the plasma frequency in the solid density target ($n_e >$10$^{23}$ cm$^{-3}$). The entire process of laser propagation into the underdense plasma, fast electron production and transport into the dense plasma, and ion acceleration from the plasma-vacuum interface can be modeled with LSP \cite{welch2}.

Detailed equation-of-state (EOS) and multi-group diffusion radiation transport modeling capability was recently implemented in the LSP framework with fluid particles \cite{thoma1}. The EOS and opacity data needed for the algorithm are pre-calculated by the Propaceos code \cite{macfarlane1}, which utilizes detailed atomic models for plasmas in local thermodynamic equilibrium (LTE) as well as non-LTE states. The EOS model is used to evolve the ion charge state and introduce non-ideal gas behavior. The radiation energy density field is calculated, which is coupled to the plasma.

\section{Review of Ignition Scale Calculations}
\label{ignscale}

This section summarizes the FI calculations carried out so far,
showing the dependence of the ignition energies on the electron beam parameters.
Electron-driven FI modelling relies on the characterization of the fast electron source, the transport from the generation zone to the compressed core and the energy deposition in the fuel. Thus, the complete description of FI requires the integration of different models/codes that deals with different spatial and temporal scales. Fully integrated calculations are not possible with the present computer resources. Here, we focus our attention on the partially integrated calculations that consist of characterizing the fast electron source via experiments or PIC simulations and using this source to perform fast electron transport calculations coupled to radiation-hydrodynamics, including fusion reactions. This 'integrated' model has been used so far to estimate the electron beam requirements in the fast ignition scenario \cite{solo08,honrubia1,strozzi-fastig-pop-2012}. Relativistic Fokker-Planck models for electron transport in sub-ignition targets have been developed also within the context of the FIREX-I project \cite{yokota1}.

\subsection{Ignition energies of perfectly collimated electron beams}
We assume for the moment that a perfectly collimated beam impinges on a DT assembly at a time close to the peak compression. Here, we do not take into account the fast electron scattering nor the EM fields generated by the electron beam. We first consider the target proposed by Solodov et al. with a DT mass of 0.5 mg compressed by a 300 kJ nanosecond laser pulse to a peak density of 500 gcm$^{-3}$ \cite{solodov2}. Assuming a mono-energetic electron beam of 20 $\mu$m radius and a pulse duration of 10 ps, the lowest ignition energy $E_{ig}$= 16.2 kJ is reached for 2 MeV electrons \cite{solodov2}. This ignition energy is much higher than the 7 kJ obtained from Eq.(1) for $\rho$= 500 g/cm$^3$ due to the target density profile and also because the beam radius and pulse duration do not have the optimal values given by Eqs.(2-4).  For the more realistic case of electrons with a relativistic Maxwellian energy spectrum, the lowest ignition energy raises to 21.5 kJ for an electron mean energy $\langle{E}\rangle$ = 1.25 MeV. This energy is substantially lower than the 2 MeV found for mono-energetic beams because Maxwellian electrons deposit their energy over a larger region. In addition, it is much lower than those obtained in experiments and PIC simulations for laser intensities around 10$^{20}$ Wcm$^{-2}$. More realistic electron energies can be obtained by using the ponderomotive scaling formula ${\langle{E}\rangle}/m_ec^2 = \left[1+ I_L\lambda^2/1.38\times10^{18}\right]^{1/2}-1$ , which relates the laser intensity $I_{L}$ (in Wcm$^{-2}$) and wavelength $\lambda$ (in $\mu$m) with the electron mean energy $\langle{E}\rangle$ \cite{wilks1}. For instance, let us consider a Gaussian laser pulse with a duration of 13.8 ps and $\lambda$ = 1.054 $\mu$m impinging on the fuel assembly mentioned above. Assuming that the electron mean energy is given by the ponderomotive scaling and a laser-to-fast electron conversion efficiency of 50\%, one obtains $\langle{E}\rangle$ = 6.3 MeV and an ignition energy $E_{ig}$= 53 kJ \cite{solodov2}. We emphasize that the ignition energies mentioned above have been obtained under the strong assumptions of electron straight path, no beam divergence and without accounting for the self-generated EM fields. Atzeni et al. estimated that the scattering effects raise the ignition energy by about 20\% \cite{atzeni-ppcf-2009}. Thus, even for the ideal conditions assumed here, electron beam energies of several tens of kJ are needed to ignite a target. Similar results have been obtained for the all-DT target design proposed for the HiPER project \cite{atze08}. 

\subsection{Ignition energies of divergent electron beams with an assumed initial distribution function}
More realistic calculations can be performed by means of the partially integrated model mentioned above. In this model, the beam parameters are estimated from experiments or PIC simulations conducted at laser intensities and/or pulse durations lower than those required for FI. The main features of the relativistic electron source considered so far in FI simulations can be summarized as follows:

{\bf i) Energy spectrum}: It is normally asummed that the electron spectrum is given by the exponential distribution obtained in PIC simulations for sub-ps pulses. Relativistic Maxwellian spectra have been used also. Both distributions depend on the fast electron temperature or electron mean energy $\langle{E}\rangle$, typically fitted to experiments. For laser intensities $<$10$^{19}$ Wcm$^{-2}$, the so-called Beg's law $\langle{E}\rangle$ = 150$(I_{17}\lambda^2)^{1/3}$ is used, where $\langle{E}\rangle$ is in keV, $I_{17}$ is the laser intensity in units of 10$^{17}$ Wcm$^{-2}$ and $\lambda$ the laser wavelength in $\mu$m \cite{beg1}. For intensities around 10$^{19}$ Wcm$^{-2}$ or higher, the ponderomotive scaling \cite{wilks1} reproduces well experiments and PIC simulations. However, it gives electron energies well over the desired values around 2 MeV for the laser intensities typical of the FI regime ($>$10$^{20}$ Wcm$^{-2}$).

 Chrisman et al. \cite{chri08}, Haines et al. \cite{haines1} and Kluge et al. \cite{kluge1} have recently reported scaling laws that provide electron energies lower than those obtained by the ponderomotive scaling and of the same order than those predicted by the Beg's law. In principle, this is very favourable for FI because the optimal electron range is about 1.2 g/cm$^2$ \cite{atzeni-fi-pop-1999}, which corresponds to an electron energy lower than 2 MeV \cite{solodov-stop-pop-2008} and a laser intensity about 2.4$\times$10$^{20}$ Wcm$^{-2}$ assuming the Beg's scaling. Unfortunately, as it is discussed in Section 5.3, recent 3D PIC simulations have revealed that the mean energy of relativistic electrons is similar to that given by the ponderomotive scaling and their spectrum differs substantially from the exponential or relativistic Maxwellian distributions mentioned above \cite{strozzi-fastig-pop-2012}. 

{\bf ii) Beam divergence}: In most of the fast electron calculations carried out so far it has been assumed that the electron divergence is given by the beam effective propagation angle measured in the experiments. However, as electrons propagate in metals or plastics, whose resistivity is several orders of magnitude higher than that of the DT fuel, resistive collimation effects can be important \cite{bell-2003}. In this case, the initial fast electron divergence turns out to be substantially higher than the effective propagation angle measured \cite{stephens1,green1,norreys1}. For instance, to reproduce the full propagation angle of 35$^\circ$ found in the experiments conducted by Green et al. \cite{green1}, an initial electron divergence half-angle as large as 50$^\circ$ has to be assumed in hybrid calculations \cite{honrubia1}. Recent PIC simulations have shown initial divergence half-angles of 50-55$^\circ$ for FI conditions \cite{strozzi-fastig-pop-2012}. Thus, guiding mechanisms should be envisioned for these highly divergent beams in order to have a good coupling efficiency with the dense fuel.

It is also important to account for the dependence of the divergence angle on the electron kinetic energy observed in PIC simulations. It can be taken into account in a simplified fashion by assuming that the initial divergence angle is given by the ponderomotive scaling $\tan\theta = \left[2/(\gamma-1)\right]^{1/2}$ \cite{quesnel-elec-pre-1998}, where $\theta$ is the polar angle and $\gamma$ the relativistic Lorentz factor. It is conjectured that an electron of energy $(\gamma-1)m_ec^2$ is emitted with a divergence half-angle randomly selected between 0 and $\theta$. Thus, high energy electrons are well collimated while low energy electrons have an almost isotropic distribution. This dependence on electron energy is important to get a reasonable high energy coupling between the beam and the dense fuel. However, recent 3D PIC simulations have shown that the energy spectrum and the angular distribution of the fast electron source are independent of each other, i.e. the divergence angle is the same for all electrons \cite{strozzi-fastig-pop-2012}. This implies a strong increase of the ignition energies when compared with those obtained from the ponderomotive scaling.

{\bf iii) Beam radius}. Experiments and PIC simulations of relativistic LPI with foil targets show that the size of the fast electron beam is greater than that of the laser beam \cite{stephens1}. This effect has to be taken into account in fast electron transport calculations \cite{honrubia2}. In the case of electron acceleration in re-entrant cones, it is found that the radius of the beam is approximately equal to the cone outer radius. This is also true in the double cones described in Section 6.7, where the vacuum layer between the two cone walls force fast electron propagation towards the cone tip \cite{nakamura1,cai1,debayle1,debayle3}.

{\bf iv) Conversion efficiency}. The laser-to-fast electron conversion efficiency obtained in PIC simulations ranges from 30 to 50\% for the electron energies relevant for FI, e.g. E $>$ 250 keV \cite{debayle2,strozzi-fastig-pop-2012}.

\begin{figure}[H]
\begin{center}
\includegraphics[width = \textwidth]{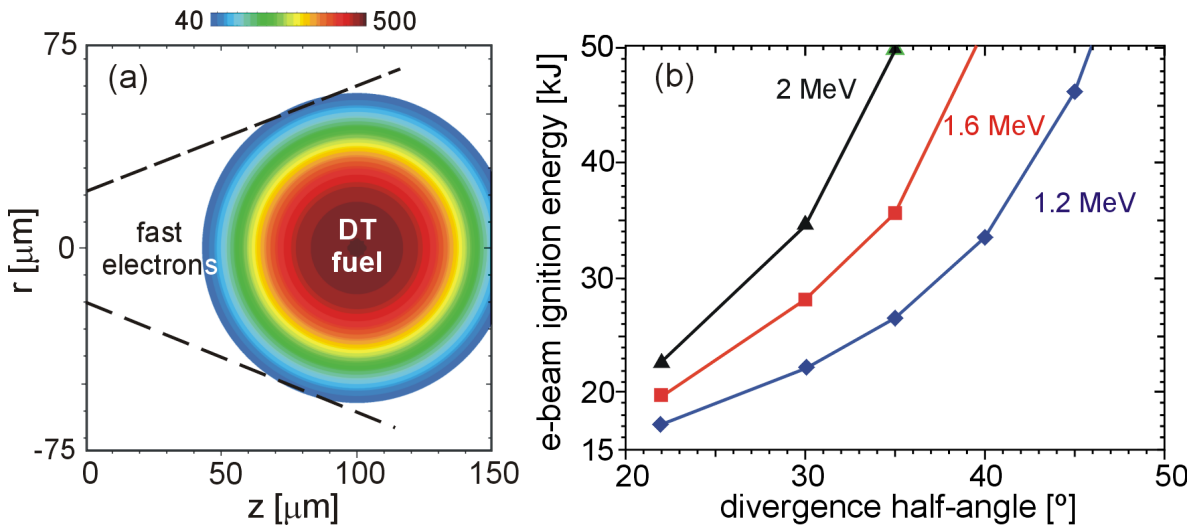}
\caption{\label{fig:ign_fig1}Left: (a) Initial target density in gcm$^{-3}$. The halo surrounding the dense core has a density of 2 gcm$^{-3}$. Right : (b) Electron beam ignition energies as a function of the divergence half-angle and the electron mean energy $\langle{E}\rangle$. The beam parameters for the case (35$^\circ$, 1.6 MeV) are 20 $\mu$m radius (HWHM) and 18 ps pulse duration (FWHM) \cite{honrubia1}}.
\end{center}
\label{fig:ignscale3}
\end{figure}

As an example of the ignition calculations assuming an initial fast electron distribution carried out so far, let us discuss the ignition energies of the idealized DT fuel configuration shown in Fig.~\ref{fig:ignscale3}(a) \cite{honrubia1}. It is assumed that the fast electron energy and divergence are given by the ponderomotive formulas with $\lambda$ = 0.527 $\mu$m multiplied by a scale factor. Despite it is technologically challenging, the 2$^{nd}$ harmonic of the Nd laser has been considered to reduce the electron energy \cite{solo08,honrubia1,strozzi-fastig-pop-2012}. The DT fuel has an initial super-Gaussian density distribution, 498$\exp(-(R/45)^4)$ gcm$^{-3}$ where $R$ is the distance to the centre in $\mu$m, sited on a density pedestal of 2  gcm$^{-3}$. The total DT mass is 0.39 mg and the initial temperature 300 eV. As the cone tip is not included in the simulation box, its effect on fast electron transport (beam filamentation, scattering and energy loss) is accounted for indirectly via the initial fast electron distribution function. Calculations have been performed with the hybrid code PETRA for fast electron transport \cite{honrubia3} coupled to the radiation-hydrodynamics code SARA \cite{honrubia4} run in 2D Eulerian mode and cylindrical r-z geometry.

Fast electron energy deposition takes place via Ohmic heating due to return currents and classical Coulomb scattering (collisional drag). Ohmic heating is important only in the low-density DT plasma, while collective behaviour is suppressed and energy deposition takes place almost exclusively by collisional drag in the dense core \cite{solo08,honrubia1,johzaki:062706,strozzi-fastig-pop-2012}. It is important to emphasize that collective effects may play a major role for core heating, but in an indirect way: self-generated B-fields may collimate the relativistic beam improving the coupling efficiency substantially. This effect appears to be very beneficial for FI: without resistive collimation there is little hope to ignite a pre-compressed target with reasonable beam energies due to the high divergence. However, as the beam collimation decreases strongly with the electron divergence angle, Eq.(9) \cite{bell-2003}, its importance in the FI scenario will depend on the full characterization of the fast electron source, which is not possible today neither by experiments nor PIC simulations. In addition to the beam collimation, resistive filamentation is the other collective effect that can play a role in electron-driven FI. It has been observed in the simulations of the ideal target shown in Fig. \ref{fig:ign_fig1}(a)\cite{honr06}, the imploded targets of Ref. \cite{solo08} and the experiments with solid and compressed plastic targets analyzed in \cite{solo09}. It was shown also in Fokker-Planck simulations of sub-ignition targets \cite{johzaki:062706}. Its effects on fuel ignition have not been studied in detail yet.

The minimum ignition energies of the target of Fig. \ref{fig:ign_fig1}(a) as a function of the initial divergence half-angle with the mean energy as a parameter are shown in Fig. \ref{fig:ign_fig1}(b). For small divergences, the beam is strongly collimated and the ignition energies are even lower than those obtained in Section 5.1 for perfectly collimated beams. For higher divergences, the beam collimation is less important and ignition energies increase more than proportionally with the divergence half-angle. It is worth noting that, even for the ideal case shown in Fig. \ref{fig:ign_fig1}, to get ignition with a 50 kJ electron beam requires a substantial reduction of the divergence half-angle, from the 50-55$^\circ$ obtained from experiments and PIC simulations to the 35$^\circ$ shown in Fig. \ref{fig:ign_fig1}(b) for 2 MeV electrons.

Similar calculations have been reported by Solodov et al. for the direct-drive capsule mentioned in Section 5.1 \cite{solodov2}. Simulations have been performed with the hybrid-PIC code LSP \cite{welch1} coupled to the radiation-hydrodynamics code DRACO \cite{radha1}. A relativistic electron beam is injected 125 $\mu$m from the target centre at a time when the maximum DT density is slightly above 500 g/cm$^3$. Assuming that beam electrons have a relativistic Maxwellian distribution with a mean energy of 2 MeV and a divergence half-angle of 20$^\circ$ (HWHM), Solodov et al. found a minimum ignition energy about 43 kJ \cite{solo08}. They also found an important resistive collimation of the fast electron beam. The ignition energy increases strongly with the divergence half-angle $\theta$, being 63 and 105~kJ for $\theta$ = 30$^\circ$ and 40$^\circ$, respectively. These energies are higher than those shown in Fig. \ref{fig:ign_fig1}(b) due to the higher stand off distance between the electron injection surface and the target centre (125 vs. 100 $\mu$m) and the higher areal density of the coronal plasma surrounding the core of the imploded target \cite{solo08}. For instance, Fig. \ref{fig:ign_fig1}(b) shows that the ignition energy for a beam with $\langle{E}\rangle$ = 2 MeV and $\theta$= 35$^\circ$ is 50 kJ while it is around 80 kJ for the target of Ref. \cite{solo08}. Assuming in this last case a laser-to-fast electron conversion efficiency of 40\%, even with the strong assumptions of no cone tip and a divergence half-angle of 35$^\circ$ only, a multi-PW laser with more than 200 kJ would be needed to ignite such a target. 

The calculations presented in Refs. \cite{solo08,honrubia1} are quite ideal because fast electrons are injected just on the coronal plasma surrounding the fuel assembly. Indeed, fast electrons have to pass through the overcritical plasma sited inside the cone, if any, and always through the cone tip before reaching the dense fuel. Johzaki et al.\cite{johzaki:062706,johzaki3} have studied the role played by the cone tip in the fast electron transport towards the dense core. They conclude that high-Z materials deteriorate substantially the quality of the electron beam even for cone tips as thin as 10 $\mu$m. This is due to the collisional drag with plasma electrons and scattering with ions, resulting in an important reduction of the energy coupling of the fast electron beam to the dense core. The use of lighter materials in the cone tip, such as CH or DLC (Diamond-Like Carbon), mitigates drag and scattering effects. These light materials produce a manageable degradation of the beam and, at the same time, can support the shock and the jet coming towards the cone at the shell collapse time. DLC is preferred to the CH for its higher density.

The effect of the cone tip material in sub-ignition targets has been quantified by Johzaki et al. within the context of the FIREX-I project \cite{johzaki:062706}. The simplified target configuration considered in this project is a CD plasma with a Gaussian density profile in radius (peak density = 200 gcm$^{-3}$, FWHM diameter = 20 $\mu$m) placed on a density pedestal of 0.2 gcm$^{-3}$. The target areal density is 0.2 gcm$^{-2}$ and the initial temperature is set to 400 eV. This imploded target configuration is heated during 5 ps by a 5~kJ, 1~MeV fast electron beam with a Gaussian radial profile of 30 $\mu$m diameter (FWHM). The beam divergence full angle is 20$^\circ$ and the electrons are injected 50 $\mu$m away from the core centre. As the electron beam size is comparable to the imploded core size and the electron range is higher than the target areal density, a hot spot is not generated and, instead, the whole core is heated. Johzaki et al. have shown a reduction of about 50\% in the core heating energy and 60\% in the peak ion temperature when a 10 $\mu$m thickness gold tip is present. This reduction is much lower, 10\% and 11\%, respectively, for lighter materials such as CH \cite{johzaki:062706}.

Another issue that can affect dramatically the fast electron beam coupling with the dense core is the existence of plasma inside the cone. As has been reported by Johzaki et al. \cite{johzaki1}, the ASE (amplified spontaneous emission) laser pre-pulse generates a long scale, low density plasma inside the cone. As a result, fast electrons are generated by the interaction of the main pulse with the pre-plasma relatively far from the cone tip and with much higher energy than without pre-plasma. This has been confirmed by PIC simulations \cite{johzaki1}, which have shown harder electron spectra and a substantial reduction of the number of electrons with energies lower than 5 MeV when a pre-plasma is present. Too hot electrons reduce substantially the electron energy deposition in the core and thus the coupling efficiency. Integrated FI experiments carried out at the ILE have evidenced this effect, observing a coupling efficiency substantially higher when the laser pre-pulse energy is reduced \cite{shiraga-gekko-ppcf-2011}
. 

An improvement of the standard field calculation model used in fast electron transport calculations has been proposed recently by Johzaki et al. \cite{johzaki:062706} and Nicolai et al. \cite{nicolai-xport-pre-2011}. They have pointed out that the generalized Ohm's law and, in particular, the pressure gradient term, which yields an azimuthal B-field proportional to $\nabla{T_e}\times \nabla{n_e}$, can play a role in transport calculations. This term is important when the directions of the gradients of electron temperature, $\nabla{T_e}$, and electron density, $\nabla{n_e}$, are not parallel, as occurs at the outer regions of the core (see Fig. 1(a)), where the electron temperature gradient is directed toward the beam axis and the density gradient is directed toward the core centre. The B-field due to  $\nabla{T_e}\times \nabla{n_e}$ has a direction opposite to the resistive collimating B-field mentioned above and its main effect is to scatter electrons away from the core. However, the growth of this B-field is relatively slow becoming important after several ps. For instance, the energy deposition in the target depicted in Fig. \ref{fig:ign_fig1}(a) heated by the beam defined by the parameters $\theta$ = 35$^\circ$ and $\langle{E}\rangle$  = 1.6 MeV is reduced by 23\% when the generalized Ohm's law is taken into account \cite{nicolai-xport-pre-2011}. A similar reduction (21\%) has been found in the calculations for the simplified FIREX-I target discussed above when the term $\nabla{T_e}\times \nabla{n_e}$ is included \cite{johzaki:062706}.

\subsection{Ignition calculations with a PIC-based electron source}
A first characterization of the fast electron source in the FI scenario via 3D PIC simulations has been reported recently \cite{strozzi-ifsa-epjwc-2012,strozzi-fastig-pop-2012,kemp1}. One of the main conclusions of this study is that the initial distribution of fast electrons can be factorized as the product of two independent functions of angle and energy. The angular distribution is super-Gaussian, $\exp\{-\left(\theta/\Delta\theta\right)^4\}$, with $\Delta\theta$ = 90$^\circ$ and a mean divergence half-angle $\langle\theta\rangle$ = 52$^\circ$. This high divergence can be explained by the curved geometry of the electron acceleration region \cite{schm12} and by the electron scattering by the oscillating magnetic field generated by the Weibel instability close to the cut-off surface \cite{adam06,debayle1}. The energy spectrum can be fitted by a quasi two-temperature profile. The first component is due to electron acceleration near the cut-off surface and has a temperature substantially lower than that given by the ponderomotive scaling $T_p$, while the second component is due to electron acceleration in the subcritical plasma and has a temperature higher than $T_p$. The overall electron mean energy is slightly higher than the ponderomotive temperature $T_p$, but only 24\% of the injected energy is carried by electrons with energies lower than $T_p$. The main differences with the electron source assumed in Section 5.2 are the much higher electron energies and divergences, and the energy independent angular spectrum. 

Strozzi et al. \cite{strozzi-ifsa-epjwc-2012,strozzi-fastig-pop-2012} have performed integrated simulations assuming the PIC-based electron source presented above. The target used is shown in Fig. 2(a), where the DT fuel has initially a super-Gaussian density distribution, $440\exp\{-\left(R/70\right)^{12}\}$ gcm$^{-3}$ where $R$ is the distance to the centre in $\mu$m, sited on a background DT plasma of 10 gcm$^{-3}$. A DLC cone is included in the simulation box. The DT mass is 0.57 mg and the initial temperature is set to 100 eV. An electron beam with quasi-uniform radial profile, $\exp\{-\ln{2}\left(r/r_{beam}\right)^8\}$ with $r_{beam}$ = 18 $\mu$m, and constant intensity in time (from 0.5 to 18.5 ps, starting from a linear ramp from 0 to 0.5 ps) impinges on the target. The transport and energy deposition of fast electrons from the cone to the core has been simulated with the hybrid-PIC code Zuma \cite{larson-zuma-dpp-2010} coupled to the radiation-hydrodynamic code Hydra \cite{marinak-hydra-pop-2001}. The Zuma code includes the generalized Ohm's law. The details are given in \cite{strozzi-ifsa-epjwc-2012,strozzi-fastig-pop-2012}. 

\begin{figure}[H]
\begin{center}
\includegraphics[width = \textwidth]{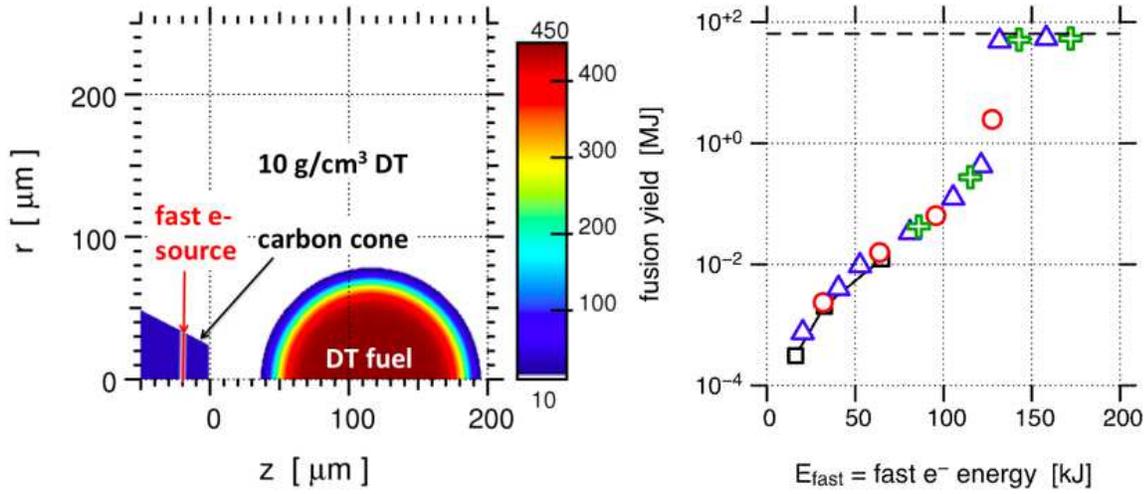}
\caption{\label{fig:ign_fig2}(a) Initial target density in gcm$^{-3}$. The red line indicates the source cylinder where fast electrons are injected. The 8 gcm$^{-3}$ (= 2.3 $\times$ solid diamond) carbon cone is coloured in blue for clarity. (b) Fusion yield vs. total injected electron energy, for Zuma-Hydra runs with an artificially collimated source $\Delta\theta$= 10$^\circ$. $r_{beam}$ = 10 $\mu$m for black squares with solid line, 14 $\mu$m for red circles, 18 $\mu$m for blue triangles and 23 $\mu$m for green crosses. The blue triangle with $E_{fast}$ = 132~kJ is the lowest value that can be deemed to be ignited \cite{strozzi-fastig-pop-2012}.}
\end{center}
\end{figure}

The fusion yield as a function of the fast electron energy for an artificially collimated electron source ($\Delta\theta$= 10$^\circ$) and for several values of $r_{beam}$ is shown in Fig. \ref{fig:ign_fig2}(b). It is worth noting that, despite the artificial source collimation, the minimum ignition energy is quite high, 132~kJ and roughly independent of the beam radius. This can be explained by taking into account that for large radii, the increase in the volume to be heated is balanced by the lower energy and penetration of fast electrons due to the lower laser intensity. Assuming the laser-to-fast electron conversion efficiency of 0.52 obtained in the PIC simulations, the 132~kJ electron beam corresponds to a laser mean intensity of about $1.4 \times 10^{21}$ Wcm$^{-2}$, which gives an electron mean energy of 8.2 MeV and a range (6.8 gcm$^{-2}$) greater than the fuel $\rho{L}$ (6 gcm$^{-2}$) and much higher than the optimal deposition range for FI (1.2 gcm$^{-2}$). This is one of the reasons why the ignition energy is so high. We can conclude that, even if a collimated electron beam could be generated, the ignition energy would be unacceptably high because fast electrons are too hot. When the fast electron divergence is included, simulations show ignition energies higher than 1 MJ \cite{strozzi-fastig-pop-2012}, making FI unfeasible, at least for the 'conventional' scheme discussed in this Section. The major limitations of the calculations presented here arise from the short time of the 3D PIC simulations ($\sim$0.36 ps), much lower than the 10 -- 20 ps pulse durations typical of FI, and from the lack of comparison with experimental data close to the true FI conditions.

\section{Concepts for Controlling Transport}
\label{control}
\subsection{Motivation}
From the preceding sections it is clear that large fast electron divergence angles can severely reduce the fast electron to hot spot coupling efficiency, thus raising the required ignitor pulse energy to levels at which FI is no longer a viable ICF scheme.  In the first instance this is just a consequence of ballistic transport.  Suppose that all other processes can be neglected, and that coupling is therefore dominated by ballistic transport with ideal stopping in the hot spot.  When the stand-off distance, $D$, is much larger than the source spot radius, $r_L$, one would therefore expect the coupling efficiency ,$\chi$, to be roughly equal to the hot spot area divided by the fast electron beam cross-section at the stand-off distance, which is,

\begin{equation}
\label{ball_couple}
\chi = \frac{r_{hs}^2}{2D^2(1-\cos\theta_{1/2})}.
\end{equation}

If $\theta_{1/2}$ is large then $\chi$ is limited to $\chi \approx r_{hs}^2/D^2$.  For $r_{hs}/D = 0.2$ this is 5\%.  Conversely, Eq. \ref{ball_couple} indicates that for $r_{hs}/D = 0.2$, one needs $\theta_{1/2} = 16.3^\circ$.  Assuming that the laser absorption physics cannot be easily engineered to produce such a low divergence angle, there is a clear need to seek additional means of controlling the transport of the fast electrons.  

In the first instance this might be `self-collimation' of the fast electron beam through the resistively generated magnetic field that should be produced around the fast electron beam (see \S \ref{basic_macro}).  The ability of this field to produce significant focussing or pinching of the fast electron beam is at least doubtful in light of experimental and simulation results that have been obtained in the last few years.

If self-collimation cannot be relied upon, then the FI scheme has to be adapted in some way so as to ensure effective transport of the fast electrons by some other means (if at all possible).  This might be possible through optical engineering of the laser pulse and a thorough understanding of the laser absorption process.  Alternatively it might be possible to exploit fast electron transport physics, and it is this that this Section is concerned with.

\subsection{Self-Collimation of the Fast Electron Beam}
In \S \ref{basic} of this review, the criteria for a fast electron beam to `self-collimate' as derived and studied in the work of Bell and Kingham.  By this, we mean the collimation of a fast electron beam even in a a homogeneous background plasma due to the resistive magnetic field that is generated due to the curl of the fast electron current density, and thus the curl of the electric field.  A simplified version of the Bell-Kingham criteria can be derived as follows.  One starts with an estimate of the magnetic field that is resistively generated,
\begin{equation}
\label{self1}
\frac{\partial{B}}{\partial{t}} \approx \frac{\eta{j_f}}{R},
\end{equation}
where $R$ is the beam radius,$j_f$ the fast electron current density,$\eta$ the resistivity, and $t$ is the time.  If we use the Spitzer resistivity ($\eta = \alpha{T^{-3/2}}$), then this can be integrated to obtain,
\begin{equation}
\label{self1b}
B \approx \frac{3n_e}{2j_fR}\left[at + T_0^{5/2}\right]^{2/5},
\end{equation}
 where $a = 2\alpha{j_f^2}/3en_e$, and $T_0$ is the initial temperature.  Next one estimates the angle ($\theta$) that a fast electron will be deflected through over the distance that the unperturbed beam takes to double its radius,

\begin{equation}
\label{self2}
\theta = \frac{eBR}{\gamma{m_e}c\tan\theta_{1/2}},
\end{equation}
where $\theta_{1/2}$ is the divergence half-angle of the beam.  If we use the fundamental Bell-Kingham criterion for collimation --- that collimation occurs when $\theta = \theta_{1/2}$ --- then we can combine Eq.\ref{self1} and \ref{self2} (and substitute power balance for $j_f$) to obtain,

\begin{equation}
\label{self3}
\theta_{1/2}\tan\theta_{1/2}=\frac{3en\bar{\varepsilon}}{\gamma{m_e}c\beta{I_L}}\left[at + T_0^{5/2}\right]^{2/5},
\end{equation}
where $\beta$ is the laser to fast electron conversion efficiency, $I_L$ is the laser intensity, and $\bar{\varepsilon}$ is the average fast electron energy.  Inserting typical numbers into Eq.\ref{self3} leads to the conclusion that self-collimation is only likely to happen for $\theta_{1/2} < $20--30$^{\circ}$.

Calculations of fast electron transport relevant to ignition scale FI indicate that this is a reasonable estimate for the regime in which self-collimation is sufficient (e.g. \cite{honrubia1};see Section \ref{ignscale}).  There is good evidence, however, that the actual fast electron divergence half-angle under ignition scale conditions will be significantly greater than 30$^{\circ}$.  Some of this evidence is experimental \cite{lancaster1,green1} coupled with theoretical and numerical interpretation \cite{honrubia1}.  Other evidence comes from large scale numerical simulations of laser absorption ,e.g. \cite{debayle1}.  With divergence half-angles that are slightly in excess of 30$^{\circ}$, magnetic field generation is still highly beneficial in terms of improving the coupling.  However the evidence suggests that the divergence half-angle could be in excess of 50$^{\circ}$, and under these conditions self-collimation does little to prevent very poor coupling to the hot spot (for typical stand-off distances).  

\subsection{Resistive Guiding of Fast Electrons}  
From the induction equation in the hybrid approximation,
\begin{equation}
\frac{\partial{\bf B}}{\partial{t}} = \eta\nabla \times {\bf j}_f + \nabla\eta \times {\bf j}_f,
\end{equation}
one can see that just as there is a term which indicates that magnetic field grows to drive fast electrons into regions of higher fast electron current density (first term), there is also a term that indicates that magnetic field grows to drive fast electrons into regions of higher resistivity (second term) \footnote{This is Davies' qualitative interpretation of this equation \cite{jrd1}.}.  The $\eta\nabla \times {\bf j}_f$ term is the effective responsible for the `self-collimation' described in the preceding section.

Resistive guiding exploits the second term, the $\nabla{\eta} \times {\bf j_f}$ term \cite{robinson-switchyard-pop-2007}.  At sufficiently high temperatures, all materials will follow a Spitzer-like resistivity in which $\eta \propto Z$.  Therefore if one structures a target by using two materials with different $Z$, in principle the higher $Z$ material should confine and guide the fast electrons as magnetic fields are generated at the interace between the two materials where the $\nabla{\eta} \times {\bf j_f}$ term will be large.

It is relatively straightforward to see that the collimating fields generated by resistivity gradients must be at least as strong as those generated by the $\eta\nabla \times {\bf j}_f$ term.  If we denote the $\eta\nabla \times {\bf j}_f$ term by $\dot{B}_1$, and $\nabla{\eta} \times {\bf j_f}$ by $\dot{B}_2$, then we can see that the magnitudes are approximately,
\begin{equation}
\dot{B}_1 \approx \frac{\eta{j_f}}{R_b},
\end{equation}

\begin{equation}
\dot{B}_2 \approx \frac{\eta{j_f}}{L_{int}},
\end{equation}
where $R_b$ is the radius of the fast electron beam, and $L_{int}$ is the scale-length associated with the transition in resistivity.  As it is possible to envisage transitions in resisitivity with scale-lengths, $L_{int} \ll R_b$, due to the sharp interfaces produced by target engineering, there will be range of circumstances in which resistivity gradients are capable of producing powerful confining magnetic fields.
 
\subsection{Preliminary Studies of Resistive Guiding}
\begin{figure}[H]
\begin{center}
\includegraphics[width = \textwidth]{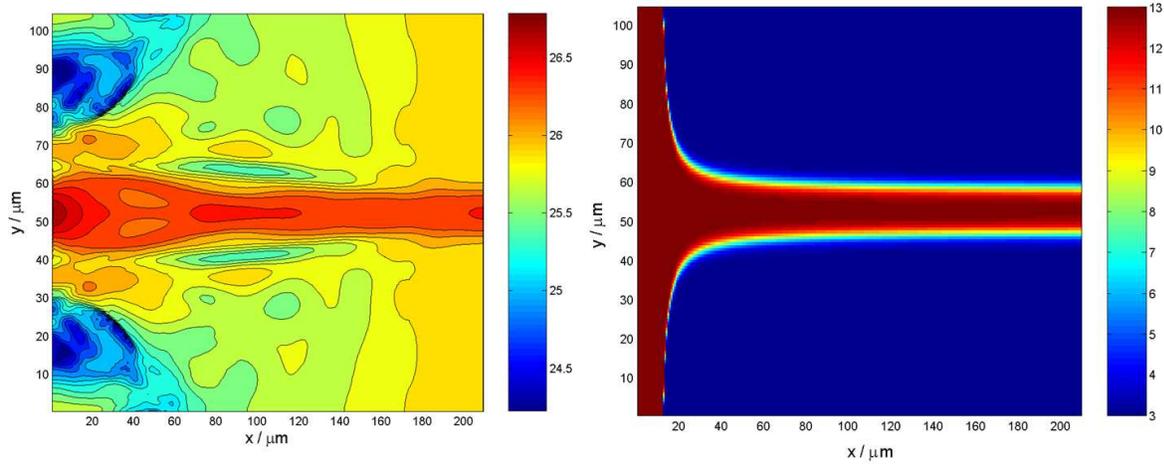}
\caption{\label{fig:con_fig6}(Left) Plot of target $Z$, and (Right) Plot of fast electron density (log$_{10}$) after several hundred fs in an early simulation carried out using the 2D hybrid-VFP code, \sc{LEDA}. }
\end{center}
\end{figure}

In the work of Robinson and Sherlock \cite{robinson-switchyard-pop-2007} this concept was explored using a `hybrid-VFP' code and very good guiding was demonstrated.  This investigation concentrated on the conditions close to those of laboratory experiments involved solid density foils and PW class lasers that deliver a few hundred Joules in about a picosecond.  The possibility that `cold target' effects might cause a problem for the concept was also investigated and it was shown that a small temperature range over which the resisitivities are inverted can be tolerated.  One might expect that once collimation has been occurred, that the $\eta\nabla \times {\bf j}_f$ will then act to reinforce this collimation and sustain it, even if the $\nabla{\eta} \times {\bf j_f}$ greatly diminishes.  Robinson and Sherlock showed some evidence for this sort of positive feedback in this concept.
These initial simulations were done in 2D Cartesian coordinates, and only considered a guiding structure that was perfectly aligned with the axis along which the fast electrons were injected.  In Fig. \ref{fig:con_fig6} results from one of these early simulations are shown, including the material composition of the target and the fast electron density after several hundred fs.

Some promising results were obtained in early experiments carried out using PW class lasers and `cold' solid-density targets \cite{kar1,ramak1}, and more simulation work was carried out using a 3D particle hybrid code to help interpret these experiments.  This combination of numerical studies and experiments gives one good reason to believe that resistive guiding is a genuinely realizable effect that {\it might} be exploited in FI.  Subsequent work in this area has therefore focussed on two major concerns:(i) Can resistive guiding work under the conditions that FI imposes on it?, and (ii) How precisely can we exploit resistive guiding in a realistic FI scenario?

\subsection{Resistive Guiding at High Energy Scales}
\label{resguiding_hienergy}
Applicability of the fast-electron collimation scheme exploiting resistivity gradients to fast ignition depends on two factors: (i) The collimating magnetic fields need to persist during the entire ignition pulse; (ii) The high resistivity path must survive the compression resulting from the implosion. In an ignition pulse, the material along the path of electron propagation is heated to keV temperatures. At such high temperatures, the resistivity of such a material can become less than the resistivity of the surrounding plasma and the resistivity gradients can be inverted. The question is whether or not the inverted resistivity gradients can cause a magnetic field reversal from collimating to de-collimating. It can also be difficult to maintain the guiding structure extending to the dense fuel up to the time of significant compression of the fast-ignition target. This problem should be addressed either by developing target designs for which the damaging effect of the implosion is minimized or placing the guiding structure inside a protective cone as was suggested in \cite{robinson2}.

Collimation of high-energy electron beams in the wire-like structures has been studied \cite{solodov5}. Simulations using the hybrid-PIC code LSP were performed for a 40 $\mu$m diameter copper wire embedded in aluminum. A 10~ps (constant in time), 2~MeV mean-energy, relativistic-Maxwellian electron beam with divergence half-angle of 67$^{\circ}$, and total energy of 20~kJ was injected into the wire. The beam was found to be effectively collimated for the whole duration of the electron pulse. About 65\% of the injected electrons were collimated on the length of the collimating structure of 150~$\mu$m. The resistivity gradient at the wire boundary was found to be inverted because of the wire heating by fast electrons in less than 0.5~ps after the beginning of the electron pulse. The collimating magnetic field, however, persisted because the magnetic field had two components: one generated by the resistivity gradients and the other by the return current gradients. Initially, the collimating magnetic field was generated by the resistivity gradients. The resulting collimation caused large current density gradients that enhanced the collimating field. The current density gradient offset the effect of the reversal of the resistivity gradient thus supporting a large saturated collimating magnetic field. Similar conclusions were obtained for lower-energy electron beams in \cite{ramak1}.

\begin{figure}[H]
\begin{center}
\includegraphics[scale=1]{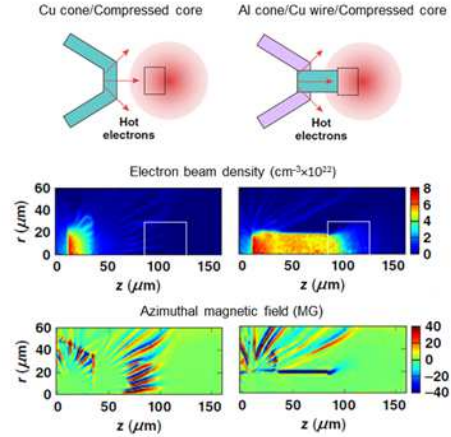}
\caption{\label{fig:con_fig1}LSP simulations \cite{solodov5} predicting significant increase in energy coupling to the compressed core by the self-generated magnetic field at the copper wire interface.}
\end{center}
\end{figure}

LSP simulations \cite{solodov5,solodov6} showed that high-energy electron beams can be guided by a mid-Z wire through the cone tip and coronal plasma of a fast-ignition target, subject to survivability of the wire during the implosion. The simulations utilized idealized cone-fuel configurations with and without a wire (Fig. \ref{fig:con_fig1}). A 75~$\mu$m long, 40~$\mu$m diameter copper wire goes through the 25~$\mu$m thick tip of aluminum cone towards the pre-compressed deuterium fuel core. The core has a super-Gaussian density distribution $400\exp(-(r/50)^4)$ gcm$^{-3}$ where $r$ is the distance from the center in $\mu$m, sited on a background deuterium plasma of 10 ~gcm$^{-3}$. The initial temperature of 100~eV was assumed, ionization and radiative cooling were modeled for copper and aluminum. The simulations used a 40~kJ, 10~ps, 1.6~MeV-mean-energy relativistic Maxwellian electron beam with initial divergence half-angle of 55$^\circ$($\propto \exp(-(\theta/\theta_0)^2)$, with $\theta_0 =$67$^\circ$), constant temporal profile, and a super-Gaussian radial profile $400exp(-(r_\perp/20)^4)$, where $r_\perp$ is the distance from the beam axis in $\mu$m, injected at the inner side of the cone tip. Comparison simulations were performed in which the cone and the wire were replaced by a single copper cone. Fast-electron energy deposition in the cylindrical region (see Fig. \ref{fig:con_fig1}) with a diameter of 60~$\mu$m and a length of 40~$\mu$m (so-called “ignition region”) was calculated. The simulations confirm that the fast-electron coupling to the core is significantly improved with the wire: 45\% coupling efficiency to the “ignition region” in the cone-wire case versus 7\% without a wire. It can be seen in Fig. \ref{fig:con_fig1} that fast electrons are effectively collimated and guided by the self-generated resistive magnetic fields at the interface of the copper wire and surrounding lower-Z plasma. Collimation of electrons to the dense fuel in a wire-like structure has been also confirmed by hybrid simulations of J. Honrubia and D. Larson using codes PETRA and ZUMA \cite{solodov6}.

The question of wire survivability was not addressed but it was noted that it may be difficult to maintain a clean high-resistivity path to the dense core at the time of peak compression in an imploded capsule. Detailed target design studies using radiation-hydrodynamics codes are required to show if such a divergence-mitigating structure can be assembled in an actual implosion.

\subsection{Advanced Uses of Resistive Guiding}
The use of resistive guiding in FI may not necessarily be restricted to the cone-wire scheme discussed above.  Other schemes have been suggested that do not require placing an element outside of the cone.  Instead, such schemes suggest putting guiding structures in an insert in the cone tip.  This may have the advantage of being more robust with respect to the hydrodynamics of fuel assembly, although detailed radiation-hydrodynamics studies are required to confirm this.

One, suggested by Schmitz \cite{schm12}, is to use a curved axisymmetric interface (ellipsoidal or paraboloidal) to produce an azimuthal magnetic field structure that acts as a curved mirror.  A divergence fast electron spray will then have its angular spread reduced by the approximately specular deflection in the strong magnetic field which is localized at this interface, in a way that is analogous to a parabolic or elliptical mirror in ray optics.  This comes at the expense (as in the case of an optical parabolic mirror) of increasing the radial extent of the beam.  As the hot spot radius and laser spot radius are likely to be comparable in size, there will be limits on how much the radial extent of the beam can be increased to reduce angular spread (the diameter of the cone apex may also have to be limited for fuel assembly).  This was not assessed in Schmitz's original proposal due to limitations on the size of the simulation domain.  Some preliminary simulation results presented by Robinson suggest that some substantial benefits might still be obtained from the elliptical mirror approach \cite{robinson3} where a simple elliptical configuration yielded about an improvement in coupling into a target hot-spot region of about 2--3 (over an unguided case).  

Another suggestion was made by Robinson and co-workers \cite{robinson2} who suggested a `magnetic switchyard' configuration: a series of concentric quasi-cylindrical guide shells immersed in a less resistive substrate.  Strong azimuthal fields grow at the interfaces, confining fast electrons into the guide shells.  Each guide shell will receive a portion of the fast electron population with limited angular spread about some mean angle.  The guide shells then curve around in an arc which re-directs this mean angle to some distant region.  As with the aforementioned mirror concept, this means that the switchyard must increase its radial extent beyond that of the source radius.  The numerical simulations presented by Robinson showed that, at least for some configurations, an improvement in coupling of about 2--3 (also compared to an unguided case).     

To the same goal, Debayle et al. \cite{debayle_xtra1} recently proposed a structured target made of narrow high- and low-$Z$ elements of density decreasing in the axial direction. The magnetic modulations developing at the filament interfaces then decay away from the surface, leading to non-specular reflections of the fast electrons trapped inside the high-$Z$ filaments. As a result, their local angular dispersion steadily decreases along their path. The capability of these targets to both guide the fast electrons and reduce their angular dispersion is, however, obtained at the cost of heavy constraints on target manufacturing.

Finally, note that transverse resistivity gradients associated to density or temperature non-uniformities in the corona also have the potential to beneficially affect the fast electron transport. This was demonstrated both experimentally and numerically in Ref. \cite{perez1} in the case of cylindrically-compressed foam targets. Insofar as they are injected shortly before the shock convergence, the fast electrons can be efficiently guided by the magnetic field growing at the shock front.  On the other hand, non-uniformities can help drive resistive filamentation \cite{robinson4}, as well as the de-collimating effects noted in Section \ref{ignscale}, so the exploitation of hydrodynamically induced non-uniformities requires careful examination.

\subsection{Double Cone Approach}
\label{doublecone}
In this concept, a vacuum gap is introduced in order to prevent those fast electrons travelling at large angles from escaping.  The vacuum gap is introduced by employing a target using two concentric cones rather than one, hence `Double Cone'. Due to the vacuum gap, the cone wall is isolated from the coronal plasma and the fast electrons are confined and guided to the tip by electrostatic and quasi-static magnetic fields formed in the vacuum gap region. 

The fast electron guiding using vacuum gap has been first proposed by Campbell, et al. \cite{campbell1} , where a collimation of high energy electrons using planner plag/gap/foil structure was numerically shown and an idea to control of fast electron using a conical plag/gap/foil structure (see \cite{campbell1}) was proposed for fast ignition application. This scheme can be applied for the beam guiding from cone tip to the core.
\begin{figure}[H]
\begin{center}
\includegraphics[width = \textwidth]{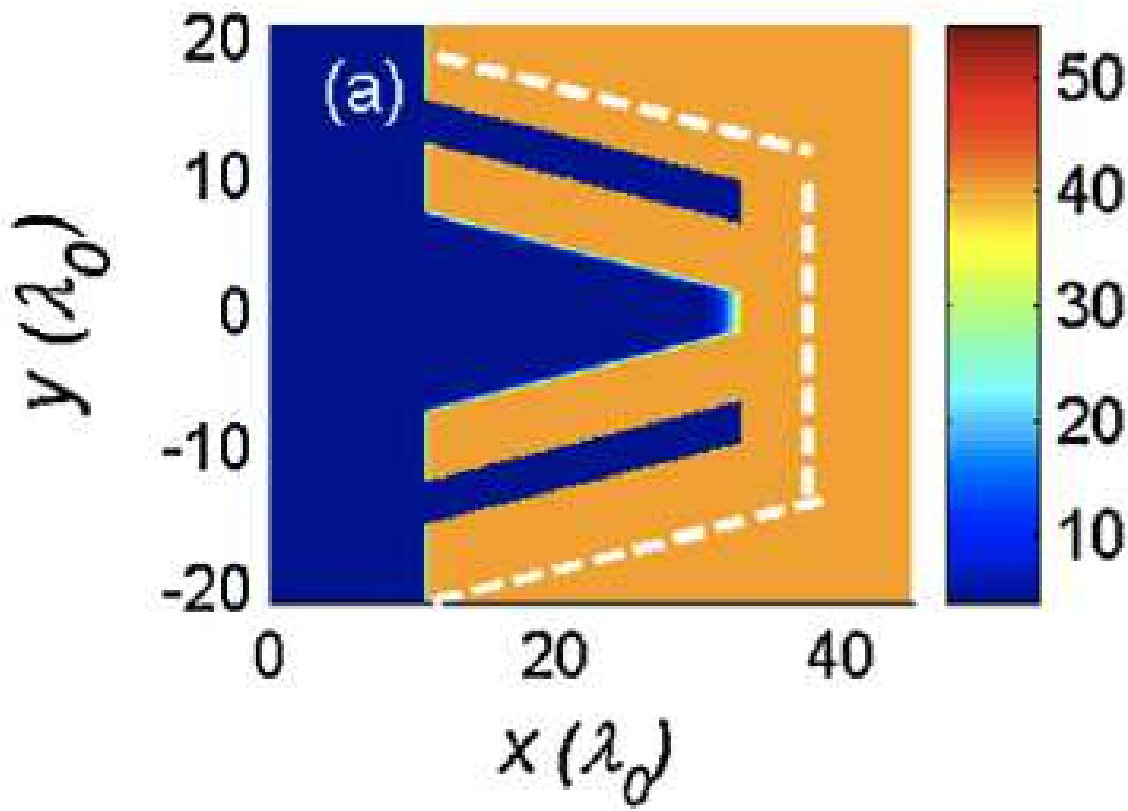}
\caption{\label{fig:con_fig5}  The double cone target employed in simulations by Cai \cite{cai1}.}
\end{center}
\end{figure}
Nakamura et al. \cite{nakamura1} have suggested a double cone target (Fig.\ref{fig:con_fig5}), where the vacuum gap is introduced into the side wall of cone target. On the basis of two-dimensional (2D) Particle-in-cell (PIC) simulation, they showed that the double-cone confines the electrons for hundreds of femtoseconds (fs) by the sheath electric field generated inside the vacuum-gap. However, the simulation time was limited to a few hundreds fs. Contrary to this, the core heating duration in practice is 10ps order. So the reduction of sheath electric field inside the gap due to the plasma expansion from the cone wall in the early phase of core heating and then the failure of confinement was feared. Later, Cai et al. \cite{cai1}, carried out ps order 2D PIC simulations and demonstrated that the double cone is still effective in confining the high-energy electrons even for the gap width of a few microns (Fig. \ref{fig:con_fig2}). 

\begin{figure}[H]
\begin{center}
\includegraphics[width = \textwidth]{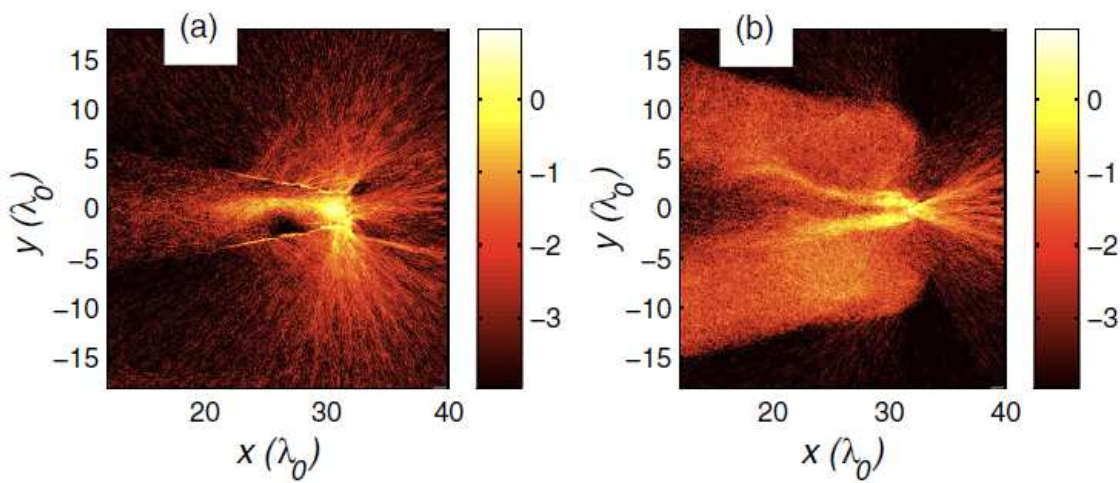}
\caption{\label{fig:con_fig2} Natural Logarithm of electron energy density for (a) single cone, and (b) double cone at 1~ps in simulations performed in \cite{cai1}}
\end{center}
\end{figure}
\begin{figure}[H]
\begin{center}
\includegraphics[width = \textwidth]{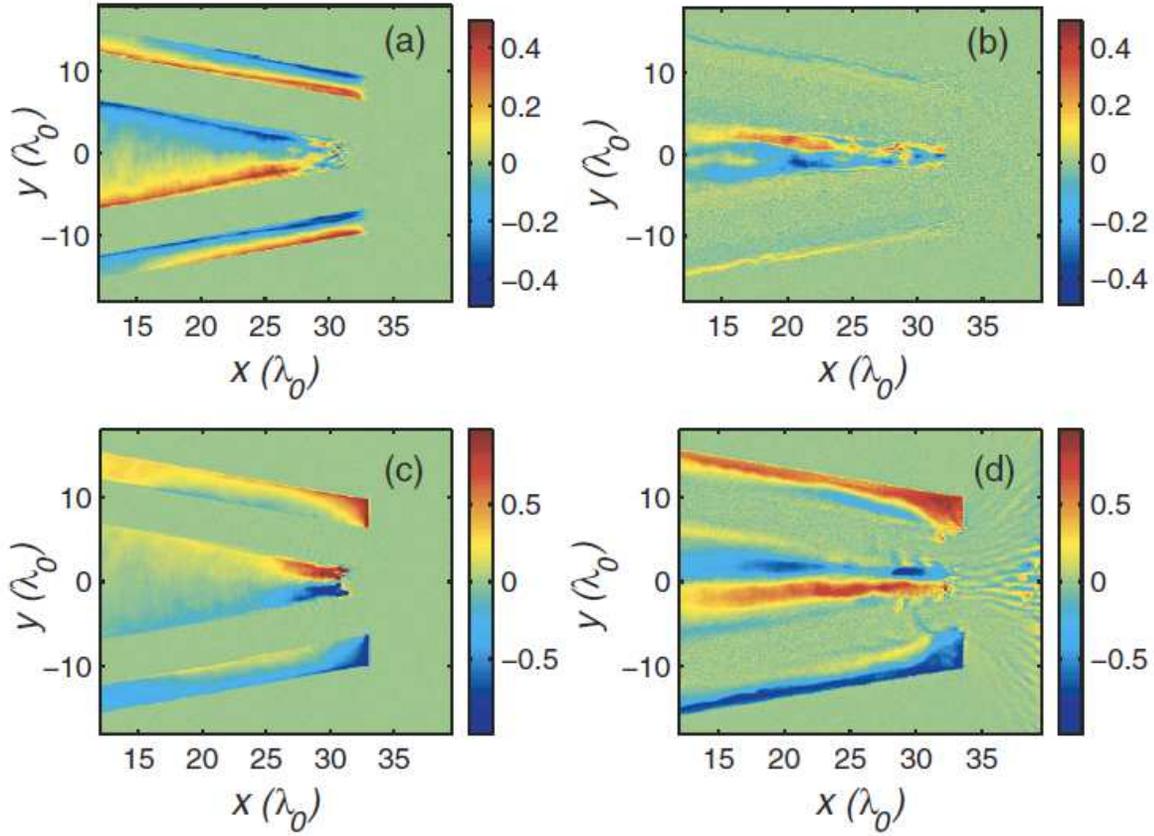}
\caption{\label{fig:con_fig3} PIC Simulations results from \cite{cai1} showing : (a,b) time-averaged sheath electric fields at 330~fs and 1500~fs, and (c,d) time-averaged magnetic fields at 330~fs and 1500~fs.  Fields are in normalized units, i.e. $m_e\omega_Lc/e$.}
\end{center}
\end{figure}

After reduction of sheath electric field due to plasma expansion, the quasi-static magnetic field works to confine the fast electrons (Fig.\ref{fig:con_fig3}). The quasi-static magnetic field has been generated due to a localized supply of high-energy electrons, originally produced at the inner-cone and the cone tip. This electron current coming from the cone tip produces a positive current inside the gap, while an opposite-directional surface-current is generated along the inner-surface of the outer-cone. The collaboration of these two currents generate a large quasi-static magnetic field inside the gap. These quasi-static fields continue to confine the high-energy electrons for longer than a few picoseconds. They showed that the double cone can reduce the beam energy loss from the side wall down to 1/3 of that for the single cone case.

\begin{figure}[H]
\begin{center}
\includegraphics[width = \textwidth]{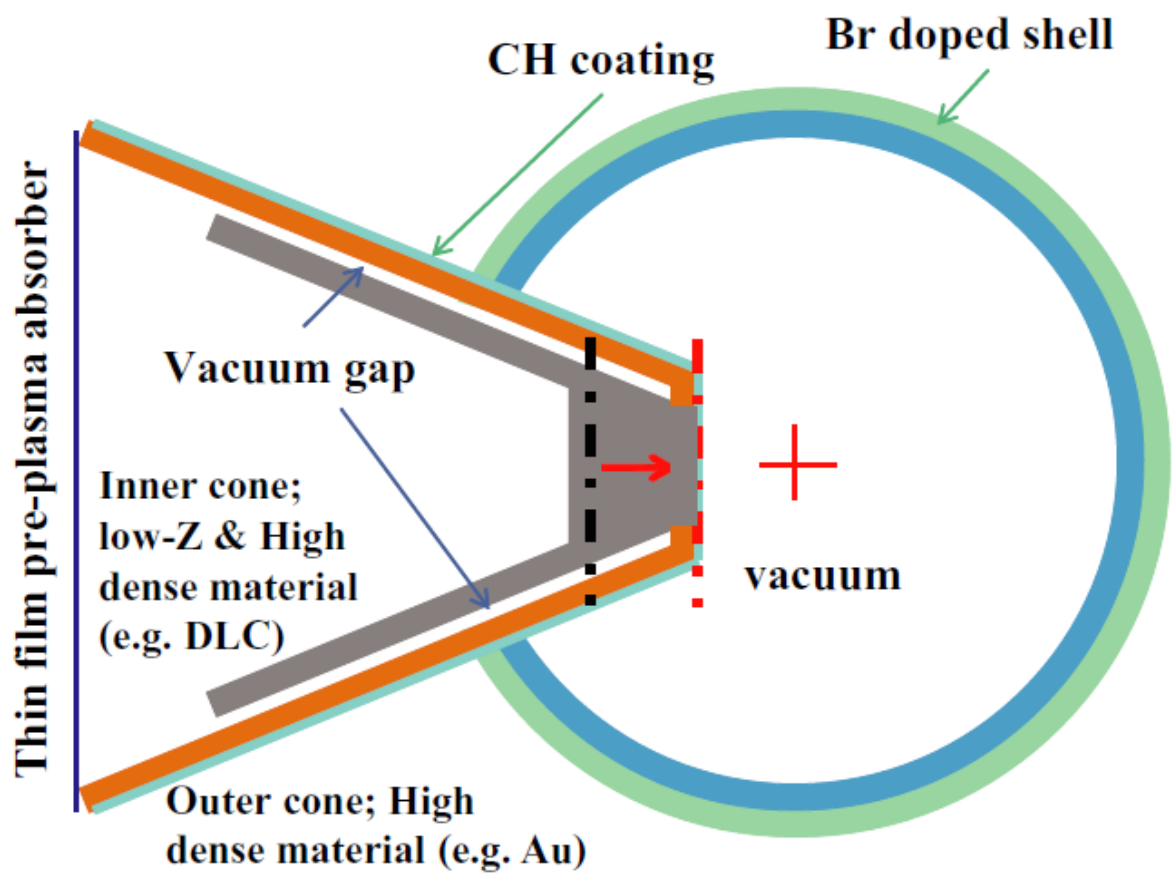}
\caption{\label{fig:con_fig4}Schematic view of the extended double cone proposed in \cite{johzaki1}.}
\end{center}
\end{figure}

Even if the double cone is used, the fast electron beam after penetrating the cone tip spatially diverges during propagation to the core due to its large divergence angle and then the enhancement of core heating rate may not be expected so much. In order to guide the fast electron beam close to the core, Johzaki, et al. \cite{johzaki1} have proposed to extend the cone tip and vacuum gap (Fig.\ref{fig:con_fig4}). They called it the ‘extended double cone’, which is a combination of the conical plag/gap/foil structure \cite{campbell1} and the double cone \cite{nakamura1,cai1}. In this case, the fast electrons travel a long distance in the extended tip region, so that low-Z, but relatively dense material (e.g. DLC) was proposed as the tip material to reduce the collisional effects \cite{johzaki:062706,johzaki3} . It was shown from the 2D PIC and FP simulations for core heating that the extended double cone with a short inner wall enhances the core heating rate more than four times when compared with the single wall cone case (see fig.\ref{fig:con_fig4} and table \ref{concepts_tbl1}).  

The preliminary experiments have been conducted by Sakawa et al. \cite{sakawa1} to prove the vacuum-shielding effect using an Al-Cu double-foil targets with and without a vacuum gap. The enhancement of the number of electrons detected in the target surface direction has been observed for the case with the vacuum gap compared to the case without vacuum gap, which demonstrates the fast electron confinement by the vacuum gap.

For more realistic evaluations of the possibilities for the extended double cone, hydrodynamic modeling of the cone’s ignition-time structural distortion must be included in the simulations. Also, the integrated experiments are indispensable to prove the guiding performance.

\begin{table}
\resizebox{\textwidth}{!}{
\begin{tabular}{|c|c|c|c|c|c|c|}
\hline
Cone Type & Pre-Plasma & $\eta_{\mbox{L}\rightarrow \mbox{fe}}$(\%) & $\eta_{L\rightarrow \mbox{fe} < 10 \mbox{MeV}}$(\%) & $\eta_{\mbox{fe}\rightarrow \mbox{core}}$(\%)  & $\eta_{\mbox{L}\rightarrow\mbox{core}}$(\%) & $\langle{T_i}\rangle_{DD}$ (keV) \\
\hline
Single Cone & None & 18 (48) & 15 (39) & 16 & 7.5 & 0.75 \\
\hline
Single Cone &  $\lambda_p =$10$\mu$m & 14 (36) & 4 (11) & 4.7 & 1.7 & 0.35 \\
\hline
Ex. Dbl. Cone & None & 31 & 27 & 62 & 19 & 1.27 \\
\hline
Ex. Dbl. Cone & $\lambda_p =$10$\mu$m & 20 & 11 & 28 & 5.5 & 0.7 \\
\hline
E.D.C. with short inner wall   & none & 41 & 35 & 79 & 32 & 1.58 \\
\hline 
E.D.C. with short inner wall  & $\lambda_p =$10$\mu$m & 23 & 14 & 36 & 8.1 & 1.01 \\
\hline
\end{tabular}
}
\caption{\label{concepts_tbl1}Summary of heating results from Extended Double Cone calculations \cite{johzaki1}. }
\end{table}

\subsection{Axial Magnetic Field Approach}
\label{axial}

There is another approach to divergence mitigation which employs
an imposed magnetic field, rather than a field self-generated by the fast electron current.  There are two main methods which have been  suggested for imposing the needed multi-MG fields: flux
compression in the fuel assembly implosion \cite{shay-fastig-pop-2012,tabak-dpp-2010}, and laser-driven coils \cite{daido-bfield-prl-1986,fujioka-gekko-ppcf-2012,fujioka_2013}.  A body of simulation work has been carried out at LLNL on the assembly of such fields, and the characterization of their advantages for electron transport, which we shall review here.

The purpose of an imposed field is to spatially confine the fast electrons to
small radius (perpendicular to the axial direction), and enhance their flux on the fuel. We distinguish confinement, or limiting the fast electrons
from spreading in space, from collimation, or reducing their velocity-space
divergence.  A confining magnetic field generally will not
collimate, so fast electrons emerge from a
confining magnetic field with their original divergence. An estimate of the product of field strength times path length needed to confine a fast electron of velocity $\mathbf{v}$ is given by \cite{robinson-switchyard-pop-2007}
\begin{equation}
  \label{eq:1}
  BL > K\gamma\beta(1-\cos\theta) \qquad K\equiv {m_ec\over e}=17.0\ \mathrm{MG}\cdot\mu\mathrm{m}.
\end{equation}
$\beta=|\mathbf{v}|/c$, $\gamma=(1-\beta^2)^{-1/2}$, and $\theta$ is the angle between $\mathbf{v}$ and the $z$ axis. For
instance, a 3 MeV electron with $\theta=45^\circ$ requires $BL>33.9$
MG$\cdot\mu$m to be confined. Keeping the field thickness smaller than the
source spot size imposes $L \le 10\ \mu$m, or $B \ge 3$ MG.  

Flux compression \cite{velikovich-flux-dpp-2012} exploits the frozen-in law of MHD, which states
that the magnetic flux $\propto \mathbf{B}\cdot d\mathbf{a}$ enclosed by a good conductor
of area $a$ remains constant.  As $a$ decreases the field strength
rises. A ``good'' conductor is one for which the resistive diffusion time $\sim \mu_0\sigma L^2$ of the
magnetic field is much longer than the implosion time ($\sigma$ is the conductivity and $L$ a field length scale). Implosions at the Omega
laser have compressed axial seed fields of $\le$0.1 MG to 20-40 MG in cylindrical
\cite{knauer-bfield-pop-2010} and spherical \cite{chang-sphere-prl-2011,hohenberger-b-omega-pop-2012} geometry.

\begin{figure}
  \centering
  \includegraphics[height=6cm]{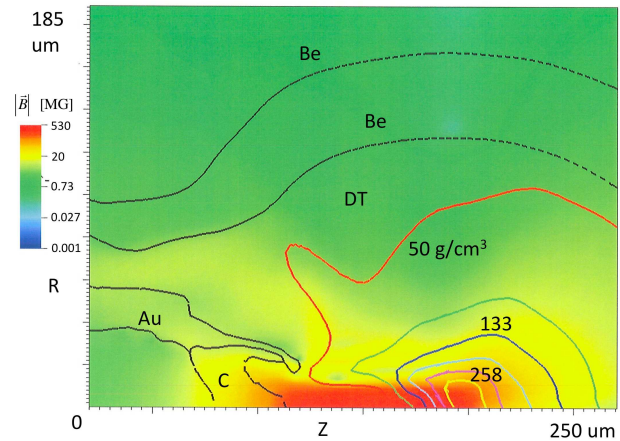}
  \caption{Magnetic field from a Hydra simulation with an initial axial field of
    0.1 MG, performed by H.\ Shay and detailed in the text. The solid or dashed
    black contours are material interfaces, the colored contours are density
    (the red, bark blue, and yellow values are indicated), and the solid
    background color is $|\mathbf{B}|$ on a log scale. Text labels indicate material.}
  \label{fig:shay}
\end{figure}

Preliminary fast ignition implosion simulations with an initial seed magnetic field have been
performed, using the MHD capabilities of Hydra \cite{shay-fastig-pop-2012} and Lasnex \cite{tabak-dpp-2010}. These have considered radiation-driven implosions around a re-entrant
cone. Figure \ref{fig:shay} shows the magnetic field in a Hydra
simulation done by H.\ Shay, starting with a uniform, axial field of 0.1 MG. It
is similar to the implosions presented in Ref.\ \cite{shay-fastig-pop-2012}, and
entails a radiation-driven beryllium ablator, a DT ice layer, and a carbon-tipped
gold cone. The plot is taken at the time of peak fuel compression, and shows
a field of 500 MG in the compressed fuel. However, the field does not diffuse far
into the cone or its interior, where the field in the critical-density
plasma is $\le 20$ MG. The fast electrons will therefore be generated by short-pulse LPI in a
region of relatively low field. This poses two challenges. First, electrons which encounter a field that
increases along field lines (e.g., an axial field that increases in the axial
direction) are subject to magnetic mirroring and reflection. In addition, there
is a standoff distance before the fast electrons reach a field strong enough to
confine them.

\begin{figure}
  \centering
  \includegraphics[height=6cm]{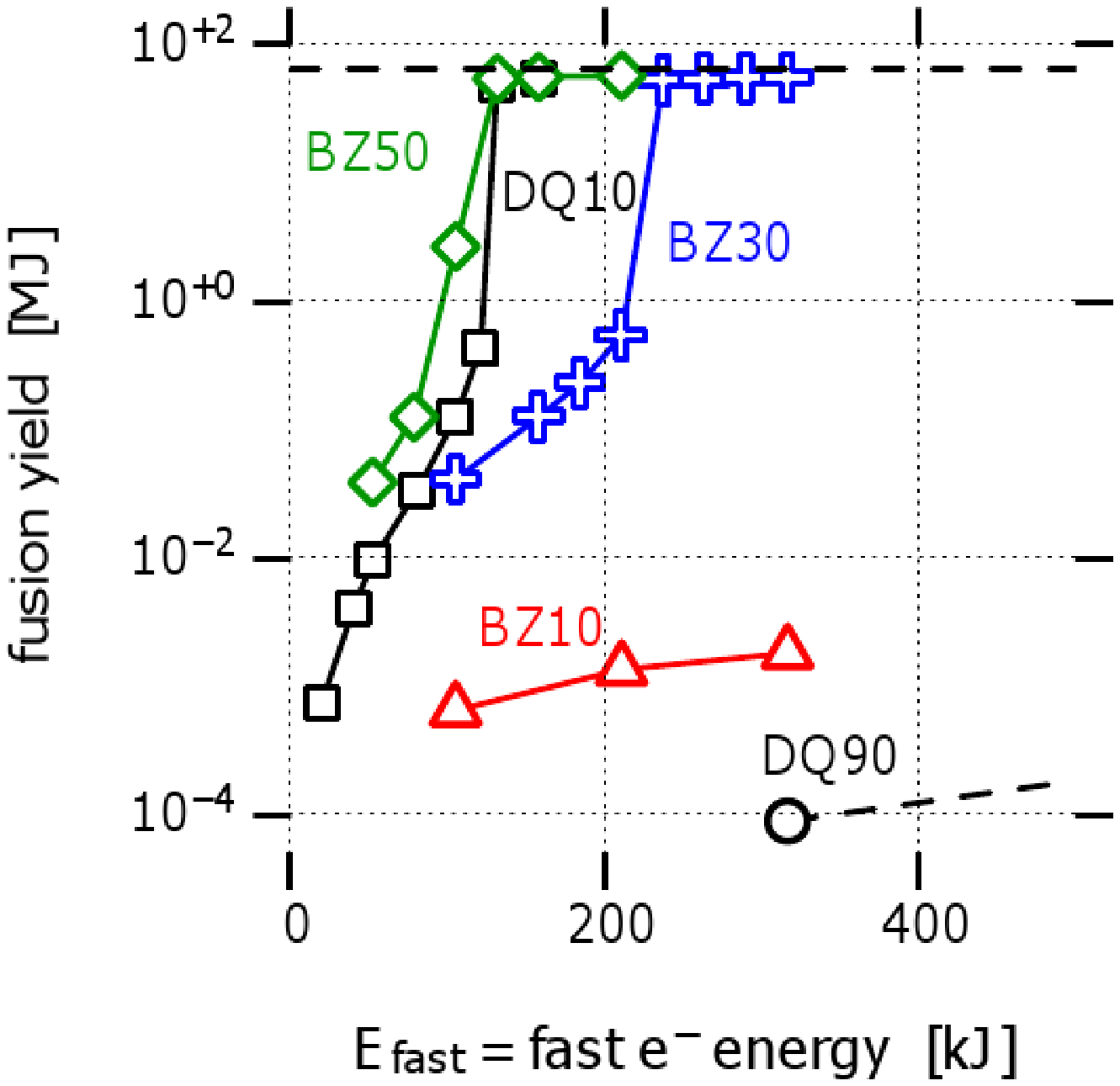}
  \includegraphics[height=6cm]{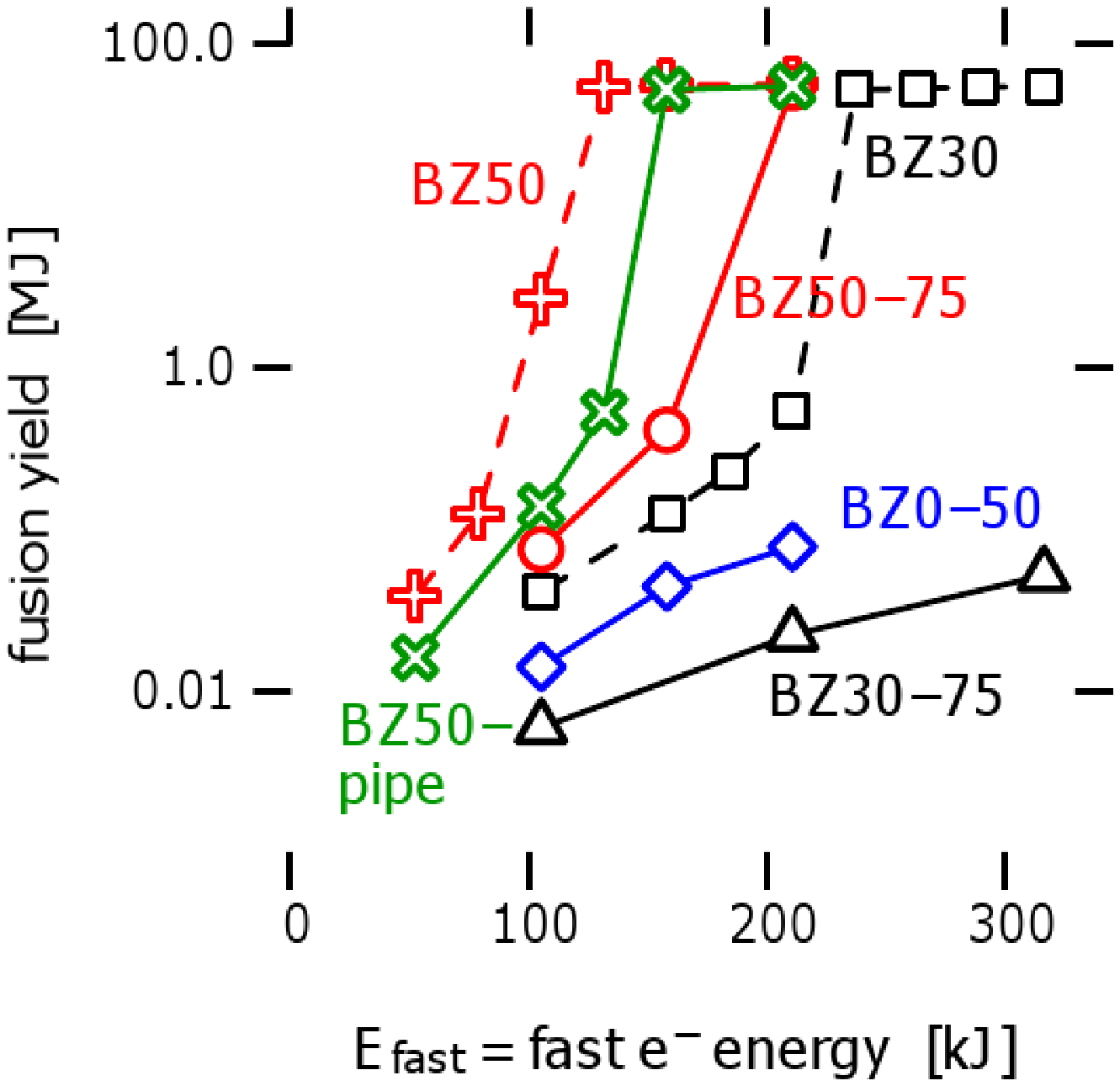}
  \caption{Reprinted from \cite{strozzi-fastig-pop-2012}: fusion yield vs.\
    added fast electron energy for Zuma-Hydra runs with PIC-based divergence
    $\Delta\theta=90^\circ$, except for one with an artificially collimated
    source $\Delta\theta=10^\circ$ and no initial B field (labeled DQ10). Other cases had
    no initial $B_z$ (DQ90), initial uniform $B_z=$ 10, 30, 50 MG (BZ10, BZ30, BZ50), non-uniform initial $B_z$ rising gradually
    from 30 to 75 MG (BZ30-75), gradually from 50 to 75 MG (BZ50-75), rapidly
    from 0.1 to 50 MG (BZ0-50), and the magnetic pipe with 50 MG peak field (BZ50-pipe).}
 \label{fig:strozzi-pop-f1213}
\end{figure}

\begin{figure}
  \centering
  \includegraphics[height=6cm]{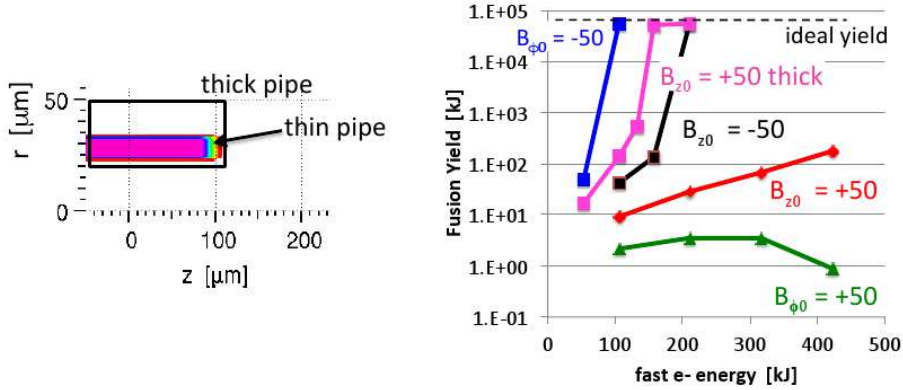}  
  \caption{Left: Initial magnetic field envelope $|B_z|$ or $|B_\phi|$ for integrated Zuma-Hydra runs with thick and
  thin ``pipe''.  Right: fusion yield vs. injected fast electron energy for
  thick pipe with $B_{z0}>0$ (taken from Ref.\ \cite{strozzi-fastig-pop-2012})
  and thin pipes with different $B_0$ signs and directions.}
  \label{fig:pipezh}
\end{figure}

Imposed multi-MG axial fields have been studied using the coupled hybrid-PIC
code Zuma and the rad-hydro code Hydra in Ref.\
\cite{strozzi-fastig-pop-2012} (see Section \ref{ignscale}, for previous description).  This work considered an idealized, spherical DT fuel
assembly with a carbon cone, and simple initial field profiles. Fusion yields
are presented in Fig.\ \ref{fig:strozzi-pop-f1213}. The electron source had a
substantial divergence based on full-PIC LPI simulations: $dN/d\Omega \propto
\exp[-(\theta/\Delta\theta)^4]$ with $\Delta\theta=90^\circ$, giving $\left<
  \theta \right>=52^\circ$; $\Omega$ is the velocity-space solid angle element. A 50 MG uniform, initial axial field performed slightly
better than an artificially collimated source with $\Delta\theta=10^\circ$ (or
$\left< \theta\right>=7^\circ$) with no imposed field. Both cases required
$\sim130$ kJ of fast electrons to ignite. A field of 30 MG needed almost twice
as much energy to ignite, while a 10 MG field performed significantly worse
(although still much better than with no initial field). This work explored the
degradation due to mirroring in non-uniform field profiles, and presented the
hollow magnetic pipe as one way to avoid mirroring. The effects on yield are presented in Fig.\ \ref{fig:strozzi-pop-f1213}. The non-uniform field cases labeled  BZ30-75, BZ50-75, and BZ0-50 demonstrate the reduced coupling due to mirroring. The magnetic pipe, case BZ50-pipe, couples almost as well as the uniform field case BZ50. Figure \ref{fig:pipezh}
shows the pipe field envelope; the thick pipe (solid black outline) is used in
Fig.\ \ref{fig:strozzi-pop-f1213}. More work on assembling pipe field structures
in fast-ignition implosions is needed to make this scheme viable.

The role of different terms in Ohm's law has been studied in several recent
works, such as \cite{johzaki:062706,nicolai-xport-pre-2011,strozzi-ifsa-epjwc-2012,strozzi-fastig-pop-2012} (models based on Ohm's law are detailed in Section \ref{hybridcodes} of the present review). All of them have shown
that the fast electron coupling to a spherical fuel region is degraded with an
extensive Ohm's law that includes terms beyond the resistive $\eta\mathbf{J}_e$,
especially $\nabla p_e$.  This produces an azimuthal $\nabla n_e \times \nabla T_e$ magnetic field at the fuel interface that pushes fast electrons to large radius and away from the hot spot.
 
Imposed magnetic pipes of differing orientation illustrate the potential benefits of axial ($B_z$) and azimuthal $(B_\phi)$ fields. Both the field direction ($z$ vs.\ $\phi$) and sign significantly affect its confinement properties, as do non-resistive terms in Ohm's law.  Electrons are confined in radius by the radial component of the $\mathbf{v}\times\mathbf{B}$ Lorentz force, which is independent of $v_r$ and $B_r$. For a simplified discussion we imagine $r$ and $\phi$ to be Cartesian, and neglect the change in $v_r$ and $v_\phi$ due to free motion. This is valid for sufficiently small Larmor radius. The expected confinement based on electron
orbits is as follows. Each sign of $B_z$ should confine electrons with one sign
of $v_\phi$ to a small radial excursion, and the other sign of $v_\phi$ to a larger one. Since the
electron source is expected to be uniform in $v_\phi$, we expect
comparable confinement from either sign of $B_z$. For  $B_\phi<0$,
however, both signs of $v_\phi$ are well confined, while both are poorly
confined for $B_\phi>0$. We therefore expect $B_\phi<0$ to confine the best,
$B_\phi>0$ to confine the worst, and both signs of $B_z$ to be intermediate and
comparable to each other.

Zuma-Hydra simulations with the same profile of initial $|\mathbf{B}|$ have
been performed for four cases all with peak magnitudes of 50 MG: $B_z>0$, $B_z<0$, $B_\phi>0$, and $B_\phi<0$. For the $B_z$ cases the magnetic field is found from a vector potential $A_\phi$, so
that $B_r$ is included to satisfy $\nabla\cdot\mathbf{B}=0$; no such $B_z$ or
$B_r$ is needed for $B_\phi(r,z)$. These calculations show a different ordering of confinement quality than the simple orbit discussion. Figure
\ref{fig:pipezh} plots the fusion yield for various initial fields. As expected from
orbits, $B_\phi<0$ performs
the best, and $B_\phi>0$ the worst.  Although both $B_z$ have intermediate
performance, $B_z<0$ confines better than $B_z>0$.  The results presented in
Ref.\ \cite{strozzi-fastig-pop-2012} unfortunately used $B_z>0$, and would be better with the opposite
choice.  Moreover, the better confinement with an imposed $B_\phi$ is promising
for self-generated field approaches, which usually give rise to a $B_\phi$.

\begin{figure}
  \centering
  \includegraphics[height=6cm]{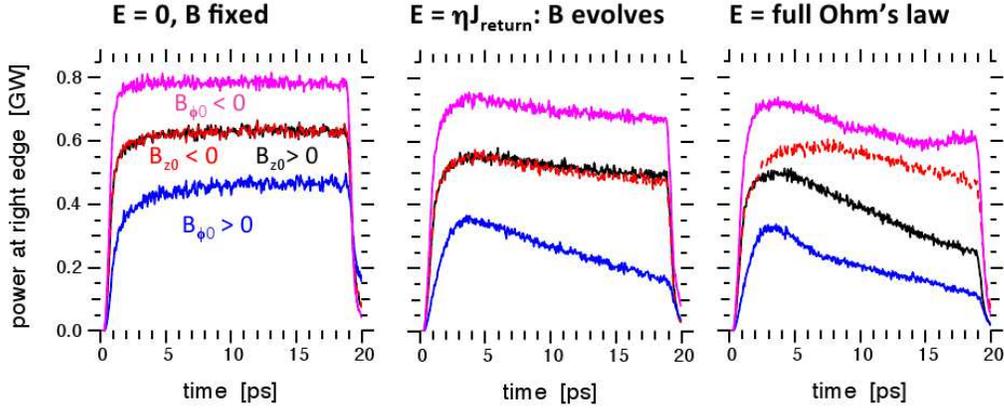}    
  \caption{Fast electron power reaching right edge in simplified Zuma-only
    simulations with initial, ``thin'' magnetic pipes as in Fig.\ \ref{fig:pipezh}.}
  \label{fig:pipezuma}
\end{figure}

Simplified Zuma-only runs (no coupling to Hydra) were performed to study
magnetic pipes of different field orientation. They indicate that the different coupling for the two signs of $B_z$ results from different magnetic field evolution, which occurs only
when non-resistive terms are included in Ohm's law. These runs used a uniform DT plasma of $\rho=10$ g/cm$^3$ and 100 eV, and a
divergent electron source with $\Delta\theta=90^\circ$. Figure
\ref{fig:pipezuma} depicts the power of fast
electrons reaching the right edge, inside the pipe radius, and counting only 1.3
MeV of kinetic energy per electron (the maximum deposited in a DT hot spot of optimal depth
$\rho \Delta z=$ 1.2 g/cm$^3$). For both the left panel ($\mathbf{E}=0$, $\mathbf{B}=$
constant) and center panel ($\mathbf{E}=\eta \mathbf{J}_e$, $\mathbf{B}$ evolves), the
ordering is as expected from orbits. However, when a more extensive Ohm's law
is used, the two signs of $B_z$ perform differently, with $B_z<0$ coupling
better than $B_z>0$. Work is ongoing to elucidate the difference in field dynamics.

Magnetic field generation by laser-driven coils \cite{daido-bfield-prl-1986} was
demonstrated experimentally in the 1980s, and has recently been suggested as a
divergence mitigation approach for fast ignition
\cite{fujioka-gekko-ppcf-2012}. In this scheme, a coil connects two plates, one
of which is struck by a kJ-class, long pulse ($\sim$ ns) laser. This generates hot ($\sim$ 10 keV) electrons by resonance absorption, which reach
the other plate and set up a potential difference between the two plates.  A
large, transient current flows through the coil and induces a large magnetic
field. Experiments at GEKKO-XII \cite{fujioka-gekko-ppcf-2012} have recently
made fields of 10 MG. Ref.\ \cite{courtois-coil-jap-2005} reports smaller-scale experiments at the VULCAN laser which generated fields of 0.1 MG, as well as a simple model of the system.  This approach eliminates the need to create large fields
in an implosion, and allows one to consider novel field configurations.  For
instance, the field could peak on the short-pulse laser side of the re-entrant
cone, so that the mirroring effect pushes fast electrons \textit{toward} the
fuel. The field is generated over a few ns, which is only a fraction of the
duration of a typical fast ignition implosion. It can therefore be timed such
that the implosion modifies the field in a limited way. Realistic rad-hydro-MHD
modeling is needed to validate the laser-driven coil approach in specific
geometries, and when coupled to specific implosions.

\section{Conclusions}
Although Fast Ignition ICF was proposed nearly 18 years ago, in-depth simulation studies of the fast electron transport aspect of the problem have only really been done over the past 7 years or so (Section \ref{ignscale}).  In this review we have looked at the following aspects : 

{\bf (i) Basic Physics}:  The fundamental physics of fast electron transport is thought to be well understood.  Where doubts do exist is in the ability to calculate the properties of dense matter, and collisional processes to very high accuracy.  Another aspect which hasn't been fully explored is the extent to which kinetic micro-instabilties affect transport, and whether significant corrections could be included in hybrid models.  These aspects need to be addressed, but it is thought that these are unlikely to radically alter the outlook for Fast Ignition.

{\bf (ii) Simulation Methods}:  In Section \ref{simmeth} it was shown that there are now a number of different simulations models and a large number of codes.  These are nearly all extendable models that can include a wide variety of physics, and most have some ``hybrid'' character.  There are a number of codes that include hydrodynamic motion, radiation transport, and thermonuclear burn --- representing a true merging of the hybrid model with standard ICF rad-hydro models.   Thus there are codes that can tackle most of the full problem.  This is subject to the inclusion of a realistic fast electron source and the approximations inherent in ``hybridization''.  Nonetheless, as this review shows, this is sufficient to use computer simulation to make a first assessment of the viability of the Fast Ignition concept.

{\bf (iii) Ignition-scale Calculations } : In recent years there have been a number of studies (see Section \ref{ignscale}) which have attempted to use the aforementioned numerical models to assess the viability of Fast Ignition at (or close to) ignition scale.  This has ranged from rather idealized calculations through to calculations which have attempted a good degree of realism.  These calculations have shown that achieving modest or ``attractive'' ignition energies ($<$100~kJ of laser energy) is difficult.  The good laser-to-fast electron conversion efficiencies that are seen in certain PIC calculations (40--50\%) are certainly beneficial, but the degradation of the coupling into the hot spot due to high fast electron divergence is considerable.

{\bf (iv) Controlling Transport} : As the divergence problem is a serious one, a number of studies have looked at modifications to the FI scheme that will allow the flow of fast electrons to be sufficiently controlled as to improve the coupling into the hot spot.  In Section \ref{control} a number of interesting possibilities were discussed including the exploitation of resistivity gradients, axial magnetic fields, and double cones.  This is a highly active area, which may eventually produce an attractive solution to the divergence problem.

The problem of simulating fast electron transport in Fast Igntion is not a trivial problem.  Nontheless, as this review shows, there has been considerable progress over the last seven years or so.  The main challenge that is now being faced is how one can modify transport so that relatively modest ignition energies can be achieved given a divergent fast electron source.  A number of concepts have been proposed, and thus it is still entirely possible that an attractive Fast Igntion point design will emerge in the coming years.

\section{Acknowledgements}
APLR and DJS (Lead Coordinators of this paper) would like to thank all the contributors for their effort, dedication, and patience.  APLR and DJS also acknowledge their particular contributions : J.R. Davies -- Drag and Scattering, \S\ref{drag+scattering}, L. Gremillet -- Beam-Plasma Instabilities \S \ref{bpinstab}, M. Sherlock -- VFP Codes \S \ref{vfpcodes}, A.A. Solodov -- Hybrid PIC Codes \S \ref{hybridpiccodes} \& Resistive Guiding at High Energy Scales \S \ref{resguiding_hienergy}, J. Honrubia -- Review of Ignition Scale Calculations \S \ref{ignscale}, T. Johzaki -- Double Cone Approach \S \ref{doublecone}, DJS --  Background Plasma Physics \S \ref{bkgdphysics} \& Axial Magnetic Field Approach \S \ref{axial}.  Remaining text prepared by APLR with help from RJK. We acknowledge the many contributions to this field by our late colleague, M.G. Haines.
\vspace{2em}
\bibliographystyle{jphysicsB}
\bibliography{fet4fi_bibl_v4}

\end{document}